\input epsf.tex
\epsfclipon
 \documentclass[11pt,a4paper,amsart]{article}
 \usepackage{jheppub}
 \usepackage{float}
 \usepackage{mathrsfs}
\usepackage{cancel}
\usepackage{makecell}
\usepackage{graphicx}

\usepackage{subcaption}
\usepackage[toc,page]{appendix}

\usepackage{hyperref} 
\usepackage{multirow,tabularx}

\usepackage{mathtools}
\usepackage{color}

\usepackage{epsfig}
\usepackage{epstopdf}
\usepackage{epsf}
\input{epsf.sty}
\usepackage{graphicx,amsmath,amssymb}

\usepackage{graphicx}
\usepackage{epsfig,multicol}
\usepackage{multirow}
\allowdisplaybreaks

\usepackage{tikz} 

\usepackage{bbm,bm,amsmath,amssymb}
\usepackage{hyperref}
\usepackage{tikz}
\usetikzlibrary{shapes.geometric, arrows}
\usetikzlibrary{shapes,positioning}
\tikzstyle{startstop} = [rectangle, rounded corners, minimum width=2cm, minimum height=1cm,text centered, text width=4.4cm, draw=black, fill=white!30]
\tikzstyle{io} = [trapezium, trapezium left angle=70, trapezium right angle=110, minimum width=2cm, minimum height=1cm, text centered, draw=black]
\tikzstyle{process} = [rectangle, minimum width=1.5cm, minimum height=1cm, text centered,, text width=4cm, draw=black, fill=white!30]
\tikzstyle{decision} = [diamond, minimum width=3cm, minimum height=1cm, text centered, text width=4.2cm, draw=black, fill=green!30]
\tikzstyle{arrow} = [thick,->,>=stealth]

\usepackage{longtable}
 \usepackage{accents}
\newlength{\dhatheight}

 \usepackage{graphicx} 
 \usepackage{array}
\newcommand{\Su}{\color{blue}}

\usepackage[normalem]{ulem}

\usepackage{comment}
\def\be{\begin{equation}}
\def\ee{\end{equation}}
\def\figs/B{B}
\def\bea{\begin{eqnarray}}
\def\eea{\end{eqnarray}}
\def\bg{\begin{eqnarray}}
\def\nd{\end{eqnarray}}

\def\log{{\rm log}}

\def\be{\begin{equation}}
\def\ee{\end{equation}}

\def\doi{http://doi.org}

\usepackage{physics}

\usepackage{tikz}
\usetikzlibrary{arrows.meta}
\usetikzlibrary{angles,quotes} 
\usetikzlibrary{decorations.markings,arrows.meta}
\tikzset{>=latex} 

\tikzset{
  midarr/.style={decoration={markings,mark=at position #1 with {\arrow{stealth}}},postaction={decorate}},
  midarr/.default=0.5
}
\colorlet{xcol}{blue!70!black}

\title{A Computational Companion to  Transient de Sitter and Quasi de Sitter States in ${\rm SO}(32)$ and ${\rm E}_8 \times {\rm E}_8$ Heterotic String Theories \textbf{I}: Formalisms}
\author{ {\Su Archana Maji} 	\vskip.03in 
	 Department of Physics, Indian Institute of Technology Bombay, Mumbai 400076, India\\
        {\tt 
        archana${}_-$phy@iitb.ac.in
       }  }

\date{\today}

\abstract{We construct four-dimensional de Sitter space as an excited state, rather than as a vacuum configuration, in type IIB, heterotic $SO(32)$, and heterotic $\mathrm{E}_8 \times \mathrm{E}_8$ string theories. This framework provides a mechanism to evade vacuum-based no-go theorems for de Sitter solutions in string theory. Starting from a generic M-theory configuration, we obtain de Sitter isometry in the dual string theories through appropriate dynamical duality sequences in the late-time limit. The excited state, identified as a Glauber--Sudarshan state, is constructed as the expectation value of the metric operator in M-theory using path-integral techniques. 

We further analyze the conditions required for the existence of a well-defined effective field theory description and show that these conditions are equivalent to the null energy condition for a $(3+1)$-dimensional FLRW cosmology. Finally, we investigate constraints arising from axionic cosmology and demonstrate how the time-dependent solutions are modified when experimental bounds on the axionic coupling constant are taken into account.

\vspace{5pt}
\noindent
This article serves as a computational companion to sections 3 and 4 of the paper 

\noindent
arXiv:2511.03798 [hep-th], where we present the detailed intermediate steps underlying the analysis in those sections.       
} 

\begin{document}

\maketitle
\section{Introduction}
The construction of de Sitter vacua within a consistent quantum theory of gravity has long remained a challenging problem. For instance, the Gibbons, Maldacena and Nuñez no-go theorem \cite{Gibbons:1984kp,Maldacena:2000mw,Gibbons_2003} has demonstrated the absence of well-defined de Sitter compactifications for a large class of supergravity theories. Following this result, numerous ingenious attempts were made to circumvent the no-go conditions~\cite{balasubramanian2005systematics,kachru2003sitter,rummel2012sufficient,cicoli2016sitter,westphal2007sitter,hertzberg2007inflationary,covi2008sitter,wrase2010classical,shiu2011stability}. Despite these efforts, a parallel set of studies~
\cite{green2012constraints,evandasgupta2014,heliudsonmirsuddho, Kutasov:2015eba} have continued to highlight persistent obstructions in constructing de Sitter compactifications within string/M-theory. It is therefore fair to say that the existence of a fully controlled classical de Sitter vacuum in these theories remains an open question~\cite{palti2019swampland}.
In this work, we take a different perspective, realising de Sitter not as a vacuum configuration but as an excited state over a time-independent supersymmetric Minkowski background. In particular, we construct de Sitter as a Glauber–Sudarshan state, which can be viewed as a statistical mixture of fundamental coherent states~\cite{Dasgupta:2025ypg}. Such a state can be constructed in M-theory, while demanding de Sitter isometry in the dual string theory. As discussed in detail in~\cite{Chakravarty_2024}, realising de Sitter in this manner resolves several conceptual issues. To mention a few, it allows for the implementation of an exact renormalisation group procedure, since the underlying spacetime is time-independent Minkowski and hence the notion of frequency remains well defined, unlike in a de Sitter background where modes are continuously redshifted. As a result, one can write down a low-energy Wilsonian effective action whose corresponding equations of motion yield the appropriate de Sitter solution. Furthermore, since the background Minkowski space is supersymmetric, zero-point energies are automatically canceled, thereby circumventing the issues associated with the zero-point energy problem.   

In M-theory, we perform a path integral analysis to evaluate the expectation value of the metric operator with respect to the aforementioned Glauber-Sudarshan state using perturbative techniques. Since the ultimate goal is to describe a universe undergoing an accelerated expansion—closely resembling our own—it becomes necessary to incorporate contributions from all orders in the perturbative series.
It is well known from quantum field theory that, except in rare cases, exact results for observables are difficult to obtain. One generally relies on perturbative analysis, which requires identifying a suitable expansion parameter that can be tuned to be small, allowing the observable of interest to be expressed as a power series in that parameter. Although perturbative methods have had notable success -- for instance, in the remarkable precision tests of the anomalous magnetic moment of the electron—they also come with shortcomings. In general, perturbative series are asymptotic with zero radius of convergence. There are two primary sources of such divergences: \textcolor{blue}{(i)} the factorial growth of the number of Feynman diagrams at higher orders in the perturbative expansion due to combinatorics~\cite{Dorigoni:2014hea}, and \textcolor{blue}{(ii)} UV and IR divergences of the so-called renormalon type, arising from the high- and low-momentum regions of phase-space integrals, respectively~\cite{Beneke_1999}.
In our analysis, we encounter divergences of the former kind. Importantly, the asymptotic nature of the perturbation series does not represent a fundamental obstacle to computing the observable of interest, but rather signals the necessity of incorporating non-perturbative contributions. These non-perturbative effects arise from non-trivial saddle points of the path integral, known as \textit{instantons}~\cite{Jentschura:2011zza}. The formal series obtained by adding instanton corrections to the original asymptotic expansion is known as a \textit{trans-series}~\cite{Marino:2023epd}. We then employ Borel-\'Ecalle resummation techniques to resum the trans-series and obtain physically meaningful results for the observables.

 The purpose of this article is to provide all intermediate steps underlying sections 3 and 4 of the paper \cite{Dasgupta:2025ypg}. The organization of the article is as follows. In Section \ref{dSspacePoincareCo}, we review some basic properties of de Sitter space in Poincaré coordinates (also referred to as flat slicing or planar coordinates). Section \ref{dSinVarSTheOrieS} presents three distinct dynamical duality sequences. Starting from a generic M-theory configuration, these sequences lead, in the late-time limit of flat slicing coordinates, to de Sitter configurations in type IIB, heterotic $SO(32)$, and heterotic $\mathrm{E}_8 \times \mathrm{E}_8$ string theory.
In Section \ref{sec:PIinMth}, we detail the path-integral approach for computing the expectation value of the metric operator with respect to the Glauber–Sudarshan state in M-theory. This computation results in a perturbative expansion which, being asymptotic in nature, necessitates Borel resummation, as discussed in Section \ref{BorelREsum}. Section \ref{sec:NEC} is devoted to the Null Energy Condition, where we explain how it effectively appears as a constraint required for the existence of a well-defined effective field theory description in type IIB, heterotic $SO(32)$, and heterotic $\mathrm{E}_8 \times \mathrm{E}_8$ string theories.
In Section \ref{sec:TCCconsTraints}, we obtain the Trans-Planckian Censorship Conjecture bound, which delineates the temporal regime over which the effective field theory description remains under control. Section \ref{sec:AxionsinE8Thoery} focuses on axions in the heterotic $\mathrm{E}_8 \times \mathrm{E}_8$ string theory. Finally, we provide a summary and discussion of the main results in Section \ref{sec:Summary&Con}. Appendices \ref{GS state}, \ref{FIntApproachPhi4}, \ref{DualitYRulES}, and \ref{sec:AppendixD}  contain technical material and supplementary computations required for the analyses presented in the main text.
\section*{Acknowledgements}
 I am deeply grateful to Keshav Dasgupta for suggesting that I work on the computational companion, and for clarifying numerous doubts while providing constant support throughout the project. I would like to acknowledge the helpful discussions and generous support of Pichai Ramadevi. I also thank Joydeep Chakravarty for several insightful discussions during an academic visit to McGill University, as well as for his constructive feedback on an earlier version of this manuscript. I appreciate the warm hospitality extended to me by Keshav Dasgupta and McGill University during my visit, where a portion of this work was carried out. This work is supported in part by the Prime Minister’s Research Fellowship, provided by the Ministry of Education, Government of India.
\section{de Sitter space in Poincare coordinates}
\label{dSspacePoincareCo}
 The Einstein's equation of motion in the presence of energy momentum tensor ($\mathrm{T}_{\mu\nu}$) and cosmological constant is given as 
 \begin{align}
      \mathrm{R}_{\mu\nu}-\frac{1}{2}g_{\mu\nu}\mathrm{R}+\Lambda g_{\mu\nu}=8\pi \mathrm{G}_{\mathrm{N}}\mathrm{T}_{\mu\nu}.\label{EEofmoTion}
 \end{align}
 de Sitter space is a solution to the above equation of motion for vanishing $\mathrm{T}_{\mu\nu}$ and $\Lambda>0$. An alternate way of realising de Sitter space is from the embedding space point of view: a $d$ dimensional de Sitter space of length scale $l$ is a set of points
 \begin{align}
(\mathrm{X}^0,\mathrm{X}^1,\cdots,\mathrm{X}^d)
\end{align}
in a flat $d+1$ dimensional Minkowski spacetime $\mathbb{R}^{1,d}$ satisfying 
\begin{align}
        -(\mathrm{X}^0)^2+(\mathrm{X}^1)^2+\cdots+(\mathrm{X}^d)^2&=l^2.\label{ConstdeSitter}
\end{align}
 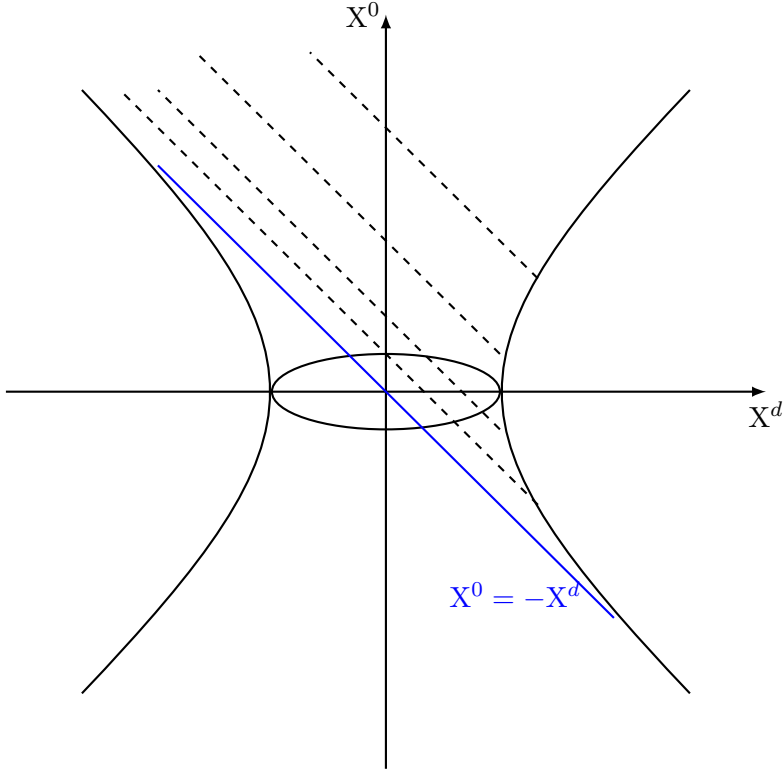
\begin{figure}[h!]
    \centering
 \begin{tikzpicture}
 \draw[-latex,thick] (-5,0) --(5,0);
 \draw[-latex,thick] (0,-5) --(0,5);
 \draw[thick] (0,0) ellipse (1.5 and 0.5);
 \draw[thick] (-4,4) .. controls (-0.7,0.5) and (-0.7,-0.5)  .. (-4,-4);
 \draw[thick] (4,4) .. controls (0.7,0.5) and (0.7,-0.5)  .. (4,-4);
 \draw[blue, thick] (3,-3) --(-3,3);
 \node at (5,-0.3) {$\mathrm{X}^d$};
 \node at (-0.3,5) {$\mathrm{X}^0$};
 \node at (1.7,-2.7) {$\textcolor{blue}{\mathrm{X}^0=-\mathrm{X}^d}$};
 \draw[dashed, thick] (2,-1.5) --(-3.5,4);
 \draw[dashed, thick] (1.5,-0.5) -- (-3,4);
 \draw[dashed, thick] (1.5,0.5) -- (-2.5,4.5);
 \draw[dashed, thick] (2,1.5) -- (-1,4.5);
 \end{tikzpicture}
    \caption{The region of embedding space $d\mathrm{S}^d$ covered by the poincare coordinates $\mathrm{X}^0\geq -\mathrm{X}^d$ (shown using dashed lines), the constant $\mathrm{X}^0$ slices are $\mathrm{S}^{d-1}$.}
    \label{FlatSlicing}
\end{figure}
At a given value of $\mathrm{X}^0$, the $d$ coordinates $(\mathrm{X}^1,\cdots, \mathrm{X}^d)$ form a sphere $\mathrm{S}^{d-1}$ 
\begin{align}
    (\mathrm{X}^1)^2+(\mathrm{X}^2)^2+\cdots+(\mathrm{X}^d)^2&=(\sqrt{l^2+(\mathrm{X}^0)^2})^2
\end{align}
of radius $\sqrt{l^2+(\mathrm{X}^0)^2}$. As the coordinate $\mathrm{X}^0$ goes from $-\infty$ to $\infty$, the radius of the sphere starts out infinitely large, shrinks to a minimum value $l$ at $\mathrm{X}^0=0$ and then expands again to infinity.  Thus, de Sitter space has the topology of $\mathbb{R}\times \mathrm{S}^{d-1}.$   It can be parameterised using different coordinate systems: static coordinates, global coordinates, flat slicing coordinates, Kruskal  coordinates etc. We will be working with the flat slicing  (also known as  poincare coordinates) in the subsequent sections. So, we elaborate on this in order to make the discussion self contained. A $d$ dimensional de Sitter space can be parameterised with poincare coordinates $(\tau,x^1,...,x^{d-1})$ as
\begin{equation}
\begin{aligned}
\mathrm{X}^0&=l\sinh\left( \frac{\tau}{l}\right)+\frac{(x^i)^2}{2l}e^{\tau/l},\\
    \mathrm{X}^i&=x^i e^{\tau/l},~~~~~(i=1,...,d-1)\\
    \mathrm{X}^d&=l\cosh\left( \frac{\tau}{l}\right)-\frac{(x^i)^2}{2l}e^{\tau/l},\label{FSlicingPar}
\end{aligned}
\end{equation}
where $\tau\in (-\infty,\infty)$ and $x^i\in (-\infty,\infty)$. 
The metric of de Sitter space obtained upon substituting (\ref{FSlicingPar}) in the metric of $d+1$ dimensional embedding Minkowski spacetime  is given by
\begin{align}
    ds^2&=-(d\mathrm{X}^0)^2+(d\mathrm{X}^1)^2+\cdots+(d\mathrm{X}^d)^2\nonumber\\
    &=-d\tau^2+e^{2\tau/l}((dx^1)^2+\cdots+(dx^{d-1})^2),\label{PlanarMet}
\end{align}
where $\sum_{i=1}^{d-1}(dx^i)^2$ is the flat metric on $\mathbb{R}^{d-1}.$  For a four dimensional de Sitter space the flat slicing metric becomes 
\begin{align}
ds^2_{d\mathrm{S}^4}&=-d\tau^2+e^{2\tau/l}((dx^1)^2+(dx^2)^2+(dx^3)^2).
\end{align}
The above metric along with the Einstein's equation of motion (\ref{EEofmoTion}) (with $\mathrm{T}_{\mu\nu}$ set equal to zero) allows us to fix the de Sitter length scale $l$ in terms of the cosmological constant: 
\begin{align}
\mathrm{R}_{11}-\frac{g_{11}}{2}\mathrm{R}+\Lambda g_{11}&=0\nonumber\\
\implies    \frac{3}{l^2}e^{2\tau/l}-\frac{e^{2\tau/l}}{2}\left(\frac{12}{l^2}\right)+\Lambda (e^{2\tau/l})&=0\nonumber\\
     \implies l=\sqrt{\frac{3}{\Lambda}}&.
    \end{align}
Expressing $\tau$ in terms of the conformal time $t= -l~e^{-\tau/l}$ give rise to the following metric of four dimensional de Sitter space in flat slicing 
\begin{align}
ds^{2}_{\mathrm{dS}}&=\frac{1}{l^{-2} t^2}(-dt^2+(dx^1)^2+(dx^2)^2+(dx^3)^2)\nonumber\\
&=\frac{3}{\Lambda t^2}(-dt^2+(dx^1)^2+(dx^2)^2+(dx^3)^2),\label{dSFlAtSlic}
\end{align}
where $t$  lies in the range  $t\in (-\infty ,0)$. From (\ref{FSlicingPar}) we note that 
\begin{align}
\mathrm{X}^0+\mathrm{X}^d&=le^{\tau/l}\geq 0 \implies \mathrm{X}^0 \ge -\mathrm{X}^d.
\end{align}
That is, unlike global coordinates, the flat slicing coordinates cover only half of the entire de Sitter space as shown in Figure \ref{FlatSlicing}.  
Similar coordinates can be chosen to cover the other half of the de Sitter space. 
\section{de Sitter isometry in various string theories: end-point of dynamical duality sequences}
\label{dSinVarSTheOrieS}
In this section, we elaborate on three distinct duality sequences to obtain a four dimensional de Sitter within type IIB, heterotic $SO(32)$ and heterotic $\mathrm{E}_8\times \mathrm{E}_8$ string theories. As we will see, starting from a generic M theory metric configuration, the duality sequence proceeds dynamically to give rise to de Sitter as the endpoint of that sequence in the late time limit i.e. in the $t\rightarrow 0^-$ limit of the Poincare patch. We emphasize that the metric configuration is to be understood as obtained by taking an expectation value of the metric operator with respect to a Glauber-Sudarshan state. In particular, if we denote the state as $|\sigma \rangle$ then the metric can be expressed as 
\begin{align}
    ds^2&=\langle \mathbf{g}_{\mathrm{AB}}\rangle_{\sigma} ~d\mathrm{Y}^{\mathrm{A}} d\mathrm{Y}^{\mathrm{B}},\label{ExpValueOfMetOp}
\end{align}
where $\mathrm{A}$ runs over the eleven (ten) coordinates in M (string) theory. In Appendix \ref{GS state} we have provided a brief discussion on the well known coherent states and outlined how the Glauber-Sudarshan states differ from them. For more details on the Glauber-Sudarshan state we refer the reader to \cite{Chakravarty_2024}. 
\subsection{de Sitter isometry in Type \texorpdfstring{$\mathrm{IIB}$}{IIB}  string theory} 
\label{typeIIBdSisom}

We begin by considering the following metric configuration of type $\mathrm{IIB}$ string theory 
\begin{align}
    ds^2_{\mathrm{IIB}} &=\frac{ {a}^{2}(t)}{ {\rm H}^2(y)}\left(-dt^2 +  \left(dx^1\right)^2 +\left(dx^2\right)^2 + \left(dx^3\right)^2\right) + {\rm H}^2(y)\mathrm{F}_{1}(t)g_{\alpha\beta}(y)dy^{\alpha}dy^{\beta}\nonumber\\
    &~~~~~~+{\rm H}^2(y)\mathrm{F}_{2}(t)g_{mn}(y) dy^m dy^n ,\label{tyPeIIBMetrIc}
\end{align}
with the type $\mathrm{IIB}$ coupling constant $g_{b}=1$, i.e. vanishing axio-dilaton\footnote{Axio-dilaton $\tau:=C_{0}+i e^{-\varphi}$ is a combination of the \textbf{R-R} field $C_{0}$ (viz. axion) and the dilaton field $\varphi$.
The type $\mathrm{IIB}$ coupling constant can be expressed in terms of the axio-dilaton field as $g_{b}=1/\mathrm{Im}(\tau)$. By vanishing axio-dilaton we mean $C_{0}=0=\varphi$ and therefore $g_{b}=1.$ 
}. Such a configuration is supported at a constant coupling point in $\mathrm{F}$ theory.

Here $a(t)$ and $\mathrm{H}(y)$ denote the scale factor and the warp factor respectively.  The compact six dimensional internal space is taken to be the product of a 2-manifold, $\mathcal{M}_{2}$ and a 4-manifold, $\mathcal{M}_{4}$. Both of these manifolds are non-k$\ddot{\text{a}}$hler and non-complex in nature. The corresponding coordinates are denoted as $y\equiv (y^{\alpha},y^{m})$ with $y^{\alpha}\in \mathcal{M}_{2}$ and $y^{m}\in \mathcal{M}_{4}$. The unwarped metric components for the two aforementioned manifolds are $g_{\alpha \beta}(y)$ and $g_{mn}(y)$. Temporal dependence of the internal space is being depicted by the warp factors $\mathrm{F}_{1}(t)$ and $\mathrm{F}_{2}(t)$. In this theory we can turn on any fluxes whose components can be along any of the internal directions. In equation (\ref{tyPeIIBMetrIc}), the following choice of  the scale factor
\begin{align}
  {a^{2}(t)}=\frac{1}{\Lambda t^2} ~~~\text{where}~~ -\infty \le t<0, \label{TheScaLeFactOr}  
\end{align}
 will allow us to obtain a four dimensional expanding space with de Sitter isometry from the compactification of type $\mathrm{IIB}$ string theory. Note that, we have absorbed the factor of 3, present in the four dimensional de Sitter metric (\ref{dSFlAtSlic}), into the definition of the warp factor $\mathrm{H}(y)$.  Further, one has to ensure that the time dependent warp factors satisfy the following constraint 
\begin{align}
    \mathrm{F}_{1}(t)\mathrm{F}^{2}_{2}(t)=1.\label{eq:IIBconst}
\end{align}
 The above constraint arises by demanding time independence of the volume of the internal space which is a necessary condition for the time independence of the four-dimensional Newton's constant\footnote{The four dimensional Newton's constant $\mathbb{G}^{(4)}$ is given in terms of the ten dimensional Newton's constant $\mathbb{G}^{(10)}$ and the volume $\mathcal{V}_{6}$ of the six dimensional internal space as $\mathbb{G}^{(4)}=\mathbb{G}^{(10)}/\mathcal{V}_{6}$. Since $\mathbb{G}^{(10)}$ is time independent to begin with, time independence of $\mathbb{G}^{(4)}$ necessitates $\mathcal{V}_{6}$ to be time independent.}. 

The construction of a Glauber-Sudarshan state  seems to be technically challenging from type $\mathrm{IIB}$ string theory \cite{Bernardo:2021rul,Brahma:2020tak,brahma2021four}. The way out is to dualize to $\mathrm{M}$-theory and then construct the state therein. A reason for the necessity of such an uplift is that there are too many degrees of freedom in type $\mathrm{IIB}$ theory as compared to those in $\mathrm{M}$ theory which makes the analysis easier when performed in the latter theory. It is evident from the duality rules, as elaborated in appendix \ref{DualitYRulES}, that most of the fluxes of $\mathrm{IIB}$ theory become four-form $\mathrm{G}$ fluxes in $\mathrm{M}$ theory. 

We now obtain the corresponding metric configuration of $\rm{M}$ theory. In order to do so we first perform a $\mathrm{T}$ duality (see Appendix \ref{DualitYRulES} for the duality rules) along $x_{3}$ direction of type $\mathrm{IIB}$ theory (\ref{tyPeIIBMetrIc}) which gives type $\mathrm{IIA}$ theory with the following metric configuration
\begin{equation}
\begin{aligned}
ds^{2}_{\mathrm{IIA}}&=\frac{a^{2}(t)}{{\rm H}^2(y)}\left(-dt^2 + \sum_{i = 1}^2 \left(dx^i\right)^2 \right) +\frac{\mathrm{H}^{2}(y)}{a^{2}(t)}(dx^{3})^2 + {\rm H}^2(y)\mathrm{F}_{1}(t)g_{\alpha\beta}dy^{\alpha}dy^{\beta}\\
    &~~~~~~~~~~~~+{\rm H}^2(y)\mathrm{F}_{2}(t)g_{mn} dy^m dy^n , \label{Type IIA MetRIc}
\end{aligned}
\end{equation}
 The type IIA coupling constant equals
\begin{align}
    {g}_{s}=\frac{\mathrm{H}(y)}{a(t)}.\label{TyPeIIA coUplInG}
\end{align}
Demanding de Sitter isometry (\ref{TheScaLeFactOr}) in type $\mathrm{IIB}$ theory (\ref{tyPeIIBMetrIc}) implies that the type $\mathrm{IIA}$ coupling constant becomes\footnote{We will see that if we demand de-Sitter isometry in either heterotic $\mathrm{SO(32)}$ or heterotic $\mathrm{E}_{8}\times\mathrm{E}_{8}$ string theory,  the type $\mathrm{IIA}$ string coupling constant will get modified.}
\begin{align}
    g_{s}&=\mathrm{H}(y)|t|\sqrt{\Lambda}.\label{IIAcouPlingdeSitterIIB}
\end{align}
 The metric of dual $\mathrm{M}$ theory has the following structure
\begin{align}
ds^{2}_{\mathrm{M}-{\rm th}}&=g_{s}^{-8/3}\left(-dt^2 + \sum_{i = 1}^2 \left({dx^i}\right)^2 \right) +g_{s}^{-2/3} {\rm H}^2(y)\left[\mathrm{F}_{1}(t)g_{\alpha\beta}dy^{\alpha}dy^{\beta}+\mathrm{F}_{2}(t)g_{mn} dy^m dy^n \right]\nonumber\\
&~~~~~~~~+g_{s}^{4/3}\delta_{ab}d\omega^{a}d\omega^{b},\label{MthEorYmeTric}
\end{align}
where $\omega^{a}\equiv (x^3,x^{11})$ are the coordinates of $\mathrm{M}$-theory \textit{torus} ${\mathbb{T}_{3,11}^{2}}/{\mathcal{G}}$, $\mathcal{G}$ is a group action without any fixed points. An important thing to note here is that the $\mathrm{M}$ theory torus is of finite size. This is precisely because a well defined effective field theory description of the underlying $\mathrm{M}$ theory action requires the type $\mathrm{IIA}$ coupling constant (\ref{IIAcouPlingdeSitterIIB}) to satisfy an upper bound $\frac {g_{s}}{\mathrm{H}(y)}<1$. This in turn implies that the de Sitter isometry in type $\mathrm{IIB}$ theory exists only for a finite time domain 
\begin{align}
    -\frac{1}{\sqrt{\Lambda}}<t<0.
\end{align}
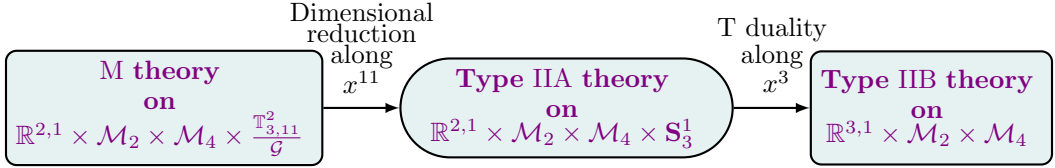
\begin{figure}[h!]
\centering
\begin{tikzpicture}[font=\small,thick]
\node[draw, rounded corners,fill=teal!10,
    minimum width=1.5cm,
    minimum height=1cm](block1) { \textcolor{violet}{
$\substack{ \textstyle \mathrm{M}~\textbf{theory} \\ \\\textstyle \textbf{on}\\ \textstyle\textbf{$\mathbb{R}^{2,1}\times\mathcal{ M}_{2}\times \mathcal{M}_{4}\times \frac{\mathbb{T}^{2}_{3,11}}{\mathcal{G}} $}}$
}};
\node[draw, rounded rectangle, fill=teal!10,
    right=of block1,
    minimum width=1.5cm,
    minimum height=1cm] (block2) { \textcolor{violet}{ 
$\substack{\textstyle\textbf{Type}~\mathrm{IIA}~\textbf{theory}\\ \\ \textstyle\textbf{on} \\\textstyle \mathbb{R}^{2,1}\times  \mathcal{M}_{2}\times \mathcal{M}_{4}\times {\mathbf{S}^{1}_{3}}}$
}};
\node[draw, rounded corners, fill=teal!10,
    right =of block2,
    minimum width=1.5cm,
    minimum height=1.5cm] (block3) { \textcolor{violet}{
$\substack{\textstyle\textbf{Type}~\mathrm{IIB}~\textbf{theory}\\ \\ \textstyle \textbf{on} \\\textstyle \mathbb{R}^{3,1}\times \mathcal{M}_{2}\times \mathcal{M}_{4}}$
     }};
\draw[-latex] (block1) edge (block2)
    (block2) edge (block3);
\node at (2.6,.8) {${\substack{\textstyle\text{Dimensional}\\ \textstyle \text{reduction} \\ \textstyle \text{along}\\\textstyle x^{11} }}$} ; 
\node at (8.0,0.7) { $\substack{\textstyle\mathrm{T}~\text{duality}\\ \textstyle \text{along}\\\textstyle~x^3}$};
\end{tikzpicture}
    \caption{Duality sequence to go from the $\mathrm{M}$ theory configuration (\ref{MthEorYmeTric}) to type $\mathrm{IIB}$ string theory (\ref{tyPeIIBMetrIc}) having de Sitter isometry. Here $\mathcal{G}$ denotes some group action without  any fixed points.} 
    \label{Type IIB tree diagram}
\end{figure}
Starting with this $\mathrm{M}$ theory set-up we can reach type $\mathrm{IIB}$ theory \textit{dynamically}. To implement this it is important to specify the scalings of the time dependent warp factors. We propose the following generic scalings
\begin{equation}
\begin{aligned}
    &\mathrm{F}_{1}(t)\equiv\mathrm{F}_{1}\left(\frac{g_{s}}{\mathrm{H}(y)}\right)=\sum_{k=0}^{\infty}{\rm A}_{k}\left(\frac{g_{s}}{\mathrm{H}(y) }\right)^{\beta_{o}+\frac{2k}{3}},\\
    &\mathrm{F}_{2}(t)\equiv \mathrm{F}_{2}\left(\frac{g_{s}}{\mathrm{H}(y)}\right)=\sum_{k=0}^{\infty}{\rm B}_{k}\left(\frac{g_{s}}{\mathrm{H}(y) }\right)^{\alpha_{o}+\frac{2k}{3}} \label{ScalingOfFi's}
\end{aligned}
\end{equation}
where $\alpha_{o}$, $\beta_{o}$ are the dominant scalings and $\mathrm{A}_{k}$, $\mathrm{B}_{k}$ are integers. Both $\alpha_{o}~\mathrm{and}~\beta_{o}$ has to satisfy the upper bound
\begin{align}
    \alpha_{o}<+\frac{2}{3},~~~\beta_{o}<+\frac{2}{3}\label{UbdofdomFactor}
\end{align}
so that only the $\mathrm{M}$ theory torus along 3,11  shrinks to zero size
taking us to the dual type $\mathrm{IIB}$ configuration (\ref{tyPeIIBMetrIc}). In fact, for the case at consideration we want the type $\mathrm{IIB}$ configuration to be the final stage of our duality sequence having de Sitter isometry.  This is possible upon letting the time dependent warp factors $\mathrm{F}_{1}(t)$ and $\mathrm{F}_{2}(t)$ to approach 1 as the conformal time $t$ tends to $0^{-}$ in the late time limit of flat slicing de Sitter space. The duality sequence is pictorially depicted in \textbf{Figure} \ref{Type IIB tree diagram}.
\subsection{de Sitter isometry in heterotic \texorpdfstring{$\mathrm{SO}(32)$}{SO(32)} string theory}\label{dSinHetSO(32)}
\begin{figure}[h!]
\centering
\begin{tikzpicture}[scale=.85,font=\small,thick]
\node at (-5.0,0) [draw, rounded corners,fill=teal!10,
    minimum width=4.5cm,
    minimum height=1cm](block1) { \textcolor{violet}{
$\substack{ \textstyle \mathrm{M}~\textbf{theory} \\ \\\textstyle \textbf{on}\\ \textstyle\textbf{$\mathbb{R}^{2,1}\times \frac{\mathbb{T}_{4,5}^{2}}{\mathcal{I}_{4}\mathcal{I}_{5}}\times \mathcal{M}_{4}\times \frac{\mathbb{T}^{2}_{3,11}}{\mathcal{I}_{3}\mathcal{I}_{11}}$}}$
}};
\node at (5.0,0) [draw, rounded rectangle, fill=teal!10,
    minimum width=2.5cm,
    minimum height=1cm] (block2) { \textcolor{violet}{ 
$\substack{\textstyle\textbf{Type}~\mathrm{IIA}~\textbf{theory}\\ \\ \textstyle\textbf{on} \\\textstyle \mathbb{R}^{2,1}\times  \frac{\mathbb{T}_{4,5}^{2}}{\mathcal{I}_{4}\mathcal{I}_{5}}\times \mathcal{M}_{4}\times \frac{\mathbf{S}^{1}_{3}}{\Omega\mathcal{I}_{3}}}$
}};
\node at (0,-3.3) [draw, rounded corners, fill=teal!10,
    minimum width=2.5cm,
    minimum height=1.5cm] (block3) { \textcolor{violet}{
$\substack{\textstyle\textbf{Type}~\mathrm{IIB}~\textbf{theory}\\ \\ \textstyle \textbf{on} \\\textstyle \mathbb{R}^{3,1}\times  \frac{\mathbb{T}_{4,5}^{2}}{\Omega(-1)^{\mathrm{F_{L}}}\mathcal{I}_{4}\mathcal{I}_{5}}\times \mathcal{M}_{4}
}$}};
\node at (0,-6.6) [draw, rounded corners, fill=teal!10,
    minimum width=2.5cm,
    minimum height=1.5cm] (block4) { \textcolor{violet}{
$\substack{\textstyle\textbf{Type}~\mathrm{IIA}~\textbf{theory}\\ \\ \textstyle \textbf{on} \\\textstyle \mathbb{R}^{3,1}\times  \frac{\mathbf{S}^{1}_{5}}{\Omega\mathcal{I}_{5}}\times \left(\mathcal{M}_{4}\rtimes \widetilde{\mathbf{S}}_{4}^{1} \right)
}$}};
\node at (5.0, -9.9) [draw, rounded rectangle, fill=teal!10,
    minimum width=2.5cm,
    minimum height=1.5cm] (block5) { \textcolor{violet}{
$\substack{\textstyle\textbf{Type}~\mathrm{I}~\textbf{theory}\\ \\ \textstyle \textbf{on} \\\textstyle \mathbb{R}^{3,1}\times   \mathcal{M}_{4}\rtimes \left( \widetilde{\mathbf{S}}_{4}^{1}\times \widetilde{\mathbf{S}}_{5}^{1} \right)
}$}};
\node at (-5.0, -9.9) [draw, rounded corners, fill=teal!10,
    minimum width=2.5cm,
    minimum height=1.5cm] (block6) { \textcolor{violet}{
$\substack{\textstyle\textbf{Heterotic}~\mathrm{SO(32)}~\textbf{theory}\\ \\ \textstyle \textbf{on} \\ \textstyle \mathbb{R}^{3,1}\times   \mathcal{M}_{4}\rtimes \left( \widetilde{\mathbf{S}}_{4}^{1}\times \widetilde{\mathbf{S}}_{5}^{1} \right)
}$}};
\draw[-latex] 
(block1) -- (block2)
(block2) edge (block3)
    (block3) edge (block4)
    (block4) edge (block5)
    (block5) edge (block6);
\node at (0,.5) {${\substack{\textstyle\text{Dimensional reduction } \\ \textstyle \text{along $x^{11}$} }}$} ; 
\node at (3.5,-2.0) { $\substack{\textstyle\mathrm{T}~\text{duality}\\ \textstyle \text{along}~x^3}$};
\node at (1.5,-4.9) { $\substack{\textstyle\mathrm{T}~\text{duality}\\ \textstyle \text{along}~y^4}$};
\node at (3.9,-8.3) { $\substack{\textstyle\mathrm{T}~\text{duality}\\ \textstyle \text{along}~y^5}$};
\node at (0,-9.7) {$\mathrm{S}$ duality};
\end{tikzpicture}
    \caption{Duality sequence to go from the $\mathrm{M}$ theory configuration (\ref{MthEorYmeTric}) to heterotic $\mathrm{SO(32)}$ string theory (\ref{HeTerotiCSO(32)MetrIc})}
    \label{Het SO(32) tree diagram}
\end{figure}
Unlike the type $\mathrm{IIB}$ scenario, we now consider the $\mathrm{M}$ theory background to be an orbifold of the form\footnote{\label{Z2 orbifold}
The $\mathbb{Z}_{2}$ operation $\mathcal{I}_{a}$ on the coordinate $x^a$ of the circle $\mathrm{S}^{1}_{a}$ converts the circle into an interval via the identification of the coordinate $~x^a \sim -x^a$, and
$\mathrm{S}^{1}_{a}/\mathcal{I}_{a}$ is an orbifold which has two boundary points viz. the end points of the interval.}  $\mathbb{R}^{2,1}\times \frac{\mathbb{T}_{4,5}^{2}}{\mathcal{I}_{4}\mathcal{I}_{5}}\times \mathcal{M}_{4}\times \frac{\mathbb{T}^{2}_{3,11}}{\mathcal{I}_{3}\mathcal{I}_{11}}$.  Such an orbifold structure is necessary in order to reach the theory of our interest $-$ the heterotic $\mathrm{SO}(32)$ theory $-$ by utilising the duality symmetries of string theory. Below, we elaborate on the duality sequence and provide the metric configuration, fluxes and coupling constant of each intermediate theory for clarity.

    [$\bullet$] The orbifold action not being reflected at the level of the metric, the $\mathrm{M}$ theory metric remains same as in equation (\ref{MthEorYmeTric})
    \begin{align}
ds^{2}_{\mathrm{M}-{\rm th}}&=g_{s}^{-8/3}\left(-dt^2 + \sum_{i = 1}^2 \left({dx^i}\right)^2 \right) +g_{s}^{-2/3} {\rm H}^2(y)\left[\mathrm{F}_{1}(t)\delta_{\alpha\beta}dy^{\alpha}dy^{\beta}+\mathrm{F}_{2}(t)g_{mn} dy^m dy^n \right]\nonumber\\
&~~~~~~~~+g_{s}^{4/3}\delta_{ab}d\omega^{a}d\omega^{b},\label{MthHetSO32}
\end{align}
where $y^{\alpha}\in \frac{\mathbb T^2_{4,5}}{\mathcal{I}_4\mathcal{I}_5},~y^{m}\in \mathcal{M}_{4}$ and $w^a\equiv (x^3,x^{11})\in \frac{\mathbb{T}^2_{3,11}}{\mathcal{I}_3\mathcal{I}_{11}}$.
   However, there is one modification that we have incorporated viz. considered the unwarpped metric of the toroidal orbifold along $4,5$ to be $\delta_{\alpha\beta}$ as opposed to an arbitrary $g_{\alpha\beta}$\footnote{Reason being the underlying duality sequences to reach the heterotic $\mathrm{SO}(32)$ theory.}. Furthermore, the warp factor $\mathrm{H}(y)$ depends only on the coordinates $y^m$ of $\mathcal{M}_{4}$.

   [$\bullet$]  Upon dimensional reduction of 11th direction we get the type $\mathrm{IIA}$ theory with metric  
   \begin{equation}
\begin{aligned}
ds^{2}_{\mathrm{IIA}}&=\frac{a^{2}(t)}{{\rm H}^2(y)}\left(-dt^2 + \sum_{i = 1}^2 \left(dx^i\right)^2 \right) +\frac{\mathrm{H}^{2}(y)}{a^{2}(t)}(dx^{3})^2 + {\rm H}^2(y)\mathrm{F}_{1}(t)\delta_{\alpha\beta}dy^{\alpha}dy^{\beta}\\
    &~~~~~~~~~~~~+{\rm H}^2(y)\mathrm{F}_{2}(t)g_{mn} dy^m dy^n .
\end{aligned}
\end{equation}
 A $\mathrm{T}$ duality along $x^{3}$ of {{type} $\mathrm{IIA}$ theory} leads to type $\mathrm{IIB}$ theory having the metric configuration as in equation (\ref{tyPeIIBMetrIc}) (with $g_{\alpha\beta}$ replaced by $\delta_{\alpha\beta}$).  Note that, the solitonic structures of both type $\mathrm{IIA}$ and $\mathrm{IIB}$ backgrounds  as shown in \textbf{Figure} (\ref{Het SO(32) tree diagram}) differ from what we had in the previous case of realising de Sitter isometry in $\mathrm{IIB}$ theory (see \textbf{Figure} (\ref{Type IIB tree diagram})). In particular, shrinking the toroidal orbifold along $3,11$ of the $\mathrm{M}$ theory configuration to zero size has led to a type $\mathrm{IIB}$ orientifold $\mathbb{R}^{3,1}\times \frac{\mathbb{T}^2_{4,5}}{\Omega (-1)^{\mathrm{F}_{\mathrm{L}}}\mathcal{I}_{4}\mathcal{I}_{5}}\times \mathcal{M}_{4}$\footnote{$\Omega$ is the world-sheet parity transformation operator which reverses the orientation of the world sheet coordinate $$\Omega: \sigma \rightarrow -\sigma.$$ It exchanges the left and right moving modes of the worldsheet fields. Since the left moving and right moving modes of the world sheet fields of type IIA theory have opposite space time chirality, the $\mathrm{IIA}$ theory is chiral on world sheet. In other words, $\Omega$ is not a symmetry of type $\mathrm{IIA}$ theory. However, the simultaneous identification of the space time coordinate $x \sim -x$ (see footnote \ref{Z2 orbifold}) restores the symmetry of the theory. For type $\mathrm{IIB}$ theory both the left moving and right moving modes transform under the same space time chirality and hence $\Omega$ is a world sheet symmetry of type $\mathrm{IIB}$ theory. 
    
    The operator $(-1)^{\mathrm{F_{L}}}$ changes the sign of all the left moving Ramond sector states. The type $\mathrm{IIB}$ theory on $\mathbb{T}^{2}_{4,5}$ modded out by the $\mathbb{Z}_{2}$ operation $\Omega (-1)^{\mathrm{F_{L}}} \mathcal{I}_{4}\mathcal{I}_{5} $ is an orientifold and this is related to type $\mathrm{I}$ theory via $\mathrm{T}$ dualities \cite{Sen_1996}.}.  Furthermore, the type $\mathrm{IIA}$ coupling constant 
\begin{align}
    g_{s}&=\frac{\mathrm{H}(y)}{a(t)}\label{TypeIIAcoUpliNg}
\end{align}
can no longer be equated to (\ref{IIAcouPlingdeSitterIIB}). Instead it will be determined by demanding de Sitter isometry at the final stage of our duality sequence i.e. heterotic $\mathrm{SO(32)}$ theory for the case at consideration.

Because of the orientifolding operation of the form  $\mathcal{I}_{4}\mathcal{I}_{5}$ on the type $\mathrm{IIB}$ solitonic configuration we can only switch on \textbf{NS-NS} and \textbf{R-R} fields with one leg along the toroidal directions $\mathbb{T}^{2}_{4,5}$ and another along $\mathcal{M}_{4}$. We denote these fields as $\mathcal{B}^{(1)}_{\alpha m}$ and ${\mathcal{B}}^{(2)}_{\alpha m}$ respectively where $y^{\alpha}\in \frac{\textstyle \mathbb{T}^{2}_{4,5}}{\textstyle \mathcal{I}_{4}\mathcal{I}_{5}}$ and $y^{m}\in \mathcal{M}_{4}$. There are $\mathrm{O}7$ planes and $\mathrm{D}7$ branes in $\mathrm{IIB}$ theory located at the four fixed points of the toroidal orbifold and extended along $\mathbb{R}^{3,1}\times \mathcal{M}_{4}$ directions. 

[$\bullet$]  A $\mathrm{T}$ duality along $y^{4}$ of $\mathrm{IIB}$ metric (\ref{tyPeIIBMetrIc}) leads to type $\mathrm{IIA}$ theory on  $\mathbb{R}^{3,1}\times \frac{\textstyle \mathbf{S}^{1}_{5}}{\textstyle \Omega \mathcal{I}_{5}}\times \left(\mathcal{M}_{4}\rtimes \mathbf{S}^{1}_{4}\right)$ $\left(\text{or, in other words type}~\mathrm{IA}~\text{theory on}~\mathbb{R}^{3,1}\times \frac{\textstyle \mathbf{S}^{1}_{5}}{ \textstyle\mathcal{I}_{5}}\times \left(\mathcal{M}_{4}\rtimes \mathbf{S}^{1}_{4}\right) \right)$ with the following metric configuration
\begin{align}
    d{s}^{2}_{\mathrm{IIA}}&= g_{s}^{-2}\left( -dt^{2}+\sum_{i=1}^{2}(dx^{i})^2 +(dx^{3})^{2}\right)+\frac{1}{\mathrm{H}^{2}(y)\mathrm{F}_{1}(t)}\left( dy^4\right)^2+\mathrm{H}^{2}(y)\mathrm{F}_{1}(t)\left( dy^{5}\right)^2\nonumber\\
    &+\left[\mathrm{H}^{2}(y)\mathrm{F}_{2}(t)g_{mn}+\frac{\mathcal{B}^{(1)}_{4m}\mathcal{B}^{(1)}_{4n}}{\mathrm{H}^{2}(y)\mathrm{F}_{1}(t)}\right]dy^{m}dy^{n}+\frac{2{\mathcal{B}}^{(1)}_{4m}}{\mathrm{H}^{2}(y)\mathrm{F}_{1}(t)}dy^{4}dy^{m}\nonumber\\
    &=g_{s}^{-2}\left(-dt^2+\sum_{i=1}^{2}(dx^{i})^2+(dx^{3})^{2} \right)+\frac{1}{\mathrm{H}^{2}(y)\mathrm{F}_{1}(t)}\left( dy^{4}+\mathcal{B}^{(1)4}_{~~~m}dy^{m}\right)^2\nonumber\\
    &~~~~~~~~~~+\mathrm{H}^{2}(y)\mathrm{F}_{1}(t)(dy^{5})^{2}+\mathrm{H}^{2}(y)\mathrm{F}_{2}(t)g_{mn}dy^{m}dy^{n},\label{SecIIAcoNfiG}
\end{align}
and the corresponding coupling constant is given as
\begin{align}
    \tilde{g}_{s}&=\frac{1}{\mathrm{H}(y)\sqrt{\mathrm{F}_{1}(t)}}.\label{TypEIIAcc}
\end{align}
The \textbf{NS-NS} fluxes on this theory are $\mathrm{B}^{}_{5m}=\mathcal{B}^{(1)}_{5m}$ and the \textbf{R-R} fluxes are ${A}_{m}, ~{C}_{45m}$. There are $\mathrm{O}8$ planes and $\mathrm{D}8$ branes located at the two fixed points of the orbifold $\frac{\mathbf{S}^1_{5}}{\Omega \mathcal{I}_{5}}$ and these are extended  along $\mathbb{R}^{3,1}\times \mathcal{M}_{4}\times \tilde{\mathbf{S}}^{1}_{4}$.

[$\bullet$] 
A further $\mathrm{T}$ duality along $y^{5}$ leads to type $\mathrm{I}$ theory on $\mathbb{R}^{3,1}\times   \mathcal{M}_{4}\rtimes \left( \widetilde{\mathbf{S}}_{4}^{1}\times \widetilde{\mathbf{S}}_{5}^{1}\right)$, metric configuration of the same being 
\begin{align}
&ds^{2}_{\mathrm{I}}=g_{s}^{-2}\left(-dt^2+\sum_{i=1}^{2}\left( dx^{i}\right)^2+\left( dx^3\right)^2 \right)+\frac{1}{\mathrm{H}^2(y)\mathrm{F}_{1}(t)}\left( dy^4\right)^2\nonumber\\
&~+\frac{1}{\mathrm{H}^{2}(y)\mathrm{F}_{1}(t)}\left(dy^5 \right)^2+\left[\mathrm{H}^2(y)\mathrm{F}_{2}(t)g_{mn}+\frac{\mathcal{B}^{(1)}_{4m}\mathcal{B}^{(1)}_{4n}}{\mathrm{H}^2(y)\mathrm{F}_{1}(t)}+\frac{\mathcal{B}^{(1)}_{5m}\mathcal{B}^{(1)}_{5n}}{\mathrm{H}^2(y)\mathrm{F}_{1}(t)}\right]dy^m dy^n\nonumber\\
    &~~~~~~~~~~+\frac{2\mathcal{B}^{(1)}_{4m}}{\mathrm{H}^2(y)\mathrm{F}_{1}(t)}dy^4dy^m+\frac{2\mathcal{B}^{(1)}_{5m}}{\mathrm{H}^2(y)\mathrm{F}_{1}(t)}dy^5dy^m\nonumber\\
    &=g_{s}^{-2}\left(-dt^2+\sum_{i=1}^{2}(dx^{i})^2+(dx^{3})^{2} \right)+\frac{1}{\mathrm{H}^{2}(y)\mathrm{F}_{1}(t)}\left( dy^{4}+\mathcal{B}^{(1)4}_{~~~m}dy^{m}\right)^2\nonumber\\
    & ~~~~~~+\frac{1}{\mathrm{H}^2(y)\mathrm{F}_{1}(t)}\left( dy^5+\mathcal{B}^{(1)5}_{~~~m}dy^m\right)^2+\mathrm{H}^{2}(y)\mathrm{F}_{2}(t)g_{mn}dy^{m}dy^{n},\label{typeImetriC}
\end{align}
and the type $\mathrm{I}$ coupling constant is given by 
\begin{align}
g^{\mathrm{I}}&=\frac{1}{\mathrm{H}^{2}(y)\mathrm{F}_{1}(t)}.\label{TypeICouplinG}
\end{align}

{This theory has $\tilde{\mathcal{B}}^{(2)}_{\alpha m}$ \textbf{R-R} fluxes  as well as space-time filling orientifold plane and D-brane.}

[$\bullet$] The $\mathrm{S}$ dual of the type $\mathrm{I}$ theory is our designated heterotic $\mathrm{SO(32)}$ theory 
\begin{align}
ds^{2}_{\text{Het}}&=g_{s}^{-2}\mathrm{H}^2(y)\mathrm{F}_{1}(t)\left(-dt^2+\sum_{i}\left(dx^{i}\right)^2+\left( dx^3\right)^2 \right) \nonumber\\
    ~+\mathrm{H}^{4}&(y)\mathrm{F}_{1}(t)\mathrm{F}_{2}(t)g_{mn}dy^{m}dy^{n}+\delta_{\alpha\beta}\left(dy^{\alpha}+\mathcal{B}^{(1)\alpha}_{~~~m} dy^{m}\right)\left(dy^{\beta}+\mathcal{B}^{(1)\beta}_{~~~n} dy^{n}\right)\nonumber\\
    &=\mathrm{F}_{1}(t)a^2(t)\left(-dt^2+\sum_{i=1}^3\left(dx^{i}\right)^2 \right) +\mathrm{H}^{4}(y)\mathrm{F}_{1}(t)\mathrm{F}_{2}(t)g_{mn}dy^{m}dy^{n}\nonumber\\
~&~~~~~~~~~~~+\delta_{\alpha\beta}\left(dy^{\alpha}+\mathcal{B}^{(1)\alpha}_{~~~m} dy^{m}\right)\left(dy^{\beta}+\mathcal{B}^{(1)\beta}_{~~~n} dy^{n}\right),\label{HeTerotiCSO(32)MetrIc}
\end{align}
with the heterotic coupling 
\begin{align}
g_{\text{Het}}&=\frac{\textstyle 1}{\textstyle g_{\mathrm{I}}}=\mathrm{H}^{2}(y)\mathrm{F}_{1}(t).\label{HeterCoupling}
\end{align}
The fluxes in this theory are $B_{\alpha m}$. Because of the presence of Wilson lines the gauge group of this theory is not $\mathrm{SO}(32)$ but is broken down to four copies of $\mathrm{SO}(8)$ group \cite{Sen_1996}.

Clearly, if we want to realise de Sitter isometry in \textit{flat slicing} on the dual heterotic theory, we have to demand 
\begin{align}
    \mathrm{F}_{1}(t)a^{2}(t)=\mathrm{F}_{1}(t)g
_{s}^{-2} \mathrm{H}^2(y)&=\frac{1}{\Lambda t^{2}}~~~\text{with}~~-\infty \le t <0 \label{dSIsometryHetSO}
\end{align}
along with 
\begin{align}
    \mathrm{F}_{1}(t)\mathrm{F}_{2}(t)=1 \label{ConstrF1F2}
\end{align}
to ensure time independence of the six dimensional internal space. Note that the internal manifold also contains the time dependent two forms $\mathcal{B}^{(1)}_{\alpha m}$ but they do not contribute to its volume for reasons as elaborated in Appendix \ref{sec:AppendixD}. Correspondingly, the type $\mathrm{IIA}$ coupling constant (\ref{TypeIIAcoUpliNg}) becomes equal to 
\begin{align}
    g_{s}&=\mathrm{H}(y)|t|\sqrt{\Lambda\mathrm{F}_{1}(t)},\label{TYpe IIA HetSO(32)}
\end{align}
which also depends on the functional form of the time dependent warp factor $\mathrm{F}_{1}(t)$ unlike the case of realising de Sitter in type IIB theory (\ref{IIAcouPlingdeSitterIIB}). 

\vspace{5pt}
The above duality sequence leading to heterotic $\mathrm{SO(32)}$ theory can in fact proceed dynamically starting with the $\mathrm{M}$ theory set up. To elaborate this statement we consider the scalings of the temporal warp factors as defined in equation (\ref{ScalingOfFi's})  with the dominant scalings $\alpha_{o}$ and $\beta_{o}$ satisfying the similar upper bound as in (\ref{UbdofdomFactor})
\begin{align}
    \alpha_{o}<+\frac{2}{3},~~~\beta_{o}<+\frac{2}{3}.
\end{align}
Just to reiterate, the reason for such upper bound is that we require only the $\mathrm{M}$ theory torus along 3,11  to shrink to zero size taking us to the dual type $\mathrm{IIB}$ configuration.  However, we no longer want the type $\mathrm{IIB}$ theory to be the final stage of our duality sequence, instead two further $\mathrm{T}$ dualities should take place along both the cycles of the toroidal orbifold $\frac{\mathrm{T}^2_{4,5}}{\mathcal{I}_4\mathcal{I}_{5}}$ of type $\mathrm{IIB}$ configuration. Hence $\beta_{o}$ must be greater than zero, implying it lies in the range 
\begin{align}
    0<\beta_{o}<+\frac{2}{3},\label{beta0Range}
\end{align}
and the condition in (\ref{ConstrF1F2}) implies $
    \alpha_{o}=-\beta_{o}$. With the above choice of $\beta_{o}$ (\ref{beta0Range}) the temporal warp factor $\mathrm{F}_{1}(t)$ appearing along the toroidal orbifold $\frac{\textstyle \mathbb{T}^{2}_{4,5}}{\textstyle \mathcal{I}_{4}\mathcal{I}_{5}}$ of type $\mathrm{IIB}$ metric (\ref{tyPeIIBMetrIc}) tends to zero in the late time limit as $t\rightarrow 0^-$. Consequently, two $\mathrm{T}$ dualities along these two cycles take place dynamically and we obtain type $\mathrm{I}$ theory with coupling (\ref{TypeICouplinG}). The strongly coupled type $\mathrm{I}$ theory goes to the $\mathrm{S}$ dual picture and gives weakly coupled (\ref{HeterCoupling}) heterotic $\mathrm{SO(32)}$ theory with de Sitter isometry (\ref{dSIsometryHetSO}) and time independent internal space (\ref{ConstrF1F2}) in the late time limit.

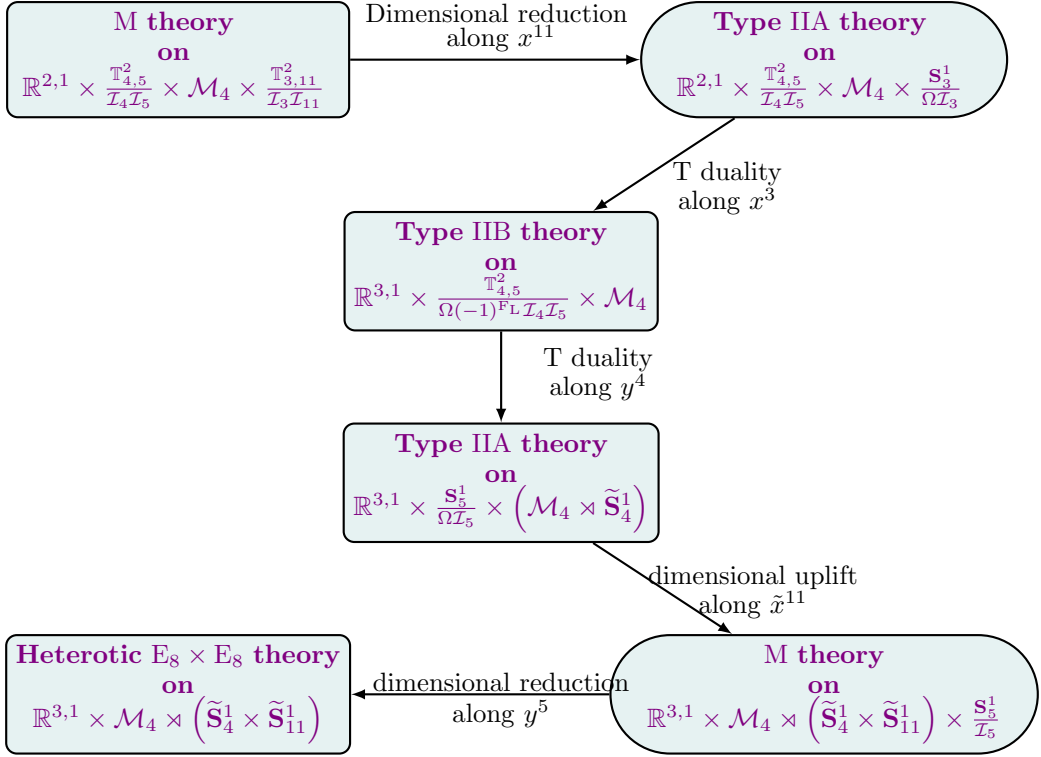
\begin{figure}[h!]
\centering
\begin{tikzpicture}[scale=0.85,font=\small,thick]
\node at (-5.0,0) [draw, rounded corners,fill=teal!10,
    minimum width=4.5cm,
    minimum height=1cm](block1) { \textcolor{violet}{
$\substack{ \textstyle \mathrm{M}~\textbf{theory} \\ \\\textstyle \textbf{on}\\ \textstyle\textbf{$\mathbb{R}^{2,1}\times \frac{\mathbb{T}_{4,5}^{2}}{\mathcal{I}_{4}\mathcal{I}_{5}}\times \mathcal{M}_{4}\times \frac{\mathbb{T}^{2}_{3,11}}{\mathcal{I}_{3}\mathcal{I}_{11}}$}}$
}};
\node at (5.0,0) [draw, rounded rectangle, fill=teal!10,
    minimum width=2.5cm,
    minimum height=1cm] (block2) { \textcolor{violet}{ 
$\substack{\textstyle\textbf{Type}~\mathrm{IIA}~\textbf{theory}\\ \\ \textstyle\textbf{on} \\\textstyle \mathbb{R}^{2,1}\times  \frac{\mathbb{T}_{4,5}^{2}}{\mathcal{I}_{4}\mathcal{I}_{5}}\times \mathcal{M}_{4}\times \frac{\mathbf{S}^{1}_{3}}{\Omega\mathcal{I}_{3}}}$
}};
\node at (0,-3.3) [draw, rounded corners, fill=teal!10,
    minimum width=2.5cm,
    minimum height=1.5cm] (block3) { \textcolor{violet}{
$\substack{\textstyle\textbf{Type}~\mathrm{IIB}~\textbf{theory}\\ \\ \textstyle \textbf{on} \\\textstyle \mathbb{R}^{3,1}\times  \frac{\mathbb{T}_{4,5}^{2}}{\Omega(-1)^{\mathrm{F_{L}}}\mathcal{I}_{4}\mathcal{I}_{5}}\times \mathcal{M}_{4}
}$}};
\node at (0,-6.6) [draw, rounded corners, fill=teal!10,
    minimum width=2.5cm,
    minimum height=1.5cm] (block4) { \textcolor{violet}{
$\substack{\textstyle\textbf{Type}~\mathrm{IIA}~\textbf{theory}\\ \\ \textstyle \textbf{on} \\\textstyle \mathbb{R}^{3,1}\times  \frac{\mathbf{S}^{1}_{5}}{\Omega\mathcal{I}_{5}}\times \left(\mathcal{M}_{4}\rtimes \widetilde{\mathbf{S}}_{4}^{1} \right)
}$}};
\node at (5.0, -9.9) [draw, rounded rectangle, fill=teal!10,
    minimum width=2.5cm,
    minimum height=1.5cm] (block5) { \textcolor{violet}{
$\substack{\textstyle \mathrm{M}~\textbf{theory}\\ \\ \textstyle \textbf{on} \\\textstyle \mathbb{R}^{3,1}\times   \mathcal{M}_{4}\rtimes \left( \widetilde{\mathbf{S}}_{4}^{1}\times \widetilde{\mathbf{S}}_{11}^{1} \right)\times \frac{\mathbf{S}^{1}_{5}}{\mathcal{I}_5}
}$}};
\node at (-5.0, -9.9) [draw, rounded corners, fill=teal!10,
    minimum width=2.5cm,
    minimum height=1.5cm] (block6) { \textcolor{violet}{
$\substack{\textstyle\textbf{Heterotic}~\mathrm{E_8\times E_8}~\textbf{theory}\\ \\ \textstyle \textbf{on} \\ \textstyle \mathbb{R}^{3,1}\times   \mathcal{M}_{4}\rtimes \left( \widetilde{\mathbf{S}}_{4}^{1}\times \widetilde{\mathbf{S}}_{11}^{1} \right)
}$}};
\draw[-latex] 
(block1) -- (block2)
(block2) edge (block3)
    (block3) edge (block4)
    (block4) edge (block5)
    (block5) edge (block6);
\node at (0,.5) {${\substack{\textstyle\text{Dimensional reduction } \\ \textstyle \text{along $x^{11}$} }}$} ; 
\node at (3.5,-2.0) { $\substack{\textstyle\mathrm{T}~\text{duality}\\ \textstyle \text{along}~x^3}$};
\node at (1.5,-4.9) { $\substack{\textstyle\mathrm{T}~\text{duality}\\ \textstyle \text{along}~y^4}$};
\node at (3.9,-8.3) { $\substack{\textstyle\mathrm{dimensional}~\text{uplift}\\ \textstyle \text{along}~\tilde{x}^{11}}$};
\node at (0,-9.7) {$\mathrm{dimensional ~reduction }$ };
\node at (0,-10.2) {$\mathrm{along}$~$y^5$ };
\end{tikzpicture}
    \caption{Duality sequence to go from the $\mathrm{M}$ theory configuration (\ref{MthEorYmeTric}) to heterotic $\mathrm{E_8}\times\mathrm{E}_8$ string theory (\ref{DiffDSE8}) with the gauge group broken down to $(SO(8))^4$ }
    \label{SubtlE8 tree diagram}
\end{figure}

    A careful look into the above described duality sequence leading to de Sitter isometry in heterotic $\mathrm{SO}(32)$, allows us to pinpoint an important subtlety. The type $\mathrm{IIA}$ theory (\ref{SecIIAcoNfiG}) that we arrived at after performing a $\mathrm{T}$ duality along the $y^4$ direction of type $\mathrm{IIB}$ orientifold $\mathbb{R}^{3,1}\times \frac{\mathbb{T}^2_{4,5}}{\Omega (-1)^{\mathrm{F}_{\mathrm{L}}}\mathcal{I}_{4}\mathcal{I}_{5}}\times \mathcal{M}_{4}$, had its coupling constant proportional to $\frac{1}{\sqrt{\mathrm{F}_{1}(t)}}$ (\ref{TypEIIAcc}). With the choice of the dominant scaling of $\mathrm{F}_{1}(t)$ as in equation (\ref{beta0Range}), $\mathrm{F}_{1}(t)\rightarrow 0$  in the late time limit of flat slicing de Sitter space. It implies that the aforementioned type $\mathrm{IIA}$ theory appears to become strongly coupled in the late time limit. Consequently, the $\mathrm{IIA}$ configuration may get uplifted to $\mathrm{M}$ theory instead of going to the $\mathrm{T}$ dual $ \mathrm{type}~\mathrm{I}~(\equiv\mathrm{type}~\mathrm{IIB}/\Omega)$ configuration (\ref{typeImetriC}). When we consider the former to be the case the duality sequence proceeds as depicted in \textbf{Figure} (\ref{SubtlE8 tree diagram}).   The metric of dimensionally uplifted $\mathrm{M}$ theory takes the following form 
\begin{align}   &{d\tilde{s}}_{\mathrm{M-th}}^{2}= \tilde{g}_{s}^{-2/3}\bigg( g_{s}^{-2}\left( -dt^{2}+\sum_{i=1}^{2}(dx^{i})^2 +(dx^{3})^{2}\right)+\frac{1}{\mathrm{H}^{2}(y)\mathrm{F}_{1}(t)}\left( dy^4\right)^2+\mathrm{H}^{2}(y)\mathrm{F}_{1}(t)\left( dy^{5}\right)^2\nonumber\\
    &+\left[\mathrm{H}^{2}(y)\mathrm{F}_{2}(t)g_{mn}+\frac{\mathcal{B}^{(1)}_{4m}\mathcal{B}^{(1)}_{4n}}{\mathrm{H}^{2}(y)\mathrm{F}_{1}(t)}\right]dy^{m}dy^{n}+\frac{2{\mathcal{B}}^{(1)}_{4m}}{\mathrm{H}^{2}(y)\mathrm{F}_{1}(t)}dy^{4}dy^{m}\bigg)+2\tilde{g}_{s}^{4/3} A_{m} dy^m d\tilde{x}^{11} \nonumber\\
    &~~~~~~~~~~~~~~~~~~~+\tilde{g}_{s}^{4/3}A_{m}A_{n}dy^m dy^n+\tilde{g}_{s}^{4/3}(d\tilde x^{11})^2.\nonumber\\
    &=\mathrm{H}^{\frac{2}{3}}(y)\mathrm{F}_{1}^{\frac{1}{3}}(t)g_{s}^{-2}\left(-dt^2+\sum_{i=1}^{2}(dx^{i})^2+(dx^{3})^{2} \right)+{\mathrm{H}^{-\frac{4}{3}}(y)\mathrm{F}_{1}^{-\frac{2}{3}}(t)}\left( dy^{4}+\mathcal{B}^{(1)4}_{~~~m}dy^{m}\right)^2\nonumber\\
&+\mathrm{H}^{\frac{8}{3}}(y)\mathrm{F}^{\frac{4}{3}}_{1}(t)(dy^{5})^{2}+\mathrm{H}^{\frac{8}{3}}(y)\mathrm{F}_{1}^{\frac{1}{3}}(t)\mathrm{F}_{2}(t)g_{mn}dy^{m}dy^{n}+\mathrm{H}^{-\frac{4}{3}}(y)\mathrm{F}^{-\frac{2}{3}}(t)\left(d\tilde x^{11}+A_{m}dy^m\right)^2.
\end{align}
Clearly, the solitonic configuration of this theory is $\mathbb{R}^{3,1}\times \mathcal{M}_{4}\rtimes (\tilde{\mathbf{S}}^1_{4}\times \tilde{\mathbf{S}}^{1}_{11})\times \frac{\mathbf{S}^1_{5}}{\mathcal{I}_{5}}$. It suggests that there are two Horava-Witten walls at the two fixed points of the orbifold $\mathbf{S}^1_{5}/\mathcal{I}_5$\footnote{In other words, the orbifold ${\textstyle \mathbf{S}^{1}_{5}}/{\textstyle \mathcal{I}_{5}}$ is a line segment between $y^{5}=0$ and $y^{5}=\pi \mathcal{R}_{y^{5}}$, where $\mathcal{R}_{y^{5}}$ is the radius of the circle $\mathbf{S}^{1}_{5}$ and there exists two ten dimensional boundaries at the two fixed points of the orbifold. These two boundaries are sometimes also referred to as end-of-the-world 9-branes.} \cite{1Horava_1996,2Horava_1996}. 
Dimensional reduction along the direction of the orbifold viz. $y^5$ give rise to the heterotic $\mathrm{E}_{8}\times \mathrm{E}_{8}$ theory which has the following metric configuration 
\begin{align}
{d{s}}_{\mathrm{E}_8\times\mathrm{E}_8 }^{2}&=\left(\mathrm{H}^{\frac{8}{3}}(y)\mathrm{F}_{1}^{\frac{4}{3}}(t) \right)^{1/2}\bigg(\mathrm{H}^{\frac{2}{3}}(y)\mathrm{F}_{1}^{\frac{1}{3}}(t)g_{s}^{-2}\left(-dt^2+\sum_{i=1}^{2}(dx^{i})^2+(dx^{3})^{2} \right)\nonumber\\
&+\mathrm{H}^{\frac{8}{3}}(y)\mathrm{F}_{1}^{\frac{1}{3}}(t)\mathrm{F}_{2}(t)g_{mn}dy^{m}dy^{n}+{\mathrm{H}^{-\frac{4}{3}}(y)\mathrm{F}_{1}^{-\frac{2}{3}}(t)}\left( dy^{4}+\mathcal{B}^{(1)4}_{~~~m}dy^{m}\right)^2
\nonumber\\
    &+\mathrm{H}^{-\frac{4}{3}}(y)\mathrm{F}^{-\frac{2}{3}}(t)\left(d\tilde x^{11}+A_{m}dy^m\right)^2 \bigg) \nonumber\\
&=\mathrm{H}^{2}(y)\mathrm{F}_{1}(t)g_s^{-2}\left(-dt^2+\sum_{i=1}^{2}(dx^{i})^2+(dx^{3})^{2} \right)+\mathrm{H}^4(y)\mathrm{F}_{1}(t)\mathrm{F}_{2}(t)g_{mn}dy^m dy^n\nonumber\\
&~~~~~~~~+\left( dy^{4}+\mathcal{B}^{(1)4}_{~~~m}dy^{m}\right)^2
+\left(d\tilde x^{11}+A_{m}dy^m\right)^2\nonumber\\
&=\mathrm{F}_{1}(t)a^{2}(t)\left(-dt^2+\sum_{i=1}^{2}(dx^{i})^2+(dx^{3})^{2} \right)+\mathrm{H}^4(y)\mathrm{F}_{1}(t)\mathrm{F}_{2}(t)g_{mn}dy^m dy^n\nonumber\\
&~~~~~~~~+\left( dy^{4}+\mathcal{B}^{(1)4}_{~~~m}dy^{m}\right)^2
+\left(d\tilde x^{11}+A_{m}dy^m\right)^2.\label{DiffDSE8} 
\end{align}
The solitonic configuration can be depicted as $\mathrm{R}^{3,1}\times \mathcal{M}_{4}\rtimes (\tilde{\mathbf S}^1_{4}\times \tilde{\mathbf{S}}^1_{11})$.
Comparing the above metric of heterotic $\mathrm{E}_{8}\times \mathrm{E}_{8}$ theory with that of the metric (\ref{HeTerotiCSO(32)MetrIc}) of heterotic $\mathrm{SO}(32)$ theory we find that they are identical. It implies that the subtlety that we were worried about has not led to a new result. Moreover, the presence of Wilson lines breaks the gauge group of the theory to four copies of $\mathrm{SO}(8)$ gauge group.

It turns out that we require three distinct time dependent warp factors in the six dimensional internal manifold in order to arrive at heterotic $\mathrm{E}_{8}\times \mathrm{E}_{8}$ theory  with unbroken gauge group in the late time limit. Before we analyse this let us consider the case of Einstein frame for the de Sitter configurations in heterotic $\mathrm{SO}(32)$ theory and heterotic $\mathrm{E}_{8}\times \mathrm{E}_{8}$ theory discussed so far.   

\subsubsection{The string frame versus the Einstein frame}
In 10 dimensions, the Einstein frame metric is related to the string frame metric as follows 
\begin{align}
\mathrm{g}_{\mu\nu}^{(E)}&=e^{-\frac{\phi}{2}}\mathrm{g}_{\mu\nu}^{(S)}.\label{EinsStriFRAme}
\end{align}
Therefore, in Einstein frame the metric configuration of heterotic $\mathrm{SO}(32)$ theory (\ref{HeTerotiCSO(32)MetrIc}) becomes
\begin{align}
ds^2_{\mathrm{Het}}&=(\mathrm{H}^2(y)\mathrm{F}_{1}(t))^{-1/2}\bigg(\mathrm{F}_{1}(t)a^2(t)\big(-dt^2+\sum_{i=1}^2\left(dx^{i}\right)^2+\left( dx^3\right)^2 \big) \nonumber\\
    ~+\mathrm{H}^{4}&(y)\mathrm{F}_{1}(t)\mathrm{F}_{2}(t)g_{mn}dy^{m}dy^{n}+\delta_{\alpha\beta}\left(dy^{\alpha}+\mathcal{B}^{(1)\alpha}_{~~~m} dy^{m}\right)\left(dy^{\beta}+\mathcal{B}^{(1)\beta}_{~~~n} dy^{n}\right)\bigg)\nonumber\\
    &=\frac{\sqrt{\mathrm{F}_{1}(t)}a^2(t)}{\mathrm{H}(y)}\left(-dt^2+\sum_{i=1}^3\left(dx^{i}\right)^2 \right)+\mathrm{H}^{3}(y)\sqrt{\mathrm{F}_{1}(t)}\mathrm{F}_{2}(t)g_{mn}dy^{m}dy^{n} \nonumber\\
    ~&~~~~~~~+\frac{1}{\mathrm{H}(y)\sqrt{\mathrm{F}_{1}(t)}}\delta_{\alpha\beta}\left(dy^{\alpha}+\mathcal{B}^{(1)\alpha}_{~~~m} dy^{m}\right)\left(dy^{\beta}+\mathcal{B}^{(1)\beta}_{~~~n} dy^{n}\right).\label{EinsFrameSO32}
\end{align}
Thus if we want to demand de Sitter isometry in Einstein frame then the following condition must hold
\begin{align}
\sqrt{\mathrm{F}_{1}(t)}a^2(t)\equiv g_{s}^{-2}\mathrm{H}^2(y)\sqrt{\mathrm{F}_{1}(t)}&=\frac{1}{\Lambda t^2}.
\end{align}
Moreover, the time independence of the volume of the six dimensional internal space $\sqrt{\mathrm{F}_{1}(t)\mathrm{F}_{2}^4(t)}$, implies the dominant scalings of $\mathrm{F}_{1}$ and $\mathrm{F}_{2}$ (\ref{ScalingOfFi's}) must satisfy $\alpha_{o}=-\frac{\beta_o}{4}$ while maintaining $0<\beta_o<\frac{2}{3}$ for the duality sequence to proceed appropriately. However, the volume of the internal four manifold $\sqrt{\mathrm{F}_1^2(t)\mathrm{F}_2^4(t)}=\left(\frac{g_{s}}{\mathrm{H}(y)}\right)^{\beta_o+2\alpha_o}=\left(\frac{g_{s}}{\mathrm{H}(y)}\right)^{\beta_o/2}$ implies that $\mathcal{M}_{4}$ shrinks to zero size as $t\rightarrow 0^-$ showing late time singularity whereas the volume of the toroidal manifold being $\frac{1}{\sqrt{\mathrm{F}_{1}(t)}}=\left(\frac{g_{s}}{\mathrm{H}(y)}\right)^{-\beta_o/2}$, it grows in the late time limit. Thus we can not get a  four dimensional de Sitter configuration in both the string frame and Einstein frame simultaneously.  To summarize, there are two possibilities: 
\begin{itemize}
    \item[$\bullet$] We can either have de Sitter isometry in string frame
    \begin{align}
&g_{s}^{-2}\mathrm{H}^2(y)\mathrm{F}_{1}(t)=\frac{1}{\Lambda t^2}\implies \left(\frac{g_s}{\mathrm{H}(y)}\right)=(\Lambda t^2)^{\frac{1}{2-\beta_o}},\label{StringCCInSF}  
\end{align}
    with time-independent six dimensional internal space and correspondingly time independent four dimensional Newton's constant 
\begin{align}
    \mathrm{F}_{1}(t)\mathrm{F}_{2}(t)&=1~~~~\implies ~~\alpha_{o}=-\beta_{o},~~\mathrm{with}~~0<\beta_o<\frac{2}{3}.\label{TimeIndpIntSp}
\end{align}
    
However, it is dual to a non de Sitter configuration in Einstein frame since 
    \begin{align}
g_{s}^{-2}\mathrm{H}^2(y)\sqrt{\mathrm{F}_{1}(t)} \neq \frac{1}{\Lambda t^2}.
    \end{align}
    Moreover, in Einstein frame the six dimensional internal space and hence the four dimensional Newton’s constant become time-dependent. In fact, the internal six dimensional manifold shows late time decompactification as
    \begin{align}
        &\sqrt{\mathrm{F}_{1}(t)\mathrm{F}_{2}^4(t)}\nonumber\\
        &=\mathrm{F}_{2}^{{3}/ {2}}(t)= \left(\frac{g_{s}}{\mathrm{H}(y)}\right)^{-\frac{3}{2}\beta_o}\rightarrow\infty ,~~~~(\mathrm{Using~equation}~(\ref{TimeIndpIntSp})).
    \end{align}
    \item Or, we can have de Sitter isometry in Einstein frame i.e.
    \begin{align}
g_{s}^{-2}\mathrm{H}^2(y)\sqrt{\mathrm{F}_{1}(t)}=\frac{1}{\Lambda t^2}\implies \left(\frac{g_s}{\mathrm{H}(y)}\right)=(\Lambda t^2)^{\frac{2}{4-\beta_o}},\label{EFStringC}
    \end{align}
    with time-independent six dimensional internal space and correspondingly time independent four dimensional Newton’s constant 
    \begin{align}
        {\mathrm{F}_{1}(t)\mathrm{F}^4_{2}(t)}=1\implies \alpha_o=-\frac{\beta_o}{4},~~\mathrm{with}~~0<\beta_o<\frac{2}{3}\label{TimeIndpEinsFrm}
    \end{align}
    along with the internal four manifold $\mathcal{M}_{4}$ showing late time singularity
    \begin{align}
\sqrt{\mathrm{F}_{1}^2(t)\mathrm{F}_{2}^4(t)}=\left(\frac{g_{s}}{\mathrm{H}(y)}\right)^{\beta_o/2}\rightarrow 0 \label{VolOfM4Mani}
    \end{align}
    and the internal two manifold $\mathcal{M}_{2}$ showing late time decompactification 
\begin{align}
\frac{1}{\sqrt{\mathrm{F}_{1}(t)}}=\left(\frac{g_{s}}{\mathrm{H}(y)}\right)^{-\beta_o/2}\rightarrow \infty.
\end{align}
This is dual to a non-de Sitter configuration in string frame since 
\begin{align}
g_{s}^{-2}\mathrm{H}^2(y)\mathrm{F}_1(t)\neq \frac{1}{\Lambda t^2}.
\end{align}
Moreover, in string frame the four dimensional internal manifold $\mathcal{M}_{4}$ shows late time singularity
    \begin{align}
        &\sqrt{\mathrm{F}_{1}(t)\mathrm{F}_{2}(t)}\nonumber\\
        &=\mathrm{F}_{2}^{-3/2}(t)=\left(\frac{g_{s}}{\mathrm{H}(y)}\right)^{\frac{3}{8}\beta_o}\rightarrow 0,~~~(\mathrm{Using ~equation}~(\ref{TimeIndpEinsFrm})). 
    \end{align}
\end{itemize}
From the above discussion it seems that string frame is preferred (as it has no pathologies) over the Einstein frame when we want to realise de Sitter isomtery in the heterotic SO(32) string theory. On the other hand, if we demand
\begin{align}
{\rm F}_1(t) = \begin{cases} \left(\Lambda t^2\right)^{ \frac{v\beta_o} { 2v -\beta_o}},~~~~~~ -{ \frac{1}{ \sqrt{\Lambda}}} < t \leq -\epsilon \\
~~~ \\
~~ 1~~,~~~~~~~~~~~~~~~ -\epsilon < t < 0
\end{cases}\label{F1inHetSo32Th}
\end{align}
where $v=1,2$ for the string and the Einstein frame respectively, $\epsilon$ is a temporal point where $\mathrm{F}_{1}(t)$ and hence $\mathrm{F}_{2}(t)$ become a constant, then in the regime $-\epsilon <t<0$ we get a consistent de Sitter isometry in either frame. To implement equation (\ref{F1inHetSo32Th}) we define 
\begin{align}
&\mathrm{F}_1(t)\equiv(\Lambda t^2)^{\frac{v\beta(t)}{2v-\beta(t)}},~~~
\mathrm{where}~~\beta(t)=\begin{cases}
       {\beta_o},~~~~~~ -{ \frac{1}{ \sqrt{\Lambda}}} < t \leq -\epsilon \\
~~~ \\
 0,~~~~~~~~~~ -\epsilon < t < 0
    \end{cases},\label{F1(t)withbeta(t)}
\end{align}
with $0<\beta_o<\frac{2}{3}$. Correspondingly, the IIA coupling constant in string frame (\ref{StringCCInSF}) and Einstein frame (\ref{EFStringC}) can be expressed as 
\begin{align}
    \frac{g_{s}}{\mathrm{H}(y)}&=\left( \Lambda t^2\right)^{\frac{v}{2v-\beta(t)}}\label{StringCCSFEF}
\end{align}
where $v=1,2$ for the string frame, the Einstein frame as earlier.  The value of $\epsilon$ can be estimated by demanding that at $|t|=\epsilon$, the Einstein frame volume of the four manifold $\mathcal{M}_{4}$ (\ref{VolOfM4Mani}) attains size $\mathrm{V}_{1}>>\alpha^{\prime 2}$. If  $\mathrm{V}_o(-1/\sqrt{\Lambda})$ is its stabilized volume at $|t|=1/\sqrt{\Lambda}$ then we obtain
\begin{align}
    \frac{\left(\frac{g_s}{{\rm H}(y)}\right)^{\beta_o/2}\bigg|_{|t|=\epsilon}}{\left(\frac{g_s}{{\rm H}(y)}\right)^{\beta_o/2}\bigg|_{|t|=1/\sqrt{\Lambda}}}
    &=\frac{\mathrm{V}_1(-\epsilon)}{\mathrm{V}_o(-1/\sqrt{\Lambda})}\nonumber\\
    \implies \left(\sqrt{\Lambda} \epsilon\right)^{\frac{2\beta_o}{4-\beta_o}}&=\frac{\rm V_1}{{\mathrm{V}}_o(-1/\sqrt{\Lambda})}~~~~~~~(\mathrm{Using~equation}~ (\ref{StringCCSFEF})~\mathrm{with}~v=2)\nonumber\\
    \implies \epsilon=\frac{1}{\sqrt{\Lambda}}&\left[\frac{\mathrm{V}_1}{\mathrm{V}_o(-1/\sqrt{\Lambda})}\right]^{\frac{4-\beta_o}{2\beta_o}}.\label{ValueofEpsilonHetSO}
\end{align}
The other time dependent warp factor is given by 
\begin{align}
    \mathrm{F}_{2}(t)&=\left( \frac{g_{s}}{\mathrm{H}(y)}\right)^{\alpha(t)}=\left(\Lambda t^2\right)^{\frac{v\alpha(t)}{2v-\beta(t)}}
\end{align}
where 
\begin{align}
    v=1,~~ \alpha(t)&=-\beta(t)~~~ \mathrm{(in~the~string~frame)}\label{alpHaBetaRelStrFr}\\
        v=2,~~ \alpha(t)&=-\frac{\beta(t)}{4}~~~ \mathrm{(in~the~Einstein ~frame)}\label{alpHaBetaRelEinstFr}
\end{align}
Now we are going to elaborate on another duality sequence that allows us to realise de Sitter space in heterotic $\mathrm{E}_8\times \mathrm{E}_8$ string theory with unbroken gauge group. Interestingly, we require to take into account three time dependent warp factors in the M-theory metric configuration to achieve the same.  
\subsection{de Sitter isometry in heterotic \texorpdfstring{$\mathrm{E}_{8}\times \mathrm{E}_{8}$}{E8XE8} string theory}
\label{E8XE8ThrdistWarpf}
[$\bullet$] We begin with the following metric configuration in $\mathrm{M}$ theory 
\begin{align}
ds^{2}_{\mathrm{M}~\text{th}}&=g_{s}^{-8/3} \left(-dt^2+ \sum_{i=1}^{2}\left(dx^{i}\right)^{2}\right)+g_{s}^{-2/3}\mathrm{H}^{2}(y)\bigg[\mathrm{F}_{1}(t)~g_{\theta_{1}\theta_{1}}(d\theta_{1})^2\nonumber\\
&~~~~~~~+\mathrm{F}_{3}(t)~g_{\theta_{2}\theta_{2}}(d\theta_{2})^2 +\mathrm{F}_{2}(t)~~g_{mn}dy^m dy^n\bigg]+g_{s}^{4/3}\delta_{ab}dw^a dw^b\label{MthThreewarpFac}
\end{align} 
with the solitonic structure as  $\mathbb{R}^{2,1}\times \mathbf{S}^{1}_{\theta_{1}}\times \frac{\textstyle \mathbf{S}^{1}_{\theta_{2}}}{\textstyle \mathcal{I}_{\theta_{2}}}\times\mathcal{M}_{4}\times \frac{\textstyle \mathbb{T}^{2}_{3,11}}{\textstyle\mathcal{G}}$, here $\mathcal{M}_{4}$ is spanned by the coordinates $(y^6,y^7,y^8,y^9)$ and $(w^{1},w^{2})$ parameterises the $\mathrm{M}$ theory torus $\frac{\mathbb{T}^2_{3,11}}{\mathcal{G}}$. $\mathcal{G}$ is a group action without any fixed points. There exists two Horava-Witten walls at the boundaries of the manifold $\mathbf{S}^{1}_{\theta_2}/\mathcal{I}_{\theta_2}$ \cite{1Horava_1996, 2Horava_1996}.

From the solitonic structure of the above $\mathrm{M}$ theory configuration one might think that the dimensional reduction along $\theta_{2}$ immediately give rise to heterotic $\mathrm{E}_{8}\times \mathrm{E}_{8}$ theory with  four dimensional de Sitter isometry. However, we encounter few subtleties  when we try to perform such a naive dimensional reduction:

 First of all for the $\theta_2$ direction to shrink faster as compared to the toroidal direction, the temporal warp factor $\mathrm{F}_{3}(t)$ must scale as $\mathrm{F}_{3}\equiv \left(\frac{g_{s}}{\mathrm{H}}\right)^{\hat{b}}$ with $\hat{b}>2$. Assuming this to be the case the metric of dimensionally reduced theory becomes 
\begin{align}
    ds^2_{\mathrm{E}_{8}\times \mathrm{E}_{8}}&=\sqrt{\mathrm{F}_{3}(t)}g_{s}^{-3}\left(-dt^2+\sum_{i=1}^{2}(dx^i)^2 \right)+\mathrm{F}_{1}(t)\sqrt{\mathrm{F}_{3}(t)}g_{s}^{-1}(d\theta_1)^2 \nonumber\\
    &+\mathrm{F}_{2}(t)\sqrt{\mathrm{F}_{3}(t)}g_{s}^{-1}g_{mn}dy^mdy^n+\sqrt{\mathrm{F}_{3}(t)}g_{s}\delta_{ab}dw^a dw^b,\label{DimRedTheta2}
\end{align}
where we have set $\mathrm{H}(y)=g_{\theta_1\theta_1}=g_{\theta_2\theta_{2}}=1$. It shows that the third direction remains compact and evolves as $\sqrt{\mathrm{F}_{3}(t)}g_{s}$ whereas $\mathbb{R}^{2,1}$ has a different time evolution given as $\sqrt{\mathrm{F}_{3}(t)}g_{s}^{-3},$ clearly we do not have a well defined de Sitter isometry on the heterotic $\mathrm{E}_{8}\times\mathrm{E}_{8}$ theory.  

In order to obtain appropriate de Sitter isometry on the dual theory  let us identify $\theta_{1}$ with the third spatial direction without worrying about its compactness. Moreover, consider the $g_{s}$ scaling of the $\mathrm{F}_{1}$ warp factor to be of the form $g_{s}^{-2}$. Consequently, the metric of dimensionally reduced heterotic $\mathrm{E}_{8}\times\mathrm{E}_{8}$ theory (\ref{DimRedTheta2}) takes the following form  
 \begin{align}
    ds^2_{\mathrm{E}_{8}\times \mathrm{E}_{8}}&=\sqrt{\mathrm{F}_{3}(t)}g_{s}^{-3}\left(-dt^2+\sum_{i=1}^{2}(dx^i)^2 +(d\theta_1)^2\right) \nonumber\\
    &+\mathrm{F}_{2}(t)\sqrt{\mathrm{F}_{3}(t)}g_{s}^{-1}g_{mn}dy^mdy^n+\sqrt{\mathrm{F}_{3}(t)}g_{s}\delta_{ab}dw^a dw^b.\label{CasebDimredE8E8}
 \end{align}
 Now, with the identification 
 \begin{align}
     \sqrt{\mathrm{F}_{3}(t)}g_{s}^{-3}\equiv\frac{1}{\Lambda t^2},~~~~-\infty<t<0 \label{deSitIsomE8th}
 \end{align}
 we obtain a de Sitter space in flat slicing in the $\mathrm{E}_{8}\times\mathrm{E}_{8}$ theory. With $\mathrm{F}_{3}=g^{\hat{b}}_{s}$, we find 
 \begin{align}
     g_{s}=(\Lambda t^2)^{\frac{2}{6-\hat{b}}},~~~2<\hat{b}<6.
 \end{align}
 For the volume of the six dimensional internal space $
\sqrt{g_{s}^{-2}\mathrm{F}_{2}^4(t)\mathrm{F}_{3}^3(t)}$, to remain time independent the warp factor $\mathrm{F}_{2}$ must scale as 
\begin{align}
    \mathrm{F}_{2}=g_{s}^{\frac{2-3\hat{b}}{4}}.
\end{align}
Therefore, the metric (\ref{CasebDimredE8E8}) can be expressed as 
\begin{align}
    ds^2_{\mathrm{E}_8\times \mathrm{E}_{8}}&=\frac{1}{\Lambda t^2}\left(-dt^2+\sum_{i=1}^{2}(dx^i)^2+(d\theta_1)^2\right)+(\Lambda t^2)^{-\frac{\hat{b}+2}{2(6-\hat{b})}}g_{mn}dy^mdy^n\nonumber\\
    &~~~~~~~~~~~~+(\Lambda t^2)^{\frac{\hat{b}+2}{6-\hat{b}}}\delta_{ab}dw^a dw^b.
\end{align}
It shows that even though the six dimensional internal space remains time independent the toroidal manifold along $3,11$ directions show late time singularity as $t\rightarrow 0$ in the late time limit of a flat slicing de Sitter space. 

 If instead we identifying one of the isometry directions of $\mathcal{M}_{4}$ (upon expressing it as a product manifold $\mathcal{M}_{3}\times\mathrm{S}^1$) with the third spatial direction, then $\mathrm{F}_{2}$ must scale as $g_{s}^{-2}$ in order to have appropriate four dimensional de Sitter isometry on the dimensionally reduced heterotic $\mathrm{E}_8\times \mathrm{E}_8$ theory  (\ref{DimRedTheta2})
\begin{align}
    ds^2_{\mathrm{E}_8\times \mathrm{E}_{8}}&=\sqrt{\mathrm{F}_{3}(t)}g_{s}^{-3}\left(-dt^2+\sum_{i=1}^{2}(dx^i)^2+(dy^6)^2\right)+\mathrm{F}_{1}(t)\sqrt{\mathrm{F}_{3}(t)}g_{s}^{-1}(d\theta_1)^2\nonumber\\
    &~~~~~~~+\sqrt{\mathrm{F}_{3}(t)}g_s^{-3}g_{pq}dy^p dy^q+\sqrt{\mathrm{F}_{3}(t)}g_{s}\delta_{ab}dw^a dw^b,\label{CasecDimRedE8}
\end{align}
where $\mathcal{M}_{3}$ is spanned by the coordinates $y^p\in(y^7,y^8,y^9)$. Again with the identification 
\begin{align}
    \sqrt{\mathrm{F}_{3}(t)}g_{s}^{-3}&=\frac{1}{\Lambda t^2},~~~~~~~-\infty <t<0
\end{align}
we get a four dimensional de Sitter space in flat slicing in $\mathrm{E}_{8}\times \mathrm{E}_{8}$ theory with the value of $g_{s}$ being $(\Lambda t^2)^{\frac{2}{6-\hat{b}}}$, $2<\hat{b}<6$. Now, the volume of the six dimensional internal space $\mathcal{M}_{3}\times \mathrm{S}^1_{\theta_1}\times \frac{\mathbb{T}^2_{3,11}}{\mathcal{G}}$ turns out to be $\sqrt{\mathrm{F}_{1}(t)\mathrm{F}^3_{3}(t)g_{s}^{-8}}.$ For this to remain time independent the warp factor $\mathrm{F}_{1}$ must scale as 
\begin{align}
    \mathrm{F}_{1}=g_{s}^{8-3\hat{b}}.\label{gsScalofF1Cc}
\end{align}
Assuming this to be the case, the metric (\ref{CasecDimRedE8}) takes the following form 
\begin{align}
    ds^2_{\mathrm{E}_8\times \mathrm{E}_8}&=\frac{1}{\Lambda t^2}\left(-dt^2+\sum_{i=1}^2 (dx^i)^2+(dy^6)^2\right)+(\Lambda t^2)^{\frac{14-5\hat{b}}{6-\hat{b}}}(d\theta_1)^2\nonumber\\
    &~~~~+\frac{1}{\Lambda t^2}g_{pq}dy^p dy^q +(\Lambda t^2)^{\frac{\hat{b}+2}{6-\hat{b}}}\delta_{ab}dw^a dw^b.\label{E8XE8Casec}
\end{align}
The $g_{s}$ scaling of $\mathrm{F}_{1}$ (\ref{gsScalofF1Cc}) suggests that for this case $\hat{b}$ must lie in the range $\frac{22}{9}<\hat{b}<6$. The lower bound is to ensure that dimensional reduction only takes place along the $\theta_{2}$ direction leading to heterotic $\mathrm{E}_{8}\times \mathrm{E}_{8}$ theory having de Sitter isometry. However, the metric configuration (\ref{E8XE8Casec}) shows that for $\frac{22}{9}<\hat{b}<\frac{14}{5}$ there is a late time singularity not only along the toroidal manifold $\frac{\mathbb{T}^2_{3,11}}{\mathcal{G}}$ but also the $\mathbf{S}^{1}_{\theta_1}$ circle shrinks to zero size as $t\rightarrow 0$ and for $\frac{14}{5}<\hat{b}<6$ there is a late time singularity for the toroidal manifold  $\frac{\mathbb{T}^2_{3,11}}{\mathcal{G}}$. 

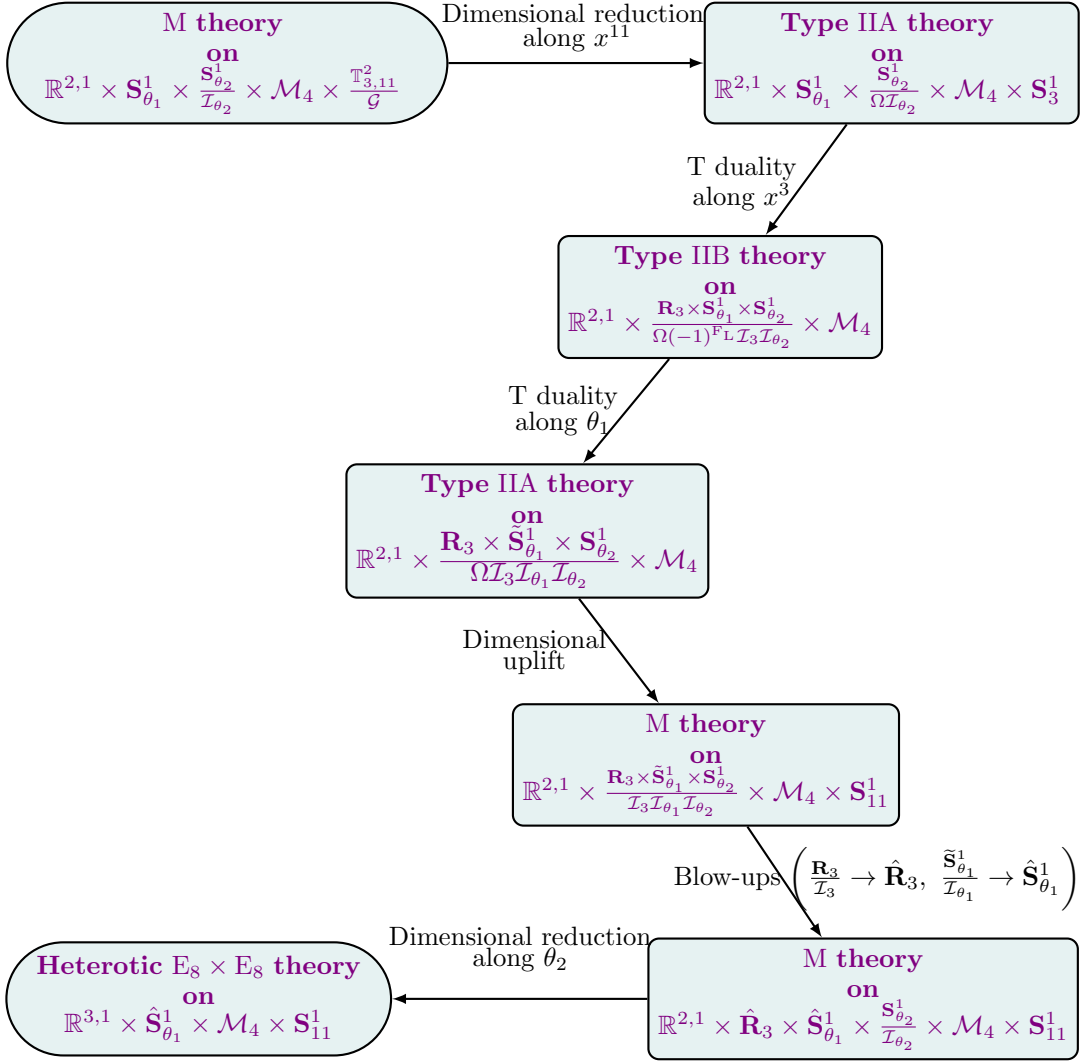
\begin{figure}[h!]
\centering
\begin{tikzpicture}[scale=0.94,font=\small,thick]
\node at (-4.9,0) [draw, rounded rectangle,fill=teal!10,
    minimum width=4.5cm,
    minimum height=1cm](block1) { \textcolor{violet}{
$\substack{ \textstyle \mathrm{M}~\textbf{theory} \\ \\\textstyle \textbf{on}\\ \textstyle\textbf{$\mathbb{R}^{2,1}\times \mathbf{S}^{1}_{\theta_1}\times\frac{\mathbf{S}^{1}_{\theta_2}}{\mathcal{I}_{\theta_{2}}}\times \mathcal{M}_{4}\times \frac{\mathbb{T}^{2}_{3,11}}{\mathcal{G}}$}}$
}};
\node at (4.4,0) [draw, rounded corners, fill=teal!10,
    minimum width=2.5cm,
    minimum height=1cm] (block2) { \textcolor{violet}{ 
$\substack{\textstyle\textbf{Type}~\mathrm{IIA}~\textbf{theory}\\ \\ \textstyle\textbf{on} \\\textstyle \mathbb{R}^{2,1}\times  \mathbf{S}^{1}_{\theta_{1}}\times \frac{\mathbf{S}^{1}_{\theta_2}}{\Omega \mathcal{I}_{\theta_2}}\times \mathcal{M}_{4}\times {\mathbf{S}^{1}_{3}}}$
}};
\node at (2.0,-3.3) [draw, rounded corners, fill=teal!10,
    minimum width=2.5cm,
    minimum height=1.5cm] (block3) { \textcolor{violet}{
$\substack{\textstyle\textbf{Type}~\mathrm{IIB}~\textbf{theory}\\ \\ \textstyle \textbf{on} \\\textstyle \mathbb{R}^{2,1}\times  \frac{\mathbf{R}_{3}\times\mathbf{S}^{1}_{\theta_1}\times \mathbf{S}^{1}_{\theta_2}}{\Omega(-1)^{\mathrm{F_{L}}}\mathcal{I}_{3}\mathcal{I}_{\theta_2}}\times \mathcal{M}_{4}
}$}};

\node at (-0.7,-6.6) [draw, rounded corners, fill=teal!10,
    minimum width=2.5cm,
    minimum height=1.5cm] (block4) { \textcolor{violet}{
$\substack{\textstyle\textbf{Type}~\mathrm{IIA}~\textbf{theory}\\ \\ \textstyle \textbf{on} \\\textstyle \mathbb{R}^{2,1}\times  \frac{\textstyle\mathbf{R}_{3}\times\tilde{\mathbf{S}}^{1}_{\theta_{1}}\times\mathbf{S}^{1}_{\theta_{2}}}{\textstyle\Omega\mathcal{I}_{3}\mathcal{I}_{\theta_{1}}\mathcal{I}_{\theta_{2}}}\times \mathcal{M}_{4} 
}$}};

\node at (1.8, -9.9) [draw, rounded corners, fill=teal!10,
    minimum width=2.5cm,
    minimum height=1.5cm] (block5) { \textcolor{violet}{
$\substack{\textstyle \mathrm{M}~\textbf{theory}\\ \\ \textstyle \textbf{on} \\\textstyle \mathbb{R}^{2,1}\times  \frac{\mathbf{R}_{3}\times\tilde{\mathbf{S}}^{1}_{\theta_{1}}\times\mathbf{S}^{1}_{\theta_{2}}}{\mathcal{I}_{3}\mathcal{I}_{\theta_{1}}\mathcal{I}_{\theta_{2}}}\times \mathcal{M}_{4}\times \mathbf{S}^{1}_{11}}$}};

\node at (4.0, -13.2) [draw, rounded corners, fill=teal!10,
    minimum width=2.5cm,
    minimum height=1.5cm] (block6) { \textcolor{violet}{
$\substack{\textstyle \mathrm{M}~\textbf{theory}\\ \\ \textstyle \textbf{on} \\\textstyle \mathbb{R}^{2,1}\times  \hat{\mathbf{R}}_{3}\times\hat{\mathbf{S}}^{1}_{\theta_{1}}\times\frac{\mathbf{S}^{1}_{\theta_{2}}}{\mathcal{I}_{\theta_{2}}}\times \mathcal{M}_{4}\times \mathbf{S}^{1}_{11}}$}};

\node at (-5.3, -13.2) [draw, rounded rectangle, fill=teal!10,
    minimum width=2.5cm,
    minimum height=1.5cm] (block7) { \textcolor{violet}{
$\substack{\textstyle\textbf{Heterotic}~\mathrm{E}_{8}\times \mathrm{E}_{8}~\textbf{theory}\\ \\ \textstyle \textbf{on} \\ \textstyle \mathbb{R}^{3,1}\times   \hat{\mathbf{S}}^{1}_{\theta_1}\times\mathcal{M}_{4}\times {\mathbf{S}}_{11}^{1}
}$}};
\draw[-latex] 
(block1) -- (block2)
(block2) edge (block3)
    (block3) edge (block4)
    (block4) edge (block5)
    (block5) edge (block6)
    (block6) edge (block7);
\node at (0,.5) {${\substack{\textstyle\text{Dimensional reduction } \\ \textstyle \text{along $x^{11}$} }}$} ; 
\node at (2.3,-1.7) { $\substack{\textstyle\mathrm{T}~\text{duality}\\ \textstyle \text{along}~x^3}$};
\node at (-0.2,-4.9) { $\substack{\textstyle\mathrm{T}~\text{duality}\\ \textstyle \text{along}~\theta_{1}}$};
\node at (-.6,-8.3) { $\substack{\textstyle\text{Dimensional}\\ \textstyle \text{uplift}}$};
\node at (2.1,-11.5) {Blow-ups};
\node at (5.0,-11.5) {$\left(\frac{\mathbf{R}_{3}}{\mathcal{I}_{3}}\rightarrow \hat{\mathbf{R}}_{3},~\frac{\widetilde{\mathbf{S}}^{1}_{\theta_1}}{\mathcal{I}_{\theta_1}}\rightarrow \hat{\mathbf{S}}^{1}_{\theta_{1}} \right)$};
\node at (-0.8,-12.5) {$\substack{\textstyle\text{Dimensional reduction}\\ \textstyle \text{along}~\theta_{2}} $ };
\end{tikzpicture}
    \caption{Duality sequence to go from the $\mathrm{M}$ theory configuration (\ref{MthThreewarpFac}) to heterotic $\mathrm{E}_{8}\times \mathrm{E}_{8}$ string theory (\ref{HetroTicE8XE8 metric})}
    \label{Het E8XE8 tree diagram}
\end{figure}

The above analysis confirms that a dimensional reduction of the orbifold coordinate ${\theta_2}$ of the $\mathrm{M}$ theory configuration (\ref{MthThreewarpFac}) fails to give rise to a well defined de Sitter configuration on the resulting  heterotic $\mathrm{E}_{8}\times \mathrm{E}_{8}$ string theory. Below we elaborate on a duality sequence (see Figure \ref{Het E8XE8 tree diagram} for pictorial demonstration) that is devoid of any such pathologies. 

[$\bullet$] On the configuration (\ref{MthThreewarpFac}), the only fluxes that we can turn on  are $\mathbf{C}_{am\theta_2}$\footnote{For example, if we turn on $\mathbf{C}_{m\theta_{1}\theta_2}$ then at the final stage of the duality sequence it would give rise to the \textbf{NS}-\textbf{NS} two form  $\mathbf{B}_{3m}$ in the $\mathrm{E}_{8}\times \mathrm{E}_{8}$ heterotic theory. However, we do not want the two form potential to have any leg along the external space time. Likewise, $\mathbf{C}_{am\theta_1}$ breaks the isometry of de Sitter space. Hence we can not turn on such fluxes.} where $a$ runs over the $\mathrm{M}$ theory torus along 3,11 and $y^{m}\in \mathcal{M}_{4}$. 

Dimensional reduction along $x^{11}$ of $\mathrm{M}$ theory leads to 
type IIA theory on $\mathbb{R}^{2,1}\times \mathbf{S}^{1}_{\theta_{1}}\times \frac{\textstyle \mathbf{S}_{\theta_{2}}^{1}}{\textstyle \Omega \mathcal{I}_{\theta_{2}}}\times \mathcal{M}_{4}\times \mathbf{S}^{1}_{3}$ with metric 
\begin{align}
ds^{2}_{\mathrm{IIA}}&=g_{s}^{-2} \left(-dt^2+ \sum_{i=1}^{2}\left(dx^{i}\right)^2\right)+\mathrm{H}^{2}(y)\bigg[\mathrm{F}_{1}(t)~(d\theta_{1})^2+\mathrm{F}_{3}(t)~(d\theta_{2})^2+\mathrm{F}_{2}(t)g_{mn}dy^m dy^n\bigg]\nonumber\\
&~~~~~~~~~~~~~~~~~+{g_{s}^{2}}~(dx^3)^2
\end{align}
and  $g_{s}=\frac{\mathrm{H}(y)}{a(t)}$ is the corresponding type $\mathrm{IIA}$ coupling constant. The \textbf{NS-NS} and \textbf{R-R} fluxes on this theory are ${B}^{}_{m\theta_2}$ and ${C}_{3m\theta_2}$ respectively. This theory has $\mathrm{O}8$ planes and $\mathrm{D}8$ branes located at the fixed points of the orientifold $\mathbf{S}^{1}_{\theta_2}/\Omega \mathcal{I}_{\theta_2}.$

[$\bullet$] $\mathrm{T}$ duality along $x^{3}$ direction would give rise to type $\mathrm{IIB}$ theory on $\mathbb{R}^{2,1}\times  \frac{\textstyle \mathbf{R}_{3}\times\mathbf{S}^{1}_{\theta_1}\times \mathbf{S}^{1}_{\theta_2}}{\textstyle\Omega(-1)^{\mathrm{F_{L}}}\mathcal{I}_{3}\mathcal{I}_{\theta_2}}\times \mathcal{M}_{4}
 $ with the following metric configuration
\begin{align}
ds^{2}_{\mathrm{IIB}}&=g_{s}^{-2} \left(-dt^2+ \sum_{i=1}^{3}\left(dx^{i}\right)^2\right)+\mathrm{H}^{2}(y)\bigg[\mathrm{F}_{1}(t)~(d\theta_{1})^2+\mathrm{F}_{3}(t)~(d\theta_{2})^2+\mathrm{F}_{2}(t)~g_{mn}dy^m dy^n\bigg].\label{DualIIBE8XE8}
\end{align}
The type $\rm{IIB}$ coupling $g_{b}=1$ and the \textbf{NS-NS} and \textbf{R-R} fluxes are $\mathcal{B}^{(1)}_{m\theta_2}$ and ${\mathcal{B}}^{(2)}_{m\theta_2}$ respectively. Clearly, the vector bundles of this theory are $\mathrm{O}7$ planes and $\mathrm{D}7$ branes along $\mathbb{R}^{2,1}\times \mathbf{S}^{1}_{\theta_1}\times\mathcal{M}_{4}$. 

[$\bullet$] A further $\mathrm{T}$ duality along $\theta_{1}$ leads to type $\mathrm{IIA}$ theory on $\mathbb{R}^{2,1}\times  \frac{\textstyle \mathbf{R}_{3}\times\tilde{\mathbf{S}}^{1}_{\theta_{1}}\times\mathbf{S}^{1}_{\theta_{2}}}{\textstyle \Omega\mathcal{I}_{3}\mathcal{I}_{\theta_{1}}\mathcal{I}_{\theta_{2}}}\times \mathcal{M}_{4}$ whose metric is as follows
\begin{align}
d\tilde{s}^{2}_{\mathrm{IIA}}&=g_{s}^{-2} \left(-dt^2+ \sum_{i=1}^{3}\left(dx^{i}\right)^2\right)+\bigg[\frac{1}{\mathrm{H}^2\mathrm{F}_{1}(t)}(d\theta_{1})^2+\mathrm{H}^{2}(y)\mathrm{F}_{3}(t)(d\theta_{2})^2\nonumber\\
&~~~~~~~~~~~~~~~~~~~~~~~~+\mathrm{H}^{2}(y)\mathrm{F}_{2}(t)g_{mn}dy^m dy^n\bigg].\label{TypeIIA2MetRic} 
\end{align}
The coupling constant of this theory is given by  
\begin{align}
    \tilde{g}_{s}=\frac{1}{{\rm H}(y)\sqrt{{\rm F}_{1}(t)~}},\label{IIA2CouplinG}
\end{align}
and the corresponding fluxes are ${C}_{\theta_1 m\theta_2}$, ${B}^{}_{m\theta_2}$. Also, there are $\mathrm{O}6$ planes and $\mathrm{D}6$ branes along $\mathbb{R}^{2,1}\times \mathcal{M}_{4}.$

[$\bullet$] Now, we perform a dimensional uplift to obtain
$\mathrm{M}$ theory on $\mathbb{R}^{2,1}\times  \frac{\textstyle\mathbf{R}_{3}\times\tilde{\mathbf{S}}^{1}_{\theta_{1}}\times\mathbf{S}^{1}_{\theta_{2}}}{\textstyle \mathcal{I}_{3}\mathcal{I}_{\theta_{1}}\mathcal{I}_{\theta_{2}}}\times \mathcal{M}_{4}\times \mathbf{S}^{1}_{11}$. This is nothing but the Taub-NUT geometry where the Taub-NUT base is the orbifold $\frac{\mathbf{R}_{3}\times\tilde{\mathbf{S}}^{1}_{\theta_1}\times\mathbf{S}^1_{\theta_2}}{\mathcal{I}_{3}\mathcal{I}_{\theta_1}\mathcal{I}_{\theta_2}}$ and $\mathbf{S}^1_{11}$ is the circle fibration. 
{
\begin{align}
    ds^{2}_{\mathrm{M}-\text{th}}&=(\tilde{g}_{s})^{-2/3}g_{s}^{-2} \left(-dt^2+ \sum_{i=1}^{3}\left(dx^{i}\right)^2\right)+(\tilde{g}_{s})^{-2/3}\bigg[\frac{1}{\mathrm{H}^2~\mathrm{F}_{1}(t)~}(d\theta_{1})^2\nonumber\\
&~~+\mathrm{H}^{2}(y)~\mathrm{F}_{3}(t)~ (d\theta_{2})^2+\mathrm{H}^{2}(y)\mathrm{F}_{2}(t)~g_{mn}dy^m dy^n\bigg] +\tilde{g}_{s}^{4/3}(d\tilde{x}_{11})^2\nonumber\\
&=\frac{\left({\rm H}^2(y)~\right)^{1/3}{\rm F}_{1}^{1/3}(t)}{g_{s}^{2}}\left(-dt^2+ \sum_{i=1}^3\left(dx^{i}\right)^2\right)\nonumber\\
&+\bigg[\frac{{\rm F}_{1}^{-2/3}(t)}{\left({\rm H}^2(y) \right)^{2/3}}(d\theta_1)^2+\left({\rm H}^{8}(y)\right)^{1/3}{\rm F}_{1}^{1/3}(t){\rm F}_{3}(t)(d\theta_2)^2\nonumber\\
+\big({\rm H}^8&(y) \big)^{1/3}{\rm F}_{1}^{1/3}(t){\rm F}_{2}(t)g_{mn}dy^{m}dy^{n}\bigg]+({\rm H}^{-2}(y)  )^{2/3}{\rm F}_{1}^{-2/3}(t) (d\tilde{x}_{11})^2.\label{Mth2ndE8E8dS}
\end{align}}
The three form potentials on this theory are $\mathbf{C}_{\theta_1 m\theta_2}$, $\mathbf{C}_{m\theta_2 r}$.

[$\bullet$] We know that $\mathrm{M}$ theory compactified on $\mathbf{S}^{1}/\mathbb{Z}_{2}$ give rise to heterotic $\mathrm{E}_{8}\times\mathrm{E}_{8}$ theory \cite{1Horava_1996}. So we perform two blow-ups on the  solitonic structure of the $\mathrm{M}$ theory configuration arrived at the previous step 
\begin{align}
    \frac{\textstyle \mathbf{R}_{3}}{\textstyle \mathcal{I}_{3}}\rightarrow \hat{\mathbf{R}}_{3},~~~\frac{\textstyle \tilde{\mathbf{S}}^{1}_{\theta_{1}}}{\textstyle \mathcal{I}_{\theta_{1}}}\rightarrow \hat{\mathbf{S}}^1_{\theta_{1}}.
\end{align}
This give rise to the $\mathrm{M}$ theory configuration  on $\mathbb{R}^{3,1}\times\hat{\mathbf{S}}^1_{\theta_{1}}\times\frac{\mathbf{S}_{\theta_2}^1}{\mathcal{I}_{\theta_2}}\times \mathcal{M}_{4}\times\mathbf{S}^{1}_{11}$. Therefore, we have two Horava-Witten walls at the two fixed points of $\frac{\textstyle \mathbf{S}^{1}_{\theta_{2}}}{\textstyle \mathcal{I}_{\theta_{2}}}$. 

[$\bullet$]  Dimensional reduction along the $\theta_{2}$ direction would then give us the {heterotic ${\rm E}_{8}\times {\rm E}_{8}$} theory on $\mathbb{R}^{3,1}\times   \hat{\mathbf{S}}^{1}_{\theta_1}\times\mathcal{M}_{4}\times {\mathbf{S}}_{11}^{1}
$. The metric configuration of the same being
\begin{align}
ds^{2}_{\text{Het}}&=\tilde{g}_{s}^{-1}g_{s}^{-2}\mathrm{H}(y)\sqrt{{\rm F}_{3}(t)} \left(-dt^2+ \sum_{i=1}^3\left(dx^{i}\right)^2\right)\nonumber\\
&+\tilde{g}_{s}^{-1}\mathrm{H}(y)\sqrt{{\rm F}_{3}(t)}\bigg[\frac{1}{\mathrm{H}^2(y)\mathrm{F}_{1}(t)~}(d\theta_{1})^2+\mathrm{H}^{2}(y)\mathrm{F}_{2}(t)~g_{mn}dy^m dy^n\bigg]\nonumber\\
&+\tilde{g}_{s}\mathrm{H}(y)\sqrt{\mathrm{F}_{3}(t)~}(d\tilde{x}_{11})^2\nonumber\\
    &=a^2(t)\mathrm{F}^{1/2}_{1}(t)\mathrm{F}_{3}^{1/2}(t)\left(-dt^2+ \sum_{i=1}^3\left(dx^{i}\right)^2\right)\nonumber\\
&+\left[\mathrm{F}_{1}^{-1/2}\mathrm{F}_{3}^{1/2} (d\theta_{1})^2+\mathrm{H}^{4}(y)\mathrm{F}^{1/2}_{1}\mathrm{F}_{3}^{1/2}\mathrm{F}_{2}g_{mn}dy^{m}dy^{n}\right]+\mathrm{F}_{1}^{-1/2}\mathrm{F}_{3}^{1/2}(d\tilde{x}_{11})^2.\label{HetroTicE8XE8 metric}
    \end{align}
Coupling constant of the heterotic $\mathrm{E}_{8}\times \mathrm{E}_{8}$ theory is as follows 
\begin{align}
g_{\text{Het}}&=\mathrm{H}^{2}(y)\mathrm{F}_{1}^{1/4}(t)\mathrm{F}_{3}^{3/4}(t),\label{CouplIngE8E8HeT}
\end{align}
and the non-vanishing two-form potentials are $\mathbf{B}_{m\theta_1},~\mathbf{B}_{mr}$.
 Now, demanding de Sitter isometry on the heterotic $\mathrm{E}_{8}\times \mathrm{E}_{8}$ theory gives the following form of the $\mathrm{IIA}$ coupling constant $g_{s}$  
\begin{align}
a^2(t){\sqrt{\mathrm{F}_{1}(t)\mathrm{F}_{3}(t)}}{}&\equiv g_{s}^{-2}\mathrm{H}^2(y)\sqrt{\mathrm{F}_{1}(t)\mathrm{F}_{3}(t)}=\frac{1}{\Lambda t^{2}}\nonumber\\
\implies \frac{g_{s}}{\mathrm{H}(y)}=&\sqrt{\Lambda}|t|\left(\mathrm{F}_{1}(t)\mathrm{F}_{3}(t)\right)^{1/4}~~\text{with}~~-\infty <t<0.\label{gsinE8thEorY}
\end{align}
Further, the time independence of the four dimensional Newton's constant implying the time independence of the volume of the internal space of $\mathrm{E}_{8}\times \mathrm{E}_{8}$ theory give rise to the following constraint between the three time dependent warp factors
    \begin{align}
\mathrm{F}_{1}(t)\mathrm{F}_{2}^{4}(t)\mathrm{F}_{3}^{3}(t)&=1.\label{ConstrF1F2F3}
    \end{align}

For the aforementioned duality sequence to proceed dynamically we have to ensure that the time dependent warp factor $\mathrm{F}_{1}(t)$ goes to zero faster as compared to $\mathrm{F}_{3}(t)$. Then the constraint (\ref{ConstrF1F2F3}) automatically implies that $\mathrm{F}_{2}(t)$ warp factor needs to grow. In order to quantify these warp factors, let us consider 
\begin{align}
    \mathrm{F}_{1}(t)=(\sqrt{\Lambda}|t|)^{p},~~~\mathrm{F}_{3}(t)=(\sqrt{\Lambda}|t|)^{q}.
\end{align}
From (\ref{gsinE8thEorY}) we obtain 
\begin{align}
    \frac{g_s}{\mathrm{H}(y)}&=(\sqrt{\Lambda}|t|)^{\frac{4+p+q}{4}}.
\end{align}
Therefore, 
\begin{align}
    \mathrm{F}_{1}(t)\equiv \mathrm{F}_{1}\left(\frac{g_s}{\mathrm{H}(y)} \right)&=\left(\frac{g_s}{\mathrm{H}(y)}\right)^{\frac{4p}{4+p+q}},\\
        \mathrm{F}_{3}(t)\equiv \mathrm{F}_{3}\left(\frac{g_s}{\mathrm{H}(y)} \right)&=\left(\frac{g_s}{\mathrm{H}(y)}\right)^{\frac{4q}{4+p+q}}.
\end{align}
Since the M theory torus along 3,11 directions (\ref{MthThreewarpFac}) should shrink to zero size taking us to the type IIB theory (\ref{DualIIBE8XE8}) we have to impose the following conditions 
\begin{align}
    0<\frac{4p}{4+p+q}<\frac{2}{3}~~,~~~0<\frac{4q}{4+p+q}<\frac{2}{3}.
\end{align}
Let 
\begin{align}
    \frac{4p}{4+p+q}=\frac{2\hat{\alpha}_o}{3},~~~\frac{4q}{4+p+q}=\frac{2\hat{\beta}_o}{3}\label{pqinalphabeta}
\end{align}
with $1>\hat{\alpha}_{o}>\hat{\beta}_o>0$ so that $\mathrm{F}_{1}(t)$ decays faster as compared to $\mathrm{F}_{3}(t)$. Solving (\ref{pqinalphabeta}) for $p$ and $q$ we get 
\begin{align}
    p=\frac{4\hat{\alpha}_{o}}{6-\hat{\alpha}_o-\hat{\beta}_o},~~~~q=\frac{4\hat{\beta}_o}{6-\hat{\alpha}_o-\hat{\beta}_o}.
\end{align}
This implies 
\begin{align}
    \frac{g_s}{\mathrm{H}(y)}&=\left(\sqrt{\Lambda}|t|\right)^{\frac{6}{6-\hat{\alpha}_o-\hat{\beta}_o}},\label{StrCCinE8E8}
\end{align}
and the two decreasing warp factors take the following form 
\begin{align}
    \mathrm{F}_{1}(t)&=\begin{cases}
    (\Lambda t^2)^{\frac{2\hat{\alpha}_o}{6-\hat{\alpha}_o-\hat{\beta}_o}},~~~~~-\frac{1}{\sqrt{\Lambda}}<t \leq -\epsilon\\
    \\
    (\Lambda t^2)^{\frac{\hat{\gamma}_o}{3-\hat{\gamma}_o}},~~~~~~~~~~~-\epsilon <t<0
    \end{cases}\label{F1(t)HeteRE8E8}\\
    \mathrm{F}_{3}(t)&=\begin{cases}
    (\Lambda t^2)^{\frac{2\hat{\beta}_o}{6-\hat{\alpha}_o-\hat{\beta}_o}},~~~~~-\frac{1}{\sqrt{\Lambda}}<t\leq -\epsilon\\
    \\
    (\Lambda t^2)^{\frac{\hat{\gamma}_o}{3-\hat{\gamma}_o}},~~~~~~~~~~~-\epsilon <t<0
    \end{cases}\label{F3(t)HeteRE8E8}
\end{align}
One might wonder the reason behind setting $\hat{\alpha}_{o}=\hat{\beta}_{o}=\hat{\gamma}_{o}$ in the temporal domain $-\epsilon <t<0$. Note that, even though the constraint (\ref{ConstrF1F2F3}) keeps the volume of the six dimensional internal space of heterotic $\mathrm{E}_8\times \mathrm{E}_{8}$ theory (\ref{HetroTicE8XE8 metric}) time independent, in the late time limit, both the circles with coordinates $\theta_{1}$ and $\tilde{x}_{11}$ grow as 
\begin{align}
\left(\frac{\mathrm{F}_{3}}{\mathrm{F}_1}\right)^{\frac{1}{4}}\equiv (\Lambda t^2)^{\frac{\hat{\beta}_o-\hat{\alpha}_o}{2(6-\hat{\alpha}_o-\hat{\beta}_o)}}\rightarrow \infty,~~~(\because 1>\hat{\alpha}_o>\hat{\beta}_o>0)
\end{align}
 and the volume of the internal four manifold $\mathcal{M}_{4}$ decays as 
 \begin{align}
\left(\frac{\mathrm{F}_{1}}{\mathrm{F}_{3}}\right)^{\frac{1}{2}}\equiv (\Lambda t^2)^{\frac{\hat{\alpha}_o-\hat{\beta}_o}{6-\hat{\alpha}_o-\hat{\beta}_o}}\rightarrow 0.   \label{VolM4E8E8ThrY}  
 \end{align}
 In the domain $-\epsilon <t<0$, once we set $\hat{\alpha}_{o}=\hat{\beta}_{o}=\hat{\gamma}_{o}$, in other words, identify $\mathrm{F}_{1}$ with $\mathrm{F}_{3}$, leading to the following metric structure 
\begin{align}
 ds^2_{\mathrm{Het}}&=   a^2(t)~\mathrm{F}_{1}(t)\left(-dt^2+ \sum_{i}\left(dx^{i}\right)^2+\left(dx^3\right)^2\right)+(d\theta_{1})^2+(d\tilde{x}_{11})^2\nonumber\\
&~~~~ +\mathrm{H}^{4}(y)\mathrm{F}_{1}(t)\mathrm{F}_{2}(t)g_{mn}dy^{m}dy^{n}
\end{align}
we obtain four dimensional de Sitter isometry by demanding
\begin{align}
    a^2(t) \mathrm{F}_{1}(t)&=\frac{1}{\Lambda t^2}.
\end{align}
 Moreover, the constraint (\ref{ConstrF1F2F3}) now becomes
 \begin{align}
     \mathrm{F}_{1}(t)\mathrm{F}_{2}(t)=1,
 \end{align}
hence in the temporal domain $-\epsilon <t<0$ we obtain time independent four dimensional Newton's constant as well as time independent internal space in the heterotic $\mathrm{E}_{8}\times \mathrm{E}_{8}$ theory with de Sitter isometry.    

To implement the behaviors of $\mathrm{F}_{1}(t)$ and $\mathrm{F}_{3}(t)$ as in equations (\ref{F1(t)HeteRE8E8}) and (\ref{F3(t)HeteRE8E8}) we define 
\begin{align}
    \mathrm{F}_{1}(t)&= (\Lambda t^2)^{\frac{2\hat{\alpha}(t)}{6-\hat{\alpha}(t)-\hat{\beta}(t)}}\equiv \left(\frac{g_s}{\mathrm{H}(y)} \right)^{\frac{2\hat{\alpha}(t)}{3}},~\mathrm{where}~\hat{\alpha}(t)=\begin{cases}
     \hat{\alpha}_o,~~~~-\frac{1}{\sqrt{\Lambda}}<t\leq -\epsilon\\
        \\
    \hat{\gamma}_{o},~~~~~-\epsilon<t<0
    \end{cases}\label{EXpofF1(t)}\\
        \mathrm{F}_{3}(t)&= (\Lambda t^2)^{\frac{2\hat{\beta}(t)}{6-\hat{\alpha}(t)-\hat{\beta}(t)}}\equiv \left(\frac{g_s}{\mathrm{H}(y)} \right)^{\frac{2\hat{\beta}(t)}{3}},~\mathrm{where}~\hat{\beta}(t)=\begin{cases}
        \hat{\beta}_o,~~~~-\frac{1}{\sqrt{\Lambda}}<t\leq -\epsilon\\
        \\
        \hat{\gamma}_{o},~~~~~~-\epsilon<t<0
    \end{cases}\label{EXpofF3(t)}
\end{align}
where $\frac{g_{s}}{\mathrm{H}(y)}$ is given as 
\begin{align}
    \frac{g_{s}}{\mathrm{H}(y)}&=\left(\Lambda t^2 \right)^{\frac{3}{6-\hat{\alpha}(t)-\hat{\beta}(t)}}.\label{TimedepgsinE8Thr}
\end{align}
 Using the constraint (\ref{ConstrF1F2F3}), we find 
\begin{align}
    \mathrm{F}_{2}(t)=\left(\frac{g_s}{\mathrm{H}(y)}\right)^{-\frac{\hat{\alpha}(t)}{6}-\frac{\hat{\beta}(t)}{2}}=\left(\sqrt{\Lambda}|t|\right)^{-\left(\frac{\hat{\alpha}(t)+3\hat{\beta}(t)}{6-\hat{\alpha}(t)-\hat{\beta}(t)}\right)}. \label{ExpofF2inE8TheorY}
\end{align}

We determine the value of $\epsilon$ by letting the volume of the four manifold $\mathcal{M}_{4}$ (\ref{VolM4E8E8ThrY}) to attain size $\mathrm{V}_{1}>>\alpha^{\prime~2}$ at $|t|=\epsilon.$ If $\mathrm{V}_{o}(-1/\sqrt{\Lambda})>>\mathrm{V}_1$ is its stabilized volume at $|t|=1/\sqrt{\Lambda}$ then  
\begin{align}
    \frac{\left(\frac{g_s}{{\rm H}(y)}\right)^{\frac{\hat{\alpha}_o-\hat{\beta}_o}{3}}\bigg|_{|t|=\epsilon}}{\left(\frac{g_s}{{\rm H}(y)}\right)^{\frac{\hat{\alpha}_o-\hat{\beta}_o}{3}}\bigg|_{|t|=1/\sqrt{\Lambda}}}
    &=\frac{\mathrm{V}_1}{\mathrm{V}_o(-1/\sqrt{\Lambda})}\nonumber\\
    \implies (\sqrt{\Lambda}\epsilon)^{\frac{2(\hat{\alpha}_o-\hat{\beta}_o)}{6-\hat{\alpha}_o-\hat{\beta}_o}}&=\frac{\mathrm{V}_1}{\mathrm{V}_{o}(-1/\sqrt{\Lambda})}\nonumber\\
    \implies \epsilon&=\frac{1}{\sqrt{\Lambda}}\left[\frac{\mathrm{V}_1}{\mathrm{V}_o(-1/\sqrt{\Lambda})} \right]^{\frac{6-\hat{\alpha}_o-\hat{\beta}_o}{2\hat{\alpha}_o-2\hat{\beta}_o}}.
\end{align}

Thus the coupling (\ref{IIA2CouplinG}) of the type IIA configuration (\ref{TypeIIA2MetRic}) approaches the strong coupling limit as  
\begin{align}
    \tilde{g}_{s}\propto \frac{1}{\sqrt{\mathrm{F}_{1}(t)}}=\left(\frac{g_{s}}{\mathrm{H}(y)}\right)^{-\hat{\alpha}(t)/3}.
\end{align}
The above coupling becomes strong faster as compared to the rate at which the $\mathbf{S}^1_{\theta_2}$ circle shrinks to zero size, viz. $\sqrt{\mathrm{F}_{3}(t)}\equiv \left( \frac{g_s}{\mathrm{H}(y)}\right)^{\hat{\beta}(t)/3},$ leading to an uplift to M-theory (\ref{Mth2ndE8E8dS}). A careful look into this metric structure of M-theory tells us that in the late time all the internal directions are growing except $\mathbf{S}^1_{\theta_2}$ and $\mathcal{M}_{4}$. This is because the volume of the four manifold scales as 
\begin{align}
\sqrt{\left(\mathrm{F}_{1}^{1/3}\mathrm{F}_{2}\right)^4}=\left( \frac{g_s}{\mathrm{H}(y)}\right)^{\frac{\hat{\alpha}(t)}{9}-\hat{\beta}(t)}\rightarrow 0 ~~~(\mathrm{as~long~as}~\frac{\hat{\alpha}_o}{9}>  \hat{\beta}_o),
\end{align}
and the circle $\mathbf{S}^1_{\theta_2}$ scales as 
\begin{align}
\sqrt{\mathrm{F}_{1}^{1/3}\mathrm{F}_{3}}&=\left(\frac{g_s}{\mathrm{H}(y)}\right)^{\frac{\hat{\alpha}(t)+3\hat{\beta}(t)}{9}}\rightarrow 0.
\end{align}
Since we demand only the circle $\mathbf{S}^1_{\theta_2}$ to shrink but not the four manifold, the two parameters $\hat{\alpha}_{o}$ and $\hat{\beta}_{o}$  must satisfy $1>\hat{\alpha}_{o}>\hat{\beta}_{o}>\frac{\hat{\alpha}_o}{9}>0$.  

\noindent
In the next section, we review the path integral analysis to compute the expectation value of the metric operator (\ref{ExpValueOfMetOp}) in the M theory \cite{Brahma:2022wdl}. 
\section{Path integral analysis }
\label{sec:PIinMth}
In the low energy limit, the effective action of M-theory becomes identical to that of the 11 dimensional supergravity theory. The field content of this theory is as follows: it contains graviton $ \mathrm{g}_{\mathrm{M}\mathrm{N}}$ - the symmetric traceless tensor of the little group $SO(9)$ - which has 44 physical degrees of freedom, the three form field $ \mathrm{C}_{\mathrm{MNP}}$ with 84 degrees of freedom and the 32 component Majorana spinor $\Psi^{\mathrm{M}}_{a~}$ ($a$ denotes the spinor index)-which has 128 degrees of freedom. Here $\mathrm{M},\mathrm{N},\mathrm{P}\in \mathbb{R}^{2,1}\times \mathcal{M}_{4}\times \mathcal{M}_{2}\times \frac{\mathbb{T}^2}{\mathcal{G}}$ of the $\mathrm{M}$-theory metric (\ref{MthEorYmeTric}). Let the Glauber-Sudarshan state be denoted as $|\sigma\rangle=|{\alpha},{\beta},{\gamma}\rangle$ where $|{\alpha}\rangle \equiv |{\alpha}_{\mathrm{MP}}\rangle$, $|{\beta} \rangle\equiv |{\beta}_{\mathrm{MNP}}\rangle$ and $|{\gamma}\rangle\equiv |{\gamma}_{\mathrm{M}}\rangle$ are the Glauber-Sudarshan states associated to the graviton, three form field and the Rarita-Schwinger fermion degrees of freedom respectively. We express the expectation value of the metric operator with respect to the aforementioned state as
\begin{align}
    \langle \hat{\mathbf{g}}_{\mathrm{MN}}\rangle_{\sigma}&=\frac{\langle \sigma |\hat{\mathbf{g}}_{\mathrm{MN}}|\sigma \rangle}{\langle \sigma|\sigma\rangle}=\frac{\langle \Omega | \hat{\mathbb{D}}^{\dagger}(\sigma){\hat{\mathbf{g}}_{\rm{MN}}}\hat{\mathbb{D}}(\sigma)|\Omega\rangle}{\langle \Omega | \hat{\mathbb{D}}^{\dagger}(\sigma)\hat{\mathbb{D}}(\sigma)|\Omega \rangle}\nonumber\\
    &=\frac{\int [{\cal D} \rm{g}_{\rm MN}] [{\cal D}{\rm C}_{\rm MNP}] [{\cal D}\overline\Psi_{\rm M}] [{\cal D}
\Psi_{\rm N}]~e^{i{\bf S}_{\rm tot}}~ \mathbb{D}^\dagger({\alpha}, {\beta}, \gamma)  g_{\rm MN}
\mathbb{D}({\alpha}, {\beta}, \gamma) }{
\int [{\cal D} g_{\rm MN}] [{\cal D}{\rm C}_{\rm MNP}] [{\cal D}\overline\Psi_{\rm M}] [{\cal D}
\Psi_{\rm N}]
~e^{i{\bf S}_{\rm tot}} ~\mathbb{D}^\dagger({\alpha}, {\beta}, \gamma) 
\mathbb{D}({\alpha}, {\beta}, \gamma)},\label{ExpofMetOp}
\end{align}
where total action, ${\bf S}_{\rm tot} \equiv {\bf S}_{\rm kinetic} + {\bf S}_{\rm interaction} + 
{\bf S}_{\rm ghost} + {\bf S}_{\rm gauge-fixing}$. Since the exact evaluation of the above path integral is difficult to perform, we consider a toy model without any loss of generality: viz. consider three scalar fields $(\varphi_{1},\varphi_{2},\varphi_{3})$ representing one of the components of graviton, three form field and the Rarita-Schwinger fermion respectively. Further, we take into account an  interaction term of the following form 
\begin{align}
\mathbf{S}_{\text{int}}&\equiv \int d^{11}x \sum_{n,...,s}c_{nmpqrs}\partial^{n} \varphi^{q}_{1}(x)\partial^{m}\varphi_{2}^{r}(x)\partial^{p}\varphi_{3}^{s}(x),\label{SinTeraCtiOn}
\end{align}
where $n+m+p\in 2 \mathbb{Z}_{+}$ 
and $(n,m,p,q,r,s)\in \mathbb{Z}_{+}$. Replacing the fields $\varphi_{1}(x),~\varphi_{2}(x)$ and $\varphi_{3}(x)$ in terms of their fourier modes, 
$\mathbf{S}_{\text{int}}$ takes the following form
{
\begin{align}
    \mathbf{S}_{\text{int}}&\equiv \int d^{11}x \sum_{n,...,s}c_{nmpqrs}~\partial^{n}\left(\int_{-\infty}^{\infty} d^{11}k~e^{ikx}~\widetilde{\varphi}_{1}(k)\right)^{q}\partial^{m}\left(\int_{-\infty}^{\infty} d^{11}l~e^{ilx}~\widetilde{\varphi}_{2}(l)\right)^{r}\nonumber\\
    &~~~~~~~~~~~
   \times \partial^{p}\left(\int_{-\infty}^{\infty} d^{11}f~e^{ifx}~\widetilde{\varphi}_{3}(f)\right)^{s}\nonumber\\
   &=\int d^{11}x \sum_{n,...,s}c_{nmpqrs} (i)^{n+m+p}\int \prod_{i=1}^{q}d^{11}k_{i} \prod_{j=1}^{r}d^{11}l_{j}\prod_{t=1}^{s}d^{11}f_{t}~\left(\sum_{i=1}^{q}k_{i} \right)^n \left(\sum_{j=1}^{r}l_{j} \right)^{m}\nonumber\\
   &\times \left(\sum_{t=1}^{s}f_{t} \right)^{p} \prod_{i=1}^{q}\widetilde{\varphi}_{1}(k_{i})\prod_{j=1}^{r}\widetilde{\varphi}_{2}(l_{j})\prod_{t=1}^{s}\widetilde{\varphi}_{3}(f_{t})\exp\left[i\left(\sum_{i=1}^{q}k_{i}+\sum_{j=1}^{r}l_{j}+\sum_{t=1}^{s}f_{t}\right)x \right]\nonumber\\
   &=\sum_{n,...,s}c_{nmpqrs} (-1)^{(n+m+p)/2}\int \prod_{i=1}^{q}d^{11}k_{i} \prod_{j=1}^{r}d^{11}l_{j}\prod_{t=1}^{s}d^{11}f_{t}~\left(\sum_{i=1}^{q}k_{i} \right)^n \left(\sum_{j=1}^{r}l_{j} \right)^{m}\nonumber\\
   &\times \left(\sum_{t=1}^{s}f_{t} \right)^{p} \prod_{i=1}^{q}\widetilde{\varphi}_{1}(k_{i})\prod_{j=1}^{r}\widetilde{\varphi}_{2}(l_{j})\prod_{t=1}^{s}\widetilde{\varphi}_{3}(f_{t})~\delta^{(11)}\left(\sum_{i=1}^{q}k_{i}+\sum_{j=1}^{r}l_{j}+\sum_{t=1}^{s}f_{t}\right)\label{DelTaFunCTionn}\\
   &= \sum_{n,...,s}c_{nmpqrs} (-1)^{(n+m+p)/2}\int \prod_{i=1}^{q}d^{11}k_{i} \prod_{j=1}^{r}d^{11}l_{j}\prod_{t=1}^{s-1}d^{11}f_{t}~\left(\sum_{i=1}^{q}k_{i} \right)^n \left(\sum_{j=1}^{r}l_{j} \right)^{m}\nonumber\\
   \times \bigg(&-\sum_{i=1}^{q}k_{i}-\sum_{j=1}^{r}l_{j} \bigg)^{p} \prod_{i=1}^{q}\widetilde{\varphi}_{1}(k_{i})\prod_{j=1}^{r}\widetilde{\varphi}_{2}(l_{j})\prod_{t=1}^{s-1}\widetilde{\varphi}_{3}(f_{t})\widetilde{\varphi}_{3}\left(-\sum_{i=1}^{q}k_{i}-\sum_{j=1}^{r}l_{j}-\sum_{t=1}^{s-1}f_{t}\right).\label{SinTerActiONinTegrAlREp}
\end{align}
}
Since the fields $\varphi(x)$ are \textit{real} valued
\begin{align}
    \varphi^{\star}(x)=\int_{-\infty}^{\infty}d^{11}k ~\widetilde{\varphi}^{\star}(k)&e^{-ikx}=\int_{-\infty}^{\infty}d^{11}k ~\widetilde{\varphi}^{\star}(-k)e^{ikx}=\varphi(x),
\end{align}
the fourier modes satisfy $\widetilde{\varphi}^{\star}(k)=\widetilde{\varphi}(-k)$. Therefore, $\mathbf{S}_{\rm{int}}$ can be expressed as 
\begin{align}
\mathbf{S}_{\text{int}}&= \sum_{n, ..., s} c_{nmpqrs} 
(-1)^{(n + m + 3p)/2} \int \prod_{i = 1}^q d^{11} {k}_i \prod_{j = 1}^r d^{11} {l}_j \prod_{t = 1}^{s - 1} d^{11} {f}_t~ \left(\sum_{i = 1}^q {k}_i\right)^n \left(\sum_{j = 1}^r {l}_j\right)^m\nonumber\\
\times  &\left(\sum_{i = 1}^q {k}_i + \sum_{j = 1}^ r {l}_j\right)^p
 \prod_{i = 1}^q \widetilde{\varphi}_1({k}_i) \prod_{j = 1}^r \widetilde{\varphi}_2({l}_j) \prod_{t = 1}^{s - 1} \widetilde{\varphi}_3({f}_t)
~\widetilde\varphi^\ast_3\left(\sum_{i = 1}^q {k}_i + \sum_{j = 1}^ r {l}_j + \sum_{t = 1}^{s-1} {f}_t\right)\nonumber\\
\equiv & \sum_{n, ..., s} c_{nmpqrs} 
(-1)^{(n + m + 3p)/2} \frac{1}{\mathrm{V}^{q+r+s-1}}\sum_{\left\{{k_{u_i}}\right\},\left\{{l_{v_j}}\right\},\left\{{f_{w_t}}\right\}}~ \left(\sum_{i = 1}^q {k}_{u_i}\right)^n \left(\sum_{j = 1}^r {l}_{v_j}\right)^m\nonumber\\
\times  \bigg(\sum_{i = 1}^q &{k}_{u_i} + \sum_{j = 1}^ r {l}_{v_j}\bigg)^p
 \prod_{i = 1}^q \widetilde{\varphi}_1({k}_{u_i}) \prod_{j = 1}^r \widetilde{\varphi}_2({l}_{v_j}) \prod_{t = 1}^{s - 1} \widetilde{\varphi}_3({f}_{w_t})
~\widetilde\varphi^\ast_3\left(\sum_{i = 1}^q {k}_{u_i} + \sum_{j = 1}^ r {l}_{v_j} + \sum_{t = 1}^{s-1} {f}_{w_t}\right).\label{FormOfSint}
\end{align}
Likewise, the kinetic part of the action expressed in terms of the fourier modes of the fields is as follows
\begin{align}
\mathbf{S}_{\mathrm{kin}}&=\int_{-\infty}^{\infty}d^{11}x~\big( \partial_{\mu}\varphi_{1}\partial^{\mu}\varphi_{1}+\partial_{\mu}\varphi_{2}\partial^{\mu}\varphi_{2}+\partial_{\mu}\varphi_{3}\partial^{\mu}\varphi_{3}\big)\nonumber\\
&=\int_{-\infty}^{\infty}d^{11}k~ k^{2}~|\widetilde{\varphi}_{1}(k)|^2+\int_{-\infty}^{\infty}d^{11}l~ l^{2}~|\widetilde{\varphi}_{2}(l)|^2+\int_{-\infty}^{\infty}d^{11}f~ f^{2}~|\widetilde{\varphi}_{3}(f)|^2.
\end{align}
There is another ingredient that we require in order to evaluate the path integral (\ref{ExpofMetOp}), viz. the displacement operator corresponding to the Glauber-Sudarshan state. As discussed in Appendix \ref{GS state}, it can be expressed in terms of the fourier modes of the fields as 
\begin{align}
\mathbb{D}(\sigma\equiv\{\alpha,\beta,\gamma\})&=\exp\bigg(\frac{1}{\mathrm{V}}\bigg(\sum_{i}\mathbf{Re}~{\alpha}(k_{i}) ~\mathbf{Re}~\widetilde{\varphi}_{1}(k_{i})+\mathbf{Im}~{\alpha}(k_{i})~\mathbf{Im}~\widetilde{\varphi}_{1}(k_{i})\nonumber\\
&~~~~~~~~~~~~~~+\sum_{j}\mathbf{Re}~{\beta}(l_{j}) ~\mathbf{Re}~\widetilde{\varphi}_{2}(l_{j})+\mathbf{Im}~{\beta}(l_{j})~\mathbf{Im}~\widetilde{\varphi}_{2}(l_{j})\nonumber\\
&~~~~~~~~~~~~~~+\sum_{t}\mathbf{Re}~{\gamma}(f_{t}) ~\mathbf{Re}~\widetilde{\varphi}_{3}(f_{t})+\mathbf{Im}~{\gamma}(f_{t})~\mathbf{Im}~\widetilde{\varphi}_{3}(f_{t})\bigg)\bigg).   
\end{align}
Now, we can write down the numerator of the path integral (\ref{ExpofMetOp}) as   
\begin{align}\label{NumPInT}
 &{\rm Num}\left[\langle  \hat{\mathbf{g}}_{\mathrm{MN}} \rangle_{{\sigma}}\right]\equiv 
\prod_{i, j, t}\int d\left({\bf Re}~\widetilde{\varphi}_1(k_i)\right) d\left({\bf Im}~\widetilde{\varphi}_1(k_i)\right) d\left({\bf Re}~\widetilde{\varphi}_2(l_j)\right) d\left({\bf Im}~\widetilde{\varphi}_2(l_j)\right)
\nonumber\\
&\times  d\left({\bf Re}~\widetilde{\varphi}_3(f_t)\right) d\left({\bf Im}~\widetilde{\varphi}_3(f_t)\right)
\cdot {1\over {\rm V}} \sum_{k_{i}^{''}} \psi_{{\bf k}''_i}({\bf x}, y, z) e^{-ik''_{i,0} t} 
\left({\bf Re}~\widetilde{\varphi}_1(k''_i) + i {\bf Im}~\widetilde{\varphi}_1(k''_i)\right) \nonumber\\
&\times  {\rm exp}\Big[{2\over {\rm V}}\sum_{i, j, t}\Big({\bf Re}~{\alpha}(k'_i)
~{\bf Re}~\widetilde{\varphi}_1(k'_i) + {\bf Im}~{\alpha}(k'_i)
~{\bf Im}~\widetilde{\varphi}_1(k'_i) + {\bf Re}~{\beta}(l'_j)
~{\bf Re}~\widetilde{\varphi}_2(l'_j) \nonumber\\
& ~~~~~+  {\bf Im}~{\beta}(l'_j)
~{\bf Im}~\widetilde{\varphi}_2(l'_j) + {\bf Re}~{\gamma}(f'_t)
~{\bf Re}~\widetilde{\varphi}_3(f'_t) + {\bf Im}~{\gamma}(f'_t)
~{\bf Im}~\widetilde{\varphi}_3(f'_t) \Big)\Big] \nonumber\\ 
& \times  {\rm exp}\Big[{i\over {\rm V}} \Big(\sum_{i} k_i^2 \vert \widetilde\varphi_1(k_i)\vert^2 + 
\sum_{j} l_j^2 \vert \widetilde\varphi_2(l_j)\vert^2 + 
\sum_{t} f_t^2 \vert \widetilde\varphi_3(f_t)\vert^2\Big)\Big] \nonumber\\
&\times  {\rm exp}\Big[ \sum_{\mathcal{S}^\prime}{i\over {\rm V}^u} c_{nmpqrs} (-1)^{(n + m + 3p)/2}
\left(\sum_{i = 1}^q {k}_{u_i}\right)^n \left(\sum_{j = 1}^r {l}_{v_j}\right)^m
\left(\sum_{i = 1}^q {k}_{u_i} + \sum_{j = 1}^ r {l}_{v_j}\right)^p\nonumber\\
& \times 
\prod_{i, j, t = 1}^{q, r, s-1}\Big({\bf Re}~\widetilde{\varphi}_1(k_{u_i}) + i {\bf Im}~\widetilde{\varphi}_1(k_{u_i})\Big) 
\Big({\bf Re}~\widetilde{\varphi}_2(l_{v_j}) + i {\bf Im}~\widetilde{\varphi}_2(l_{v_j})\Big) 
\Big({\bf Re}~\widetilde{\varphi}_3(f_{w_t}) + i {\bf Im}~\widetilde{\varphi}_3(f_{w_t})\Big) 
\nonumber\\
&\times 
\Big({\bf Re}~\widetilde{\varphi}_3\Big(\sum_{i = 1}^q {k}_{u_i} + \sum_{j = 1}^ r {l}_{v_j} + \sum_{t = 1}^{s-1}{f}_{w_t}\Big) - i {\bf Im}~\widetilde{\varphi}_3\Big(\sum_{i = 1}^q {k}_{u_i} + \sum_{j = 1}^ r {l}_{v_j} + \sum_{t = 1}^{s-1}{f}_{w_t}\Big)\Big)\Big], \end{align}
where $u=q+r+s-1$ and ${\cal S}^\prime= (\left\{{k_{u_i}}\right\},\left\{{l_{v_j}}\right\},\left\{{f_{w_t}}\right\},n,m,p,q,r,s)$. For simplification, we set the imaginary parts of the fourier modes to zero. Introducing the variables
\begin{equation}
\begin{aligned} (a_{j},b_{j},c_{j})\equiv & (a(k_j),b(l_j),c(f_j))=-\frac{i}{\mathrm{V}}(k_{j}^{2},l_{j}^{2},f_{j}^{2}),\\ (\overline{\alpha}_{i},\overline{\beta}_{i},\overline{\gamma}_{i})\equiv &(\overline{\alpha}(k_{i}),\overline{\beta}(l_{i}),\overline{\gamma}(f_{i}))=\frac{1}{\mathrm{V}}(\alpha(k_i),\beta(l_i),\gamma(f_i)), \label{abcDef}
\end{aligned}    
\end{equation}
we  re-write (\ref{NumPInT}) in the following way
{\begin{align}
     &{\rm Num}[\langle \hat{\mathbf{g}}_{\mathrm{MN}}\rangle_{\overline\sigma}]  =  \mathbb{C}(\overline\alpha_i, a_j, {\rm V}, ..) 
\int d\widetilde\varphi_1(k_1) 
~{\rm exp}\Big[-a_1\Big(\widetilde\varphi_1(k_1) - \frac{\overline{\alpha}_1}{a_1}  \Big)^2\Big]
~\cdots\nonumber\\ 
& ~~~~ \int d\widetilde\varphi_2(l_1) 
~{\rm exp}\Big[-b_1\Big(\widetilde\varphi_2(l_1) - \frac{\overline{\beta}_1}{b_1} \Big)^2\Big] \;\int d\widetilde\varphi_2(l_2) 
~{\rm exp}\Big[-b_2\Big(\widetilde\varphi_2(l_2) - \frac{\overline{\beta}_2}{b_2} \Big)^2\Big]\cdots\nonumber\\
& ~~~~\int d\widetilde\varphi_3(f_1) 
~{\rm exp}\Big[-c_1\Big(\widetilde\varphi_3(f_1) - \frac{\overline{\gamma}_1}{c_1} \Big)^2\Big]\int d\widetilde\varphi_3(f_2) 
~{\rm exp}\Big[-c_2\Big(\widetilde\varphi_3(f_2) - \frac{\overline{\gamma}_2}{c_2} \Big)^2\Big]\cdots \nonumber \\
& \times   {1\over {\rm V}}\Big(\widetilde\varphi_1(k_1)\psi_{{\bf k}_1}({\bf x}, y, z)e^{-ik_{0, 1}t} + 
\widetilde\varphi_1(k_2)\psi_{{\bf k}_2}({\bf x}, y, z)e^{-ik_{0, 2}t} + 
\widetilde\varphi_1(k_3)\psi_{{\bf k}_3}({\bf x}, y, z)e^{-ik_{0, 3}t} +\cdots \Big)\nonumber \\
&\times \Big(1 + {i}\sum_{{\cal S}}{\rm V}^{-u} c_{nmpqrs} (-1)^{v}
\Big(\begin{matrix} p \\ e_o \\ \end{matrix}\Big)(k_{u_1} + k_{u_2} +\cdots + k_{u_q})^{n + e_o}(l_{v_1} + l_{v_2} +\cdots + l_{v_r})^{m + p - e_o}\nonumber\\
 &\times \widetilde\varphi_1(k_{u_1})\widetilde\varphi_1(k_{u_2})... \widetilde\varphi_1(k_{u_q}) \widetilde\varphi_2(l_{v_1})
\widetilde\varphi_2(l_{v_2})... \widetilde\varphi_2(l_{v_r}) \widetilde\varphi_3(f_{w_1}) \widetilde\varphi_3(f_{w_2}) ... \widetilde\varphi_3(f_{w_{s-1}})
\widetilde\varphi_3(f_{w_s})\nonumber\\
& ~~ + {\cal O}(c^2_{nmpqrs})\Big), 
\label{NumErAtorOfExpValuE}
\end{align}}
where 
\begin{align}
{\cal S}&= \left(\left\{{k_{u_i}}\right\},\left\{{l_{v_j}}\right\},\left\{{f_{w_t}}\right\},n,m,p,q,r,s,e_o\right),~~    v=\frac{1}{2}(n+m+3p),\nonumber\\
&~~~~~~~~~~~~~~~~~~~~f_{w_{s}}=\sum_{i=1}^{q}k_{u_{i}}+\sum_{j=1}^{r}l_{v_{j}}+\sum_{t=1}^{s-1}f_{w_{t}},\label{fwsRelation}
\end{align}
and $\mathbb{C}(\overline{\alpha}_{i},a_{i},\mathrm{V},...)$ is an overall constant which is unimportant in our analysis.  
Note that, in (\ref{NumErAtorOfExpValuE}) we have called $\mathbf{Re}~{\xi}\equiv\xi$ for $\xi=(\widetilde{\varphi}_{1},\widetilde{\varphi}_{2},\widetilde{\varphi}_{3},\overline{\alpha}_{i},\overline{\beta}_{i},\overline{\gamma}_{i})$. In the following subsections we'll analyse in detail the above path integral for various cases. We follow the functional integral approach for computing path integrals as elaborated in Appendix \ref{FIntApproachPhi4}. 
\subsection{Contribution to the path integral at the tree level}
\label{subsec:TreeLevel}
Taking $c_{nmpqrs}$ to be zero we obtain  the tree level contribution to the path integral (\ref{NumErAtorOfExpValuE})
\pgfmathsetmacro\MathAxis{height("$\vcenter{}$")}
\begin{align}
{\rm Num}[\langle \hat{\mathbf{g}}_{\mathrm{MN}}\rangle_{\overline\sigma}]\bigg|_{\mathrm{Tree~level}}&=  \frac{1}{\mathrm{V}}\sum_{k_i}\int d\widetilde\varphi_1(k_i) 
~{\rm exp}\Big[-a_i\Big(\widetilde\varphi_1(k_i) - {\overline\alpha_i\over a_i}\Big)^2\Big]
\widetilde\varphi_1(k_i)\psi_{{\bf k}_i}({\bf x}, y, z)e^{-ik_{0, i}t}\nonumber\\
\prod_{j \ne i} 
\int d\widetilde\varphi_1(k_j) 
~&{\rm exp}\Big[-a_j\Big(\widetilde\varphi_1(k_j) - {\overline\alpha_j\over a_j}\Big)^2\Big]\times \prod_{u} 
\int d\widetilde\varphi_2(l_u) 
~{\rm exp}\Big[-b_u\Big(\widetilde\varphi_2(l_u) - {\overline\beta_u\over b_u}\Big)^2\Big] \nonumber\\
&~~\times \prod_{v} 
\int d\widetilde\varphi_3(f_v) 
~{\rm exp}\Big[-c_v\Big(\widetilde\varphi_3(f_v) - {\overline\gamma_v\over c_v}\Big)^2\Big]  .
\label{TrEEleVeldiAgNum}
\end{align}
Using the standard Gaussian integration results:
\begin{equation}
\begin{aligned}
    \int d\widetilde{\varphi}_{1}(k_{j})\exp\Big[-a_j\Big(\widetilde\varphi_1(k_j) - {\overline\alpha_j\over a_j}\Big)^2\Big]&= \left(\frac{\pi}{a_{j}}\right)^{1/2}\\
\int d\widetilde\varphi_1(k_i) 
~{\rm exp}\Big[-a_i\Big(\widetilde\varphi_1(k_i) - {\overline\alpha_i\over a_i}\Big)^2\Big]\widetilde{\varphi}_{1}(k_{i})&=\frac{\overline{\alpha}_{i}}{a_{i}}\left(\frac{\pi}{a_{i}}\right)^{1/2},\label{GInteGrals1}
\end{aligned}
\end{equation}
and taking the continuum limit we arrive at 
\begin{align}
    {\rm Num}[\langle \hat{\mathbf{g}}_{\mathrm{MN}}\rangle_{\overline\sigma}]\bigg|_{\mathrm{Tree~level}}&=\left(\prod_{j} \Big({\pi_j\over a_j}\Big)^{1/2} 
\prod_{u} \Big({\pi_u\over b_u}\Big)^{1/2} \prod_{v} \Big({\pi_v\over c_v}\Big)^{1/2} \right)\nonumber\\
&~~~\times\int d^{11}k ~{\overline\alpha(k)\over a(k)}~\psi_{{\bf k}}({\bf x}, y, z)e^{-ik_{0}t}.
\end{align}
where $\pi_{j}=\pi_{u}=\pi_{v}=\pi$. Note that, a similar tree level contribution for the denominator of $\langle \hat{\mathbf{g}}_{\mathrm{MN}}\rangle_{\overline{\sigma}}$ (\ref{ExpofMetOp}) give rise to
\begin{align}
   {\rm Den}[\langle \hat{\mathbf{g}}_{\mathrm{MN}}\rangle_{\overline\sigma}]\bigg|_{\mathrm{Tree~level}}&= \prod_{j} \Big({\pi_j\over a_j}\Big)^{1/2} 
\prod_{u} \Big({\pi_u\over b_u}\Big)^{1/2} \prod_{v} \Big({\pi_v\over c_v}\Big)^{1/2} ,
\end{align}
which implies the overall factor gets canceled out when we take into account the denominator as well. 
\subsection{First order quantum correction}
We now analyze the contribution to the path integral (\ref{NumErAtorOfExpValuE}) at first order in $c_{nmpqrs}$ due to the three field sectors ${\varphi}_{1},~{\varphi}_{2}$ and $~{\varphi}_{3}$ individually. Let us begin with the $\varphi_{2}$ sector.
\subsubsection{First order quantum correction due to \texorpdfstring{$\varphi_{2}$}{varphi2} field sector} The contribution to (\ref{NumErAtorOfExpValuE}) at $\mathcal{O}(c_{nmpqrs})$ due to $\varphi_{2}$ field is given by 
\begin{align}
    {\rm Num}[\langle \hat{\mathbf{g}}_{\mathrm{MN}}\rangle_{\overline\sigma}]\bigg|_{\mathcal{O}(c_{nmpqrs}),\varphi_2}&=\int d\widetilde{\varphi}_{2}(l_1){\rm exp}\Big[-b_1\Big(\widetilde\varphi_2(l_1) - \frac{\overline{\beta}_1}{b_1} \Big)^2\Big] \nonumber\\
    \times\int d\widetilde{\varphi}_{2}(l_2){\rm exp}\Big[-b_2&\Big(\widetilde\varphi_2(l_2) - \frac{\overline{\beta}_2}{b_2} \Big)^2\Big]\int d\widetilde{\varphi}_{2}(l_3){\rm exp}\Big[-b_3\Big(\widetilde\varphi_2(l_3) - \frac{\overline{\beta}_3}{b_3} \Big)^2\Big] \cdots \nonumber\\
\times\frac{1}{\mathrm{V}^{r}}\sum_{\left\{{l_{v_i}}\right\}}(l_{v_1}+l_{v_2}+&l_{v_3}+\cdots+l_{v_r})^{m+p-e_o}\widetilde\varphi_2(l_{v_1})
\widetilde\varphi_2(l_{v_2})\widetilde\varphi_2(l_{v_3})\cdots \widetilde\varphi_2(l_{v_r}).\label{O(c)phi2Case1}
\end{align}
To proceed further we consider various cases for the momenta $l_{v_i}$ of the $r$ copies of the field $\widetilde{\varphi}_{2}(l_{v_i})$ appearing in $\mathbf{S}_{\mathrm{int}}$ (\ref{FormOfSint}). 
\vskip.2in
\noindent{\textbf{Case 1:} {$\boldsymbol{l_{v_1} = l_{v_2} = l_{v_3} = \cdots = l_{v_r} \equiv l_i}$}}
\vskip.1in
For this case (\ref{O(c)phi2Case1}) becomes 
\begin{align}
    {\rm Num}[\langle \hat{\mathbf{g}}_{\mathrm{MN}}\rangle_{\overline\sigma}]\bigg|^{\mathrm{Case~1}}_{\mathcal{O}(c_{nmpqrs}),\varphi_2}&=\frac{1}{\mathrm{V}^r}\sum_{l_i}(rl_{i})^{m+p-e_o}\int d\widetilde{\varphi}_{2}(l_{i})\exp\left[-b_i\Big(\widetilde\varphi_2(l_i) - \frac{\overline{\beta}_i}{b_i} \Big)^2\right]\widetilde{\varphi}_{2}^{r}(l_i)\nonumber\\
   & ~~~~\prod_{j\neq i}\int d\widetilde{\varphi}_{2}(l_j){\rm exp}\Big[-b_j\Big(\widetilde\varphi_2(l_j) - \frac{\overline{\beta}_j}{b_j} \Big)^2\Big].
\end{align}
To evaluate the first Gaussian integration we make a change of variable $W=\widetilde\varphi_2(l_i) - \frac{\overline{\beta}_i}{b_i}$ and obtain the following 
\begin{align}
&\int d\widetilde{\varphi}_{2}(l_{i})\exp\left[-b_i\Big(\widetilde\varphi_2(l_i) - \frac{\overline{\beta}_i}{b_i} \Big)^2\right](rl_{i})^{m+p-e_o}\widetilde{\varphi}_{2}^{r}(l_i)\nonumber\\
    &=\int d W e^{-b_i W^2} (rl_i)^{m + p - e_o} \; \left(W + \frac{\overline{\beta}_{i}}{b_i} \right)^r \nonumber\\
    &= \int dW e^{-b_{i}W^2} \; (rl_{i})^{m+p-e_o} \sum_{e_{p}\in 2\mathbb{Z}_{+}}^{r}\begin{pmatrix}
    r\\
    e_{p}\end{pmatrix} \bigg(\frac{\overline{\beta}_{i}}{b_{i}}\bigg)^{r-e_{p}} W^{e_{p}}\nonumber\\
    &= \sqrt{\frac{\pi}{b_{i}}}   \sum_{e_p \in 2\mathbb{Z}_+}^r \left(\begin{matrix} r \\ e_p \\ \end{matrix}\right) 
\left({\overline\beta_i\over b_i}\right)^{r - e_p} {(rl_i)^{m+p-e_o}(e_p - 1)!!\over (2b_i)^{e_p/2}}.\label{GInteGrals2}
\end{align}
Therefore, in the continuum limit we get 
\begin{align}
        {\rm Num}&[\langle \hat{\mathbf{g}}_{\mathrm{MN}}\rangle_{\overline\sigma}]\bigg|^{\mathrm{Case~1}}_{\mathcal{O}(c_{nmpqrs}),\varphi_2}\nonumber\\
        &=\prod_{j} \Big({\pi_j\over b_j}\Big)^{1/2} \frac{1}{\mathrm{V}^{r-1}}\sum_{e_p \in 2\mathbb{Z}_+}^r \left(\begin{matrix} r \\ e_p \\ \end{matrix}\right)
\int d^{11} l ~(rl)^{m+p-e_o}  
\left({\overline\beta(l)\over b(l)}\right)^{r - e_p} {(e_p - 1)!!\over \left(2b(l)\right)^{e_p/2}}.
\end{align}
where $\pi_{j}=\pi,\forall ~j$\footnote{This notation should be understood from now on.}. The reason for not taking the continuum limit of the overall factor $ \prod_{j}\left(\frac{\pi_{j}}{b_{j}}\right)^{1/2}$ is that they cancel out when we take into account the denominator of the path integral. Note that, this case has led to a volume suppression of the form $\frac{1}{\mathrm{V}^{r-1}}$. As we proceed we'll ignore such volume suppressed terms. 
\vskip.2in
\noindent {\textbf{Case 2}: $\boldsymbol{l_{v_1} = l_{v_2} = l_{v_3} = ..... = l_{v_{r-1}} = l_i, l_{v_r} = l_j,~\text{with}~ l_i \neq l_j }$}
\vskip.1in
For this case (\ref{O(c)phi2Case1}) evaluates to 
\begin{align}
     {\rm Num}[\langle \hat{\mathbf{g}}_{\mathrm{MN}}\rangle_{\overline\sigma}]\bigg|^{\mathrm{Case~2}}_{\mathcal{O}(c_{nmpqrs}),\varphi_2}&=\frac{1}{\mathrm{V}^{r}}\sum_{{l_i,l_j}}\int d \widetilde{\varphi}_{2}(l_{i})\exp\bigg[-b_i \left( \widetilde{\varphi}_{2}(l_i)-\frac{\overline{\beta}_{i}}{b_{i}}\right)^2\bigg]\widetilde{\varphi}_{2}^{r-1}(l_{i})\nonumber\\
    \times \int d &\widetilde{\varphi}_{2}(l_{j})\exp\bigg[-b_j \left( \widetilde{\varphi}_{2}(l_j)-\frac{\overline{\beta}_{j}}{b_{j}}\right)^2\bigg]\widetilde{\varphi}_{2}(l_{j})\left((r-1)l_{i}+l_{j} \right)^{m+p-e_o}\nonumber\\
     &~~\times\prod_{k\neq i,j}\int d\widetilde{\varphi}_{2}(l_k){\rm exp}\Big[-b_k\Big(\widetilde\varphi_2(l_k) - \frac{\overline{\beta}_k}{b_k} \Big)^2\Big]\nonumber\\
     =\frac{1}{\mathrm{V}^r}\sum_{e_1=0}^{m + p - e_o} \left(\begin{matrix} m + p - e_o \\ e_1 \\ \end{matrix}\right) & (r - 1)^{e_1} \sum_{l_i}l^{e_1}_i\int d\widetilde{\varphi}_{2}(l_{i}) \exp\bigg[-b_i \left( \widetilde{\varphi}_{2}(l_i)-\frac{\overline{\beta}_{i}}{b_{i}}\right)^2\bigg]\widetilde{\varphi}_{2}^{r-1}(l_{i}) \nonumber\\
     \times\sum_{l_j}l^{m+p-e_o-e_1}_j&\int d\widetilde{\varphi}_{2}(l_{j}) \exp\bigg[-b_j \left( \widetilde{\varphi}_{2}(l_j)-\frac{\overline{\beta}_{j}}{b_{j}}\right)^2\bigg]\widetilde{\varphi}_{2}(l_{j})\nonumber\\
    \times\prod_{k\neq i,j}&\int d\widetilde{\varphi}_{2}(l_k){\rm exp}\Big[-b_k\Big(\widetilde\varphi_2(l_k) - \frac{\overline{\beta}_k}{b_k} \Big)^2\Big]\nonumber\\
    =\frac{1}{\mathrm{V}^r}\sum_{e_1=0}^{m + p - e_o} \left(\begin{matrix} m + p - e_o \\ e_1 \\ \end{matrix}\right)&  (r - 1)^{e_1} \sum_{l_i}l^{e_1}_i\left( \frac{\pi_{i}}{b_{i}}\right)^{1/2}\sum_{e_{p}\in 2\mathbb{Z}_{+}}^{r-1} \left(\begin{matrix} r-1 \\ e_p \\ \end{matrix}\right) 
\left({\overline\beta_i\over b_i}\right)^{r-1 - e_p} {(e_p - 1)!!\over (2b_i)^{e_p/2}} \nonumber\\
     \times\sum_{l_j}l^{m+p-e_o-e_1}_j&\left({\frac{\pi_{j}}{b_{j}}}\right)^{1/2}\frac{\overline{\beta}_{j}}{b_{j}}\prod_{k\neq i,j}\left({\frac{\pi_{k}}{b_{k}}}\right)^{1/2}.
\end{align}
To arrive at the last step we have used the Gaussian integration results (\ref{GInteGrals1}) and (\ref{GInteGrals2}). Now, taking the continuum limit we get 
\begin{align}
    {\rm Num}&[\langle \hat{\mathbf{g}}_{\mathrm{MN}}\rangle_{\overline\sigma}]\bigg|^{\mathrm{Case~2}}_{\mathcal{O}(c_{nmpqrs}),\varphi_2}\nonumber\\
    &=\prod_{j}\left( \frac{\pi_{j}}{b_{j}}\right)^{1/2}\frac{1}{\mathrm{V}^{r-2}}\sum_{e_1=0}^{m + p - e_o} \left(\begin{matrix} m + p - e_o \\ e_1 \\ \end{matrix}\right)  (r - 1)^{e_1}\sum_{e_{p}\in 2\mathbb{Z}_{+}}^{r-1} \left(\begin{matrix} r-1 \\ e_p \\ \end{matrix}\right)\frac{(e_{p}-1)!!}{2^{e_{p}/2}}\nonumber\\
&~~~~\times \int d^{11} l \frac{l^{e_1}}{b^{e_p/2}(l)} \left( \frac{\overline{\beta}(l)}{b(l)} \right)^{r-1-e_{p}} \; \int d^{11}  l^{\prime} \;{l^{\prime }}^{m+p-e_o-e_1}   \frac{\overline{\beta}(l^{\prime} )}{b(l^{\prime} )} 
\end{align}
This result is again volume suppressed due to the presence of the factor $\frac{1}{\mathrm{V}^{r-2}}$. In fact, such a volume suppression will always be there unless all the $r$ momentum values $l_{v_{i}}$ differ from each other. Let us now analyse such a case.
\vskip.2in
\noindent {\textbf{Case 3}: $\boldsymbol{l_{v_1} \neq l_{v_2} \neq l_{v_3} \neq ..... \neq l_{v_{r-1}}\neq l_{v_r} }$}
\vskip.1in
For the case at consideration (\ref{O(c)phi2Case1}) evaluates to
\begin{align}
    &{\rm Num}[\langle \hat{\mathbf{g}}_{\mathrm{MN}}\rangle_{\overline\sigma}]\bigg|^{\mathrm{Case~3}}_{\mathcal{O}(c_{nmpqrs}),\varphi_2}\nonumber\\
    &=\frac{1}{\mathrm{V}^r} \sum_{e_1,...,e_{r-1}}\begin{pmatrix}
    m+p-e_o\\
    e_{1}\end{pmatrix}\begin{pmatrix}
    m+p-e_o-e_{1}\\
e_{2}\end{pmatrix}\cdots\begin{pmatrix}
    m+p-e_o-\sum_{i=1}^{r-2}e_{i}\\
    e_{r-1}\end{pmatrix} \nonumber\\
    &~~~~~\times\sum_{l_1,...,l_r}l_{1}^{e_{1}}l_{2}^{e_2}\cdots l_{r}^{e_r}\int d\widetilde{\varphi}_{2}(l_{1})\exp\bigg[-b_{1}\left( \widetilde{\varphi}_{2}(l_1)-\frac{\overline{\beta}_{1}}{b_{1}}\right)^2\bigg]\widetilde{\varphi}_{2}(l_1)\nonumber\\
    &\times\int d\widetilde{\varphi}_{2}(l_{2})\exp\bigg[-b_{2}\left( \widetilde{\varphi}_{2}(l_2)-\frac{\overline{\beta}_{2}}{b_{2}}\right)^2\bigg]\widetilde{\varphi}_{2}(l_2)\cdots \int d\widetilde{\varphi}_{2}(l_{r})\exp\bigg[-b_{r}\left( \widetilde{\varphi}_{2}(l_r)-\frac{\overline{\beta}_{r}}{b_{r}}\right)^2\bigg]\widetilde{\varphi}_{2}(l_r)\nonumber\\
&~~~~~~~~~~~~~~~~~~~~\times\prod_{j\neq 1,...,r}\int d\widetilde{\varphi}_{2}(l_{j})\exp\bigg[-b_{j}\left( \widetilde{\varphi}_{2}(l_j)-\frac{\overline{\beta}_{j}}{b_{j}}\right)^2\bigg]\nonumber\\
    &=\frac{1}{\mathrm{V}^r} \sum_{e_1,...,e_{r-1}}\begin{pmatrix}
    m+p-e_o\\
    e_{1}\end{pmatrix}\begin{pmatrix}
    m+p-e_o-e_{1}\\
e_{2}\end{pmatrix}\cdots\begin{pmatrix}
    m+p-e_o-\sum_{i=1}^{r-2}e_{i}\\
    e_{r-1}\end{pmatrix} \nonumber\\
    &~~~~~\times\sum_{l_{1},...,l_r}l_{1}^{e_{1}}l_{2}^{e_2}\cdots l_{r}^{e_r}\frac{\overline{\beta}_{1}}{b_{1}}\left( \frac{\pi}{b_1}\right)^{1/2}\frac{\overline{\beta}_{2}}{b_{2}}\left( \frac{\pi}{b_2}\right)^{1/2}\cdots \frac{\overline{\beta}_{r}}{b_{r}} \left( \frac{\pi}{b_r}\right)^{1/2}\prod_{j\neq 1,...,r}\left(\frac{\pi_{j}}{b_{j}} \right)^{1/2},
\end{align}
where $e_{r}=m+p-e_o-\sum_{i=1}^{r-1}e_{i}$. Therefore, if we take the continuum limit we arrive at the following result
\begin{align}
     &{\rm Num}[\langle \hat{\mathbf{g}}_{\mathrm{MN}}\rangle_{\overline\sigma}]\bigg|^{\mathrm{Case~3}}_{\mathcal{O}(c_{nmpqrs}),\varphi_2}\nonumber\\
     &=\prod_{i}\left( \frac{\pi_{i}}{b_{i}}\right)^{1/2} \sum_{e_1,...,e_{r-1}}\begin{pmatrix}
    m+p-e_o\\
    e_{1}\end{pmatrix}\begin{pmatrix}
    m+p-e_o-e_{1}\\
e_{2}\end{pmatrix}\cdots\begin{pmatrix}
    m+p-e_o-\sum_{i=1}^{r-2}e_{i}\\
    e_{r-1}\end{pmatrix} \nonumber\\
    &~~~~~~~~~~~~~~~~~~~~\times\int d^{11}l_1 \frac{\overline{\beta}_{1}}{b_{1}}l_{1}^{e_{1}} \int d^{11}l_2 \frac{\overline{\beta}_{2}}{b_{2}}l_{2}^{e_{2}}\cdots\int d^{11}l_r \frac{\overline{\beta}_{r}}{b_{r}}l_{r}^{e_{r}} . \label{DomVarphi2}
\end{align}
So this does not have any volume suppression as per our expectation.  Let us now analyse $\varphi_{3}$ sector before getting onto $\varphi_{1}$.
\subsubsection{First order quantum correction due to \texorpdfstring{$\varphi_{3}$}{varphi3} field sector}
The contribution to (\ref{NumErAtorOfExpValuE}) at $\mathcal{O}(c_{nmpqrs})$ due to $\varphi_{3}$ field sector is given by 
\begin{align}
      {\rm Num}[\langle \hat{\mathbf{g}}_{\mathrm{MN}}\rangle_{\overline\sigma}]\bigg|_{\mathcal{O}(c_{nmpqrs}),\varphi_3}&=\int d\widetilde{\varphi}_{3}(f_1){\rm exp}\Big[-c_1\Big(\widetilde\varphi_3(f_1) - \frac{\overline{\gamma}_1}{c_1} \Big)^2\Big] \nonumber\\
    \times\int d\widetilde{\varphi}_{3}(f_2){\rm exp}\Big[-c_2&\Big(\widetilde\varphi_3(f_2) - \frac{\overline{\gamma}_2}{c_2} \Big)^2\Big]\int d\widetilde{\varphi}_{3}(f_3){\rm exp}\Big[-c_3\Big(\widetilde\varphi_2(f_3) - \frac{\overline{\gamma}_3}{c_3} \Big)^2\Big] \cdots \nonumber\\
\times\frac{1}{\mathrm{V}^{s-1}}\sum_{\left\{{f_{w_t}}\right\}}&\widetilde\varphi_3(f_{w_1})
\widetilde\varphi_3(f_{w_2})\widetilde\varphi_3(f_{w_3})\cdots  \widetilde\varphi_3(f_{w_{s-1}}) \widetilde\varphi_3(f_{w_s}).\label{O(c)phi3Case1}
\end{align}
Recall that, the momentum $f_{w_s}$ depends on the momenta values of the other two field sectors via the relation (\ref{fwsRelation}). It will make the following analysis a bit different from that of the $\varphi_{2}$ sector.  We will study few cases with different momentum values $f_{w_{t}}$ of the $s-1$ number of $\widetilde{\varphi}_{3}(f_{w_{t}})$ fields. 
\vskip.2in
\noindent{\textbf{Case 1:} {$\boldsymbol{f_{w_1} = f_{w_2} = f_{w_3} = \cdots = f_{w_{s-1}} \equiv f_k}$}}
\vskip.1in
This case implies:
\begin{align}   f_{w_s}&=\sum_{i=1}^{q}k_{u_i}+\sum_{j=1}^{r}l_{v_j}+(s-1)f_{k}.
\end{align}
As an example we consider $k_{u_1}=k_{u_{2}}=\cdots=k_{u_{q}}= k_{i}$ and $l_{v_1}=l_{v_2}=\cdots=l_{v_{r}}= l_{j}$ which leads to $f_{w_s}=qk_{i}+rl_{j}+(s-1)f_{k}$. Thus from (\ref{O(c)phi3Case1}) we obtain 
\begin{align}
    {\rm Num}&[\langle \hat{\mathbf{g}}_{\mathrm{MN}}\rangle_{\overline\sigma}]\bigg|^{\mathrm{Case~1}}_{\mathcal{O}(c_{nmpqrs}),\varphi_3}=\frac{1}{\mathrm{V}^{s-1}}\sum_{f_k}\int d\widetilde{\varphi}_{3}(f_{k})\exp\bigg[-c_k\left( \widetilde{\varphi}_{3}(f_k)-\frac{\overline{\gamma}_{k}}{c_{k}}\right)^2\bigg]\widetilde{\varphi}^{s-1}_{3}(f_{k})\nonumber\\
    \int d&\widetilde{\varphi}_{3}(f_{w_s})\exp\bigg[-c_{w_s}\left( \widetilde{\varphi}_{3}(f_{w_s})-\frac{\overline{\gamma}_{w_s}}{c_{w_s}}\right)^2\bigg]\widetilde{\varphi}_{3}(f_{w_s})\prod_{t\neq k,w_{s}}\int d\widetilde{\varphi}_{3}(f_{t})\exp\bigg[-c_t\left( \widetilde{\varphi}_{3}(f_t)-\frac{\overline{\gamma}_{t}}{c_{t}}\right)^2\bigg]\nonumber\\
    &=\frac{1}{\mathrm{V}^{s-1}}\sum_{f_k}\left( \frac{\pi}{c_{k}}\right)^{1/2}\sum_{e_{p}\in 2\mathbb{Z}_{+}}\begin{pmatrix}
        s-1\\
        e_{p}
    \end{pmatrix} \left(\frac{\overline{\gamma}_{k}}{c_{k}} \right)^{s-1-e_{p}}\frac{(e_{p}-1)!!}{(2c_{k})^{e_p/2}}\frac{\overline{\gamma}_{w_{s}}}{c_{w_s}} \left( \frac{\pi}{c_{w_s}}\right)^{1/2}\prod_{t\neq k,w_s}\left( \frac{\pi_{t}}{c_{{t}}}\right)\label{Case1phi3Sec}
\end{align}
The momentum conservation condition (\ref{fwsRelation}) can simplify the factor $\frac{\overline{\gamma}_{w_s}}{c_{w_{s}}}$ appearing in the above expression. From (\ref{abcDef}) we know that $c_{w_{s}}\sim f_{w_s}^2$. Thus we get
\begin{align}
    \frac{1}{c_{w_s}}\sim \frac{1}{f_{w_s}^2}&=\frac{1}{\left(\sum_{i=1}^{q}k_{u_i}+\sum_{j=1}^r l_{v_j}+\sum_{t=1}^{s-1}f_{w_t} \right)^2}\nonumber\\
    &=\frac{1}{\left(\sum_{t=1}^{s-1}f_{w_t}\right)^2} \left(1+\frac{\sum_{i=1}^{q}k_{u_i}+\sum_{j=1}^{r}l_{v_{j}}}{\sum_{t=1}^{s-1}f_{w_t}}\right)^{-2}.\label{Inverseofcws}
\end{align}
If we can impose the following condition for the values of the momentum 
\begin{align}
    \zeta\equiv \sum_{t=1}^{s-1}f_{w_{t}}\ge \sum_{i=1}^{q}k_{u_{i}}+\sum_{j=1}^{r}l_{v_j},\label{MomentumConst}
\end{align}
(\ref{Inverseofcws}) can be simplified further as
\begin{align}
    \frac{1}{c_{w_s}}&\sim\frac{1}{\left(\sum_{t=1}^{s-1}f_{w_t}\right)^2} \left(1-\frac{\sum_{i=1}^{q}k_{u_i}+\sum_{j=1}^{r}l_{v_{j}}}{\sum_{t=1}^{s-1}f_{w_t}}\right)^{2}\nonumber\\
    &=\frac{1}{\left(\sum_{t=1}^{s-1}f_{w_t}\right)^2} \sum_{n_1=0}^{2}\begin{pmatrix}
        2\\
        n_1
    \end{pmatrix}(-1)^{n_1}\left(\frac{\sum_{i=1}^{q}k_{u_i}+\sum_{j=1}^{r}l_{v_{j}}}{\sum_{t=1}^{s-1}f_{w_t}}\right)^{n_1}\nonumber\\
    &=\sum_{n_1=0}^{2}\begin{pmatrix}
        2\\
        n_1
    \end{pmatrix}(-1)^{n_1}\sum_{e_{n_1}=0}^{n_1}\begin{pmatrix}
        n_1\\
        e_{n_1}
    \end{pmatrix}\left( \sum_{i=1}^{q}k_{u_i}\right)^{e_{n_1}}\left(\sum_{j=1}^{r}l_{v_j} \right)^{n_{1}-e_{n_1}}\frac{1}{ \zeta^{2+n_1}}.\label{1/cws}
\end{align}
Taylor expansion of $\overline{\gamma}(f_{w_s})$ around $\zeta$ gives
\begin{align}
 &\overline{\gamma} ( f_{\omega_s}) =\overline{\gamma}(\zeta)+(f_{w_s}-\zeta)\; \overline{\gamma}^{'}(\zeta)+\frac{\left(f_{w_{s}}-\zeta\right)^2}{2!}\overline{\gamma}^{''}(\zeta)+\cdots\nonumber\\
&=\overline{\gamma}(\zeta) + \left(\sum_{i = 1}^q k_{u_i} + \sum_{j = 1}^r l_{v_j} \right) \frac{\partial \overline{\gamma}}{\partial \zeta} (\zeta) + \frac{1}{2!} \left(\sum_{i = 1}^q k_{u_i} + \sum_{j = 1}^r l_{v_j} \right)^2 \frac{\partial^2 \overline{\gamma}}{\partial \zeta^2} ( \zeta) + \cdots\nonumber
\\
& = \sum_{n_2=0}^{\infty} \frac{1}{n_2 !} \left(\sum_{i = 1}^q k_{u_i} + \sum_{j = 1}^r l_{v_j} \right)^{n_2} \frac{\partial^{n_2} \overline{\gamma}}{\partial \zeta^{n_2}} (\zeta) \nonumber
\\ 
& = \sum_{n_2=0}^{\infty} \frac{1}{n_2 !} \sum_{e_{n_2}=0}^{n_2}  \left(\begin{matrix} n_2 \\ e_{n_2} \\ \end{matrix}\right) \left(\sum_{i = 1}^q k_{u_i} \right)^{e_{n_2}} \left ( \sum_{j = 1}^r l_{v_j} \right)^{n_2 - e_{n_2}} \frac{\partial^{n_2} \overline{\gamma}}{\partial \zeta^{n_2}} (\zeta).\label{gammafws}
\end{align}
This indicates that the corresponding results of $\varphi_{1}$ sector and $\varphi_{2}$ sector should be modified as $p\rightarrow p+{n_1}+{n_2}$ and $e_{o}\rightarrow e_{o}+e_{n_1}+e_{n_2}$. However, we keep this modification implicit. After incorporating the relevant factors from (\ref{1/cws}) and (\ref{gammafws}) we write down (\ref{Case1phi3Sec}):
\begin{align}
   &{\rm Num}[\langle \hat{\mathbf{g}}_{\mathrm{MN}}\rangle_{\overline\sigma}]\bigg|^{\mathrm{Case~1}}_{\mathcal{O}(c_{nmpqrs}),\varphi_3}\nonumber\\
    &=\prod_{t}\left( \frac{\pi_{t}}{c_{t}}\right)^{1/2}\frac{1}{\mathrm{V}^{s-1}}\sum_{e_p\in 2\mathbb{Z}_{+}}\begin{pmatrix}
        s-1\\
        e_p
    \end{pmatrix}\frac{(e_p-1)!!}{2^{e_p/2}}\sum_{f_{k}}\left(\frac{\overline{\gamma}_{k}}{c_{k}}\right)^{s-1-e_{p}}\frac{1}{\left((s-1)f_{k} \right)^{2+n_1}}\frac{\partial^{n_2}\overline{\gamma}((s-1)f_{k})}{\partial \zeta^{n_2}}\nonumber
\end{align}
where the summation over $n_{1},n_2$ has been kept implicit. Taking the continuum limit we arrive at 
\begin{align}
   &{\rm Num}[\langle \hat{\mathbf{g}}_{\mathrm{MN}}\rangle_{\overline\sigma}]\bigg|^{\mathrm{Case~1}}_{\mathcal{O}(c_{nmpqrs}),\varphi_3}\nonumber\\
    =\prod_{t}&\left( \frac{\pi_{t}}{c_{t}}\right)^{1/2}\frac{1}{\mathrm{V}^{s-2}}\sum_{e_p\in 2\mathbb{Z}_{+}}\begin{pmatrix}
        s-1\\
        e_p
    \end{pmatrix}\frac{(e_p-1)!!}{2^{e_p/2}(s-1)^{2+n_1+n_2}}\int d^{11}f\left(\frac{\overline{\gamma}(f)}{c(f)}\right)^{s-1-e_{p}}\frac{1}{f^{2+n_1}}\frac{\partial^{n_2}\overline{\gamma}((s-1)f)}{\partial f^{n_2}}.
\end{align}
As we could infer the result is volume suppressed due to the presence of the factor $\frac{1}{\mathrm{V}^{s-2}}$. So let us consider the case that will be of  importance to us viz. all the $s-1$ momenta different  from each other.
\vskip.2in
\noindent{\textbf{Case 2:} {$\boldsymbol{f_{w_1} \neq f_{w_2} \neq f_{w_3} \neq \cdots \neq f_{w_{s-1}} }$}}
\vskip.1in
For this case (\ref{O(c)phi3Case1}) evaluates to
\begin{align}
&{\rm Num}[\langle \hat{\mathbf{g}}_{\mathrm{MN}}\rangle_{\overline\sigma}]\bigg|^{\mathrm{Case~2}}_{\mathcal{O}(c_{nmpqrs}),\varphi_3}\nonumber\\
&=\frac{1}{\mathrm{V}^{s-1}}\sum_{f_{1},...,f_{s-1}}\int d\widetilde{\varphi}_{3}(f_{1})\exp\bigg[-c_{1}\left( \widetilde{\varphi}_{3}(f_1)-\frac{\overline{\gamma}_{1}}{c_{1}}\right)^2\bigg]\widetilde{\varphi}_{3}(f_{1})\nonumber\\
&\times\int d\widetilde{\varphi}_{3}(f_{2})\exp\bigg[-c_{2}\left( \widetilde{\varphi}_{3}(f_2)-\frac{\overline{\gamma}_{2}}{c_{2}}\right)^2\bigg]\widetilde{\varphi}_{3}(f_{2})\cdots\int d\widetilde{\varphi}_{3}(f_{w_s})\exp\bigg[-c_{w_s}\left( \widetilde{\varphi}_{3}(f_{w_s})-\frac{\overline{\gamma}_{w_s}}{c_{w_s}}\right)^2\bigg] \nonumber\\
&~~~~~~~~~~~~~~~\times \prod_{t\neq {1},...,{s-1}}\int d\widetilde{\varphi}_{3}(f_{t})\exp\bigg[-c_{t}\left( \widetilde{\varphi}_{3}(f_{t})-\frac{\overline{\gamma}_{t}}{c_{t}}\right)^2\bigg]\widetilde{\varphi}_{3}(f_{t})\nonumber\\
&=\prod_{i}\left( \frac{\pi_{i}}{c_{i}}\right)^{1/2}\frac{1}{\mathrm{V}^{s-1}}\sum_{f_1,...,f_{s-1}}\frac{\overline{\gamma}_{1}}{c_1}\cdots \frac{\overline{\gamma}_{s-1}}{c_{s-1}}\frac{\overline{\gamma}_{w_{s}}}{c_{w_s}}.
\end{align}
Incorporating the relevant factors from (\ref{1/cws}) and (\ref{gammafws}) we can express the above expression as
\begin{align}
  {\rm Num}&[\langle \hat{\mathbf{g}}_{\mathrm{MN}}\rangle_{\overline\sigma}]\bigg|^{\mathrm{Case~2}}_{\mathcal{O}(c_{nmpqrs}),\varphi_3}  =\prod_{i}\left( \frac{\pi_{i}}{c_{i}}\right)^{1/2}\frac{1}{\mathrm{V}^{s-1}}\sum_{f_1,...,f_{s-1}}\frac{\overline{\gamma}_{1}}{c_1}\cdots \frac{\overline{\gamma}_{s-1}}{c_{s-1}}\frac{1}{\zeta^{2+n_1}}\frac{\partial^{n_{2}}\overline{\gamma}}{\partial \zeta^{n_2}}(\zeta)\nonumber\\
  &=\prod_{i}\left( \frac{\pi_{i}}{c_{i}}\right)^{1/2} \frac{1}{\mathrm{V}^{s-1}}\sum_{f_{1},...,f_{s-1}}\frac{\overline{\gamma}_{1}}{c_1}\cdots \frac{\overline{\gamma}_{s-1}}{c_{s-1}}\frac{1}{(f_{1}+\cdots+f_{s-1})^{2+n_1}}\nonumber\\
&~~~~~~~~~~~~~~~~~~~~~~~~~~~~~~~~~~~~~~~~\times\left(\frac{\partial}{\partial f_{1}}+\cdots+\frac{\partial}{\partial f_{s-1}}\right)^{n_{2}}\overline{\gamma}(f_{1}+\cdots+f_{s-1})\nonumber\\
&=\prod_{i}\left( \frac{\pi_{i}}{c_{i}}\right)^{1/2} \int d^{11}f_1\frac{\overline{\gamma}_{1}}{c_1}\cdots\int d^{11}f_{s-1} \frac{\overline{\gamma}_{s-1}}{c_{s-1}}\frac{1}{(f_{1}+\cdots+f_{s-1})^{2+n_1}}\nonumber\\
&~~~~~~~~~~~~~~~~~~~~~~~~~~~~~~~~~~~~~~~~\times\left(\frac{\partial}{\partial f_{1}}+\cdots+\frac{\partial}{\partial f_{s-1}}\right)^{n_{2}}\overline{\gamma}(f_{1}+\cdots+f_{s-1}),
\end{align}
where in the last line we have taken the continuum limit. Now we analyse the source sector in the following subsection.
\subsubsection{First order quantum correction due to \texorpdfstring{$\varphi_{1}$}{varphi1} field sector}
As mentioned earlier $\varphi_{1}$ represents one of the degrees of freedom out of the 44 metric degrees of freedom.
The contribution due to $\varphi_{1}$ sector in the path integral (\ref{NumErAtorOfExpValuE}) at $\mathcal{O}(c_{nmpqrs})$ is as follows
\begin{align}
&    {\rm Num}[\langle \hat{\mathbf{g}}_{\mathrm{MN}}\rangle_{\overline\sigma}]\bigg|_{\mathcal{O}(c_{nmpqrs}),\varphi_1}=\int d\widetilde{\varphi}_{1}(k_1){\rm exp}\Big[-a_1\Big(\widetilde\varphi_1(k_1) - \frac{\overline{\alpha}_1}{a_1} \Big)^2\Big] \nonumber\\
   & \times\int d\widetilde{\varphi}_{1}(k_2){\rm exp}\Big[-a_2\Big(\widetilde\varphi_1(k_2) - \frac{\overline{\alpha}_2}{a_2} \Big)^2\Big]\int d\widetilde{\varphi}_{1}(k_3){\rm exp}\Big[-a_3\Big(\widetilde\varphi_1(k_3) - \frac{\overline{\alpha}_3}{a_3} \Big)^2\Big] \cdots \nonumber\\
    &\times   {1\over {\rm V}}\Big(\widetilde\varphi_1(k_1)\psi_{{\bf k}_1}({\bf x}, y, z)e^{-ik_{0, 1}t} + 
\widetilde\varphi_1(k_2)\psi_{{\bf k}_2}({\bf x}, y, z)e^{-ik_{0, 2}t} + 
\widetilde\varphi_1(k_3)\psi_{{\bf k}_3}({\bf x}, y, z)e^{-ik_{0, 3}t} +\cdots \Big)\nonumber\\
&
\times\frac{1}{\mathrm{V}^{q}}\sum_{\left\{{k_{u_i}}\right\}}(k_{u_1}+k_{u_2}+k_{u_3}+\cdots+k_{u_q})^{n-e_o}\widetilde\varphi_1(k_{u_1})
\widetilde\varphi_1(k_{u_2})\widetilde\varphi_1(k_{u_3})\cdots \widetilde\varphi_1(k_{u_q}).\label{O(c)phi1Case1}
\end{align}
By now it is clear to us how to evaluate such an integral for a given distribution of the momenta $\{k_{u_i}\}$. For completeness we provide the explicit result for the case when all the $q$ momentum values differ from each other. 
\vskip.2in
\noindent{\textbf{Case 1:} {$\boldsymbol{k_{u_1} \neq k_{u_2} \neq k_{u_3} \neq \cdots \neq k_{u_{q}}}$}}
\vskip.1in
We also consider the source momentum to be different from all the $\{k_{u_i}\}$ momentum values. For this case (\ref{O(c)phi1Case1}) evaluates to
\begin{align}
    {\rm Num}&[\langle \hat{\mathbf{g}}_{\mathrm{MN}}\rangle_{\overline\sigma}]\bigg|^{\mathrm{Case}~1}_{\mathcal{O}(c_{nmpqrs}),\varphi_1}\nonumber\\
    &=\frac{1}{\mathrm{V}^q}\sum_{e_{1},...,e_{q-1}}\begin{pmatrix}
        n-e_o\\
        e_1
    \end{pmatrix}\begin{pmatrix}
        n-e_o-e_1\\
        e_2
    \end{pmatrix}\cdots\begin{pmatrix}
        n-e_o-\sum_{i=1}^{q-1}e_i\\
        e_{q-1}
    \end{pmatrix}\nonumber\\
&~~~~~~\times\sum_{k_1,...,k_q}k_{1}^{e_1}k_{2}^{e_2}\cdots k_{q}^{e_q}\int d\widetilde{\varphi}_{1}(k_1){\rm exp}\Big[-a_1\Big(\widetilde\varphi_1(k_1) - \frac{\overline{\alpha}_1}{a_1} \Big)^2\Big] \widetilde{\varphi}_{1}(k_1)\nonumber\\
    \times\int &d\widetilde{\varphi}_{1}(k_2){\rm exp}\Big[-a_2\Big(\widetilde\varphi_1(k_2) - \frac{\overline{\alpha}_2}{a_2} \Big)^2\Big]\widetilde{\varphi}_{1}(k_2)\cdots\int d\widetilde{\varphi}_{1}(k_q){\rm exp}\Big[-a_q\Big(\widetilde\varphi_1(k_q) - \frac{\overline{\alpha}_q}{a_q} \Big)^2\Big] \widetilde{\varphi}_{1}(k_{q}) \nonumber\\
    &\times   {1\over {\rm V}}\sum_{k_i\neq k_1,...,k_q}\int d\widetilde{\varphi}_{1}(k_{i})\exp\bigg[-a_{i}\left( \widetilde{\varphi}_{1}(k_{i})-\frac{\overline{\alpha}_{i}}{a_{i}}\right)^2\bigg]\Big(\widetilde\varphi_1(k_i)\psi_{{\bf k}_i}({\bf x}, y, z)e^{-ik_{0, i}t}\Big)\nonumber\\
    &~~~~~~~~~~\times\prod_{j\neq i,1,...,q}\int d\widetilde{\varphi}_{1}(k_{j})\exp\bigg[-a_{j}\left( \widetilde{\varphi}_{1}(k_{j})-\frac{\overline{\alpha}_{j}}{a_{j}}\right)^2\bigg]\nonumber\\
    &=\frac{1}{\mathrm{V}^{q}}\sum_{e_{1},...,e_{q-1}}\begin{pmatrix}
        n-e_o\\
        e_1
    \end{pmatrix}\begin{pmatrix}
        n-e_o-e_1\\
        e_2
    \end{pmatrix}\cdots\begin{pmatrix}
        n-e_o-\sum_{i=1}^{q-1}e_i\\
        e_{q-1}
    \end{pmatrix}\nonumber\\
&~~~~~~\times\sum_{k_1,...,k_q}k_{1}^{e_1}k_{2}^{e_2}\cdots k_{q}^{e_q} \frac{\overline{\alpha}_{1}}{a_{1}}\left(\frac{\pi}{a_{1}} \right)^{1/2}\frac{\overline{\alpha}_{2}}{a_{2}}\left(\frac{\pi}{a_{2}} \right)^{1/2}\cdots \frac{\overline{\alpha}_{q}}{a_{q}}\left(\frac{\pi}{a_{q}} \right)^{1/2}\nonumber\\
    &~~~~~~~~~\times   {1\over {\rm V}}\sum_{k_i\neq k_1,...,k_q}\frac{\overline{\alpha}_{i}}{a_{i}}\left( \frac{\pi}{a_{i}}\right)^{1/2}\psi_{{\bf k}_i}({\bf x}, y, z)e^{-ik_{0, i}t}~~\times\prod_{j\neq i,1,...,q}\left( \frac{\pi_{j}}{a_{j}}\right)^{1/2},
\end{align}
where $e_{q}=n-e_{o}-\sum_{i=1}^{q-1}e_{i}$. Taking the continuum limit we get 
\begin{align}
    {\rm Num}&[\langle \hat{\mathbf{g}}_{\mathrm{MN}}\rangle_{\overline\sigma}]\bigg|^{\mathrm{Case}~1}_{\mathcal{O}(c_{nmpqrs}),\varphi_1}\nonumber\\
    &=\prod_{i}\left(\frac{\pi_{i}}{a_{i}} \right)^{1/2}\sum_{e_{1},...,e_{q-1}}\begin{pmatrix}
        n-e_o\\
        e_1
    \end{pmatrix}\begin{pmatrix}
        n-e_o-e_1\\
        e_2
    \end{pmatrix}\cdots\begin{pmatrix}
        n-e_o-\sum_{i=1}^{q-1}e_i\\
        e_{q-1}
    \end{pmatrix}\nonumber\\
&~~\times\int d^{11}k_{1} \frac{\overline{\alpha}_{1}}{a_{1}} k_{1}^{e_1}\int d^{11}k_{2}\frac{\overline{\alpha}_{2}}{a_{2}}k_{2}^{e_2}\cdots\int d^{11}k_{q}\frac{\overline{\alpha}_{q}}{a_{q}}k_{q}^{e_q}\int d^{11}k \frac{\overline{\alpha}(k)}{a(k)}\psi_{\mathbf{k}}(\mathrm{X})e^{-ik_{0}t}\label{DomVaRphi1}
\end{align}
As mentioned in the introduction,  to  evaluate the expectation value of the metric operator (\ref{ExpofMetOp}) we can not just rely on the path integral analysis upto some finite orders in the coupling constant but we have to take into consideration all order contributions to the path integral. To achieve that goal in the following section we  analyse the contribution to (\ref{NumErAtorOfExpValuE}) at $\mathrm{N}$th order in coupling constant. 
\subsection{Quantum correction at \texorpdfstring{$\mathrm{N}$}{N}th order in coupling constant}
We begin with the $\varphi_{2}$ sector at $\mathrm{N}$th order in the coupling constant $c_{nmpqrs}$. We will be considering only the most \textit{dominant} contribution, in other words the contribution that is not volume suppressed. From the analysis of the previous section we know that, such a contribution is attributed to all the momentum values of the $\widetilde{\varphi}_{2}(l_{v_j})$ fields being distinct. Therefore, from (\ref{NumErAtorOfExpValuE}) we can write 
\begin{align}
    {\rm Num}&[\langle \hat{\mathbf{g}}_{\mathrm{MN}}\rangle_{\overline\sigma}]\bigg|^{\mathrm{dominant}}_{\mathcal{O}(c^{\mathrm{N}}_{nmpqrs}),\varphi_2}\nonumber\\
    &~~~~~~~=\left(\frac{1}{\mathrm{V}^{\mathrm{N}r}}\right)  \int d \widetilde{\varphi}_{2}(l_{1})\exp \left[-b_{1}\left( \widetilde{\varphi}_{2}(l_{1})-\frac{\overline{\beta}_{1}}{b_{1}}\right)^2 \right]\widetilde{\varphi}_{2}(l_{1})\nonumber\\
&~~~~~~~\times\int d\widetilde{\varphi}_{2}(l_{2})\exp \left[-b_{2}\left( \widetilde{\varphi}_{2}(l_{2})-\frac{\overline{\beta}_{2}}{b_{2}}\right)^2 \right]\widetilde{\varphi}_{2}(l_{2})\cdots\nonumber\\
&~\times\sum_{\{l_{v_j}\}}(l_{v_1}+\cdots+l_{v_r})^{m+p-e_{o}}(l_{v_{r+1}}+\cdots+l_{v_{2r}})^{m+p-e_{o}}\cdots(l_{v_{(\mathrm{N}-1)r+1}}\cdots+l_{v_{\mathrm{N}r}})^{m+p-e_o}\nonumber\\
    =\prod_{s}&\left( \frac{\pi_{s}}{b_{s}}\right)^{1/2}\sum_{\{e_i\}}\mathbb{C}^{\varphi_{2}}\left(e_1,...e_{r-1},e_{r+1},..., e_{{\rm N}r-1}\right)\ \int d^{11}l_1~{\overline\beta(l_1) l_1^{e_1}\over b(l_1)} \cdots \int d^{11}l_r~{\overline\beta(l_r) l_r^{e_r}\over b(l_r)} \nonumber\\
\times& \int d^{11}l_{r+1}~{\overline\beta(l_{r+1}) l_{r+1}^{e_{r+1}}\over b(l_{r+1})} \cdots \int d^{11}l_{{\rm N}r - 1}~{\overline\beta(l_{{\rm N}r - 1}) l_{{\rm N}r - 1}^{e_{{\rm N}r - 1}}\over b(l_{{\rm N}r - 1})}
\int d^{11}l_{{\rm N}r}~{\overline\beta(l_{{\rm N}r}) l_{{\rm N}r}^{e_{{\rm N}r}}\over b(l_{{\rm N}r})}.\label{AmpLiTudEofPhi2SecToR}
\end{align}
where $e_{\mathrm{N}r}=m+p-e_{o}-\sum^{\mathrm{N}r-1}_{i=(\mathrm{N}-1)r+1}e_{i}$ and $\mathbb{C}^{\varphi_{2}}\left(e_1,...e_{r-1},e_{r+1},..., e_{{\rm N}r-1}\right)$ encodes the combinatorial factors involved which grows as  $\mathrm{N}!$. This growth is due to the combinatorics of the $\mathrm{N}$ sets of momenta of the form $(\sum_{i}l_{v_{i}})^{m+p-e_o}$ \cite{Brahma:2022wdl}. A similar factorial growth happens due to $\varphi_{1}$ sector
\begin{align}
     {\rm Num}&[\langle \hat{\mathbf{g}}_{\mathrm{MN}}\rangle_{\overline\sigma}]\bigg|^{\mathrm{dominant}}_{\mathcal{O}(c^{\mathrm{N}}_{nmpqrs}),\varphi_1}\nonumber\\
    &~~~~~~~=\left(\frac{1}{\mathrm{V}^{\mathrm{N}q}}\right)  \int d \widetilde{\varphi}_{1}(k_{1})\exp \left[-a_{1}\left( \widetilde{\varphi}_{1}(k_{1})-\frac{\overline{\alpha}_{1}}{a_{1}}\right)^2 \right]\widetilde{\varphi}_{1}(k_{1})\nonumber\\
&~~~~~~~\times\int d\widetilde{\varphi}_{1}(k_{2})\exp \left[-a_{2}\left( \widetilde{\varphi}_{1}(k_{2})-\frac{\overline{\alpha}_{2}}{a_{2}}\right)^2 \right]\widetilde{\varphi}_{1}(k_{2})\cdots\nonumber\\
&~\sum_{\{k_{u_i}\}}\times(k_{u_1}+\cdots+k_{u_r})^{n-e_{o}}(k_{u_{r+1}}+\cdots+k_{u_{2r}})^{n-e_{o}}\cdots(k_{u_{(\mathrm{N}-1)q+1}}\cdots+k_{u_{\mathrm{N}r}})^{n-e_o}\nonumber\\
&~~~~\times {1\over {\rm V}}\sum_{k_i\neq \{k_{u_i}\}}\int d\widetilde{\varphi}_{1}(k_{i})\exp\bigg[-a_{i}\left( \widetilde{\varphi}_{1}(k_{i})-\frac{\overline{\alpha}_{i}}{a_{i}}\right)^2\bigg]\Big(\widetilde\varphi_1(k_i)\psi_{{\bf k}_i}({\bf x}, y, z)e^{-ik_{0, i}t}\Big)\nonumber\\
    =\prod_{s}&\left( \frac{\pi_{s}}{b_{s}}\right)^{1/2}\sum_{\{e_i\}}\mathbb{C}^{\varphi_1}\left(e_1,...e_{q-1},e_{q+1},..., e_{{\rm N}q-1}\right)\ \int d^{11}k_1~{\overline\alpha(k_1) k_1^{e_1}\over a(k_1)} \cdots \int d^{11}k_q~{\overline\alpha(k_q) k_q^{e_q}\over a(k_q)} \nonumber\\
\times& \int d^{11}k_{q+1}~{\overline\alpha(k_{q+1}) k_{q+1}^{e_{q+1}}\over a(k_{q+1})} \cdots \int d^{11}k_{{\rm N}q}~{\overline\alpha(k_{{\rm N}q}) k_{{\rm N}q}^{e_{{\rm N}q}}\over a(k_{{\rm N}q})}\int d^{11}k \frac{\overline{\alpha}(k)}{a(k)}\psi_{\mathbf{k}}(\mathrm{X})e^{-ik_{0}t},
\end{align}
as well as the $\varphi_{3}$ sector. It suggests that the total contribution, when we take into account all three field sectors, at $\mathrm{N}$th order in coupling constant grows as $(\mathrm{N}!)^2$. Refer to \cite{Dasgupta:2025ypg} for more discussion on the growth of the perturbation series. Thus (\ref{NumErAtorOfExpValuE}) can be expressed as 
\begin{align}
    \mathrm{Num}[\langle \hat{\mathbf{g}}_{\mu \nu}\rangle_{\overline{\sigma}}]&=\sum_{\mathrm{N}=0}^{\infty}g^{\mathrm{N}} \mathcal{C}_{\mathrm{N}}\int  d^{11}k \frac{\overline{\alpha}_{\mu\nu}(k)}{a(k)}\psi_{\mathbf{k}}(\mathrm{X})e^{-ik_0 t}\label{PerSeries}
\end{align}
where 
\begin{align}
&~~~~~~~~~~~~~~~~~\mathcal{C}_{\mathrm{N}}\equiv \mathcal{A}^{\mathrm{N}}(\mathrm{N}!)^{2},\nonumber\\
&~\text{with}~
    \mathcal{A}\sim \sum_{\{n_{i}\},\{m_{j}\}}{\mathbb{B}_1(n_1, .., n_{q-1})\mathbb{B}_2(m_1,..,m_{r-1})}
\prod_{i = 1}^q \int d^{11} k_i{\overline{\alpha}(k_i)k_i^{n_i}\over a(k_i)}\prod_{j = 1}^r \int d^{11} l_j{\overline{\beta}(l_j)l_j^{m_j}\over b(l_j)}\nonumber\\
&~~~~~~~~~~~~~~~~~~~~~~~~~~~~~~~~~\times\prod_{t = 1}^{s-1} 
\int d^{11} f_t{\overline{\gamma}(f_t)\over c(f_t)}\frac{{\overline{\gamma}}({f_{w_s}})}{c(f_{w_{s}})},
\end{align}
which is asymptotic in nature. The coefficients $\mathbb{B}_{1}$ and $\mathbb{B}_{2}$  can be read off from  (\ref{DomVaRphi1}) and (\ref{DomVarphi2})  respectively.  So in order to get a sensible answer it requires us to incorporate the Borel resummation techniques which is the theme of the following section. 
\section{Resummation of the asymptotic series}
\label{BorelREsum}
We generalise equation (\ref{PerSeries}) as 
\begin{align}
    \mathrm{Num}[\langle \hat{\mathbf{g}}_{\mu \nu}\rangle_{\overline{\sigma}}]&=\sum_{\mathrm{N}=0}^{\infty}g^{\mathrm{N}} \mathcal{A}^{\mathrm{N}}(\mathrm{N}!)^{\alpha}\int  d^{11}k \frac{\overline{\alpha}_{\mu\nu}(k)}{a(k)}\psi_{\mathbf{k}}(\mathrm{X})e^{-ik_0 t},\label{ExpValMOp}
\end{align}
where the $\mathrm{N}$th coefficient of the series grows as $(\mathrm{N}!)^{\alpha}$, this is  termed as a Gevrey-$\alpha$ series. It is important to note that the above series with $(\mathrm{N}!)^{\alpha}$ growth can equivalently be expressed as a series with $(\alpha \mathrm{N})!$ growth. This can be understood in the  following manner: using Stirlings approximation in the limit $\mathrm{N}\rightarrow \infty$ 
\begin{align}
    \log (\mathrm{N}!)^{\alpha}&\approx\alpha (\mathrm{N}~\log \mathrm{N}-\mathrm{N})=\alpha \mathrm{N}\log(\alpha \mathrm{N})-\alpha \mathrm{N}-\alpha \mathrm{N}\log\alpha=\log \left(\frac{(\alpha \mathrm{N})!}{\alpha^{\alpha \mathrm{N}}} \right)\nonumber\\
    \therefore~&~~~(\mathrm{N}!)^{\alpha}\sim \frac{(\alpha \mathrm{N})!}{\alpha^{\alpha \mathrm{N}}}.
\end{align}
Hence we have a series
\begin{align}
\mathscr{F}(g)&= \sum_{\mathrm{N}=0}^{\infty}g^{\mathrm{N}}a_{\mathrm{N}},~~~~\mathrm{where}~~a_{\mathrm{N}}\sim \mathcal{A}^{\mathrm{N}}(\alpha \mathrm{N})! .
\label{SeRiesphI(g)}
\end{align}
Here we have absorbed a constant $\alpha^\alpha$ into the definition of the parameter $g$. To resum this series  we introduce a factor of $(\alpha\mathrm{N})!$ in its numerator and denominator as:
 \begin{align}
   \mathscr{F}(g)&= \sum_{\mathrm{N}=0}^{\infty}g^{\mathrm{N}}a_{\mathrm{N}}\frac{\int_{0}^{\infty} d\mathrm{S}\exp\left(-\mathrm{S}\right)\mathrm{S}^{\alpha \mathrm{N}}}{(\alpha \mathrm{N})!}~.
 \end{align}
 Interchanging the order of summation and integral we get 
 \begin{align}
     \mathscr{F}(g)&\rightarrow \int_{0}^{\infty}d\mathrm{S}\exp(-\mathrm{S})\left[\sum_{\mathrm{N}=0}^{\infty}g^{\mathrm{N}}a_{\mathrm{N}}\frac{\mathrm{S}^{\alpha\mathrm{N}}}{(\alpha \mathrm{N})!}\right],
 \end{align}
The quantity inside the square bracket is termed as \textit{Borel transform} of  $\mathscr{F}(g)$ 
 \begin{align}
\mathcal{B}[\mathscr{F}](g\mathrm{S})&=\sum_{\mathrm{N}=0}^{\infty}g^{\mathrm{N}}a_{\mathrm{N}}\frac{\mathrm{S}^{\alpha\mathrm{N}}}{(\alpha \mathrm{N})!},
 \end{align}
which in fact has a finite radius of convergence. Thus, the \textit{Borel resummation} of the formal power series $\mathscr{F}(g)$ is basically the Laplace transform of the Borel transform 
 \begin{align}
\mathfrak{s}(\mathscr{F})(g)&= \int_{0}^{\infty}d\mathrm{S}\exp \left(-\mathrm{S} \right) \mathcal{B}[\mathscr{F}] (g\mathrm{S}).
 \end{align}
 Further simplification of this result yields
 \begin{align}
     \mathfrak{s}(\mathscr{F})(g)&= \int_{0}^{\infty}d\mathrm{S}~\exp\left(-\mathrm{S}\right) \sum_{\mathrm{N}=0}^{\infty}g^{\mathrm{N}}\mathcal{A}^{\mathrm{N}}\mathrm{S}^{\alpha \mathrm{N}}\nonumber\\
     &= \left(\frac{1}{g^{1/\alpha}}\right)\int_{0}^{\infty}d\mathrm{S}~\exp\left(-\frac{\mathrm{S}}{g^{1/\alpha}} \right)\sum_{\mathrm{N}=0}^{\infty}\mathcal{A}^{\mathrm{N}}\mathrm{S}^{\alpha \mathrm{N}}\nonumber\\
     &= \left(\frac{1}{g^{1/\alpha}} \right)\int_{0}^{\infty} d\mathrm{S}\exp\left(-\frac{\mathrm{S}}{g^{1/\alpha}} \right) \frac{1}{1-\mathcal{A}\mathrm{S}^{\alpha}},\label{BoReLreSumMedAnsWer}
 \end{align}
 where to go from 1st line to the 2nd line we have made a change of integration variable $\mathrm{S}\rightarrow \frac{\mathrm{S}}{g^{1/\alpha}}$ and to arrive at the last line we have performed the sum over $\mathrm{N}$. We note that the Borel resummation (\ref{BoReLreSumMedAnsWer}) has led to non-perturbative terms of the form $\exp \left(-\frac{\mathrm{S}}{g^{1/\alpha}}\right)$. The zeros of the following polynomial  
\begin{align}
    \mathrm{S}^{\alpha}-\frac{1}{\mathcal{A}}&\equiv \prod_{i=1}^{\alpha} \left(\mathrm{S}-a_{i}\right)
\end{align}
are nothing but the poles in the Borel plane. 
Substituting the Borel resummed series (\ref{BoReLreSumMedAnsWer}) in (\ref{ExpValMOp}) we arrive at 
\begin{align}
    \mathrm{Num}[\langle \hat{\mathbf{g}}_{\mu \nu}\rangle_{\overline{\sigma}}]&=\left[\frac{1}{g^{1/\alpha}} \int_{0}^{\infty} d\mathrm{S}\exp\left(-\frac{\mathrm{S}}{g^{1/\alpha}} \right) \frac{1}{1-\mathcal{A}\mathrm{S}^{\alpha}}\right]_{\mathrm{P.V.}}\int  d^{11}k \frac{\overline{\alpha}_{\mu\nu}(k)}{a(k)}\psi_{\mathbf{k}}(\mathrm{X})e^{-ik_0 t}.\label{BORELRSummed}
\end{align}
In the following section, we discuss the effective field theory criteria to obtain four-dimensional de Sitter space in type IIB, heterotic $SO(32)$ and heterotic $\mathrm{E}_8\times \mathrm{E}_8$ theory.
\section{Null Energy Condition}
\label{sec:NEC}
There are various energy conditions that put restrictions on the stress energy tensor. For instance, according to the null energy condition (NEC), the stress energy tensor has to satisfy the following constraint
\begin{align}
T_{\mu\nu}k^{\mu}k^{\nu}\geq 0, \label{NECConstraint}
\end{align}
for any null vector $k^{\mu}$. For a perfect fluid in 4 dimensional spacetime, the stress energy tensor is of the form 
\begin{align}
    T_{\mu\nu}&=(\rho +p)u_{\mu}u_{\nu}+pg_{\mu \nu}\label{StressEnergyTenPF}
\end{align}
where $\rho$ denotes the energy density, $p$ denotes the momentum density and $u_{\mu}$ is the velocity 4 vector. Substituting (\ref{StressEnergyTenPF}) in (\ref{NECConstraint}) we obtain 
\begin{align}
    T_{\mu \nu}k^{\mu }k^{\nu}&=( \rho +p)u_{\mu }k^\mu u_{\nu}k^\nu +pg_{\mu\nu}k^\mu k^\nu \geq 0\nonumber\\
   & \implies (\rho+ p)(u_{\mu}k^\mu)^2\geq 0.
\end{align}
Since the factor $(u_{\mu}k^{\mu})^2$ is non-negative, the NEC for a perfect fluid boils down to 
\begin{align}
    (\rho+p)\geq 0.\label{NECforPF} 
\end{align} 
For a flat 3+1 dimensional FLRW spacetime with metric configuration
\begin{align}
    ds^2=-d\tau^2+ a^2(\tau) ((dx^1)^2+(dx^2)^2+(dx^3)^2)
\end{align}
the Friedman equations take the form
\begin{align}
 (H(\tau))^2&=\frac{8 \pi G_N}{3}\rho,~~~~\dot{H}(\tau)=-4\pi G_{N}(\rho + p),
\end{align}
where $H(\tau)=\dot{a}(\tau)/a(\tau)$ is the Hubble parameter and $\tau\in (-\infty,\infty)$ is the time coordinate. Thus, the NEC in equation (\ref{NECforPF}) implies $\dot{H}(\tau)\leq 0$. If we assume $a(\tau)\propto \tau^\delta$ then the null energy condition becomes $\delta \geq 0$. In terms of the conformal time coordinate $t$, the FLRW metric takes the form 
\begin{align}
    ds^2&=a^2(t)\left(-dt^2+ (dx^1)^2+(dx^2)^2+(dx^3)^2\right)
\end{align}
The relation between $\tau$ and $t$ can be determined by noting that 
\begin{align}
    \int\tau^{-\delta} d\tau &= \int dt \nonumber\\
    \implies \tau =( (&1-\delta)t)^{{1}/{(1-\delta)}}
\end{align}
Therefore,
\begin{align}
    a(t)\propto t^{\delta/(1-\delta)}:=t^n.\label{ScaleFacFLRW}
\end{align}
Consequently, the null energy condition $\delta \geq 0$ becomes 
\begin{align}
    \delta&=\frac{1+n}{n}\geq 0\implies \frac{1}{n}\geq -1.\label{NECforFLRWcos}
\end{align}
Now, a well defined effective field theory description of the excited state configuration requires us to put the following constraint on the type IIA coupling constant   \cite{Bernardo_2021}
\begin{align}
    \frac{\partial g_s}{\partial t}\propto g_{s}^{+\mathrm{ve}}.\label{EFTconditION}
\end{align}
Let us analyze this for the case of de Sitter in type IIB theory as studied in Section \ref{typeIIBdSisom}. For convenience, below we write down the metric of IIB theory from equation (\ref{tyPeIIBMetrIc})
\begin{align}
    ds^2_{\mathrm{IIB}} &=\frac{ {a}^{2}(t)}{ {\rm H}^2(y)}\left(-dt^2 +  \left(dx^1\right)^2 +\left(dx^2\right)^2 + \left(dx^3\right)^2\right) + {\rm H}^2(y)\mathrm{F}_{1}(t)g_{\alpha\beta}(y)dy^{\alpha}dy^{\beta}\nonumber\\
    &~~~~~~+{\rm H}^2(y)\mathrm{F}_{2}(t)g_{mn}(y) dy^m dy^n.  
\end{align}
We consider the scale factor to be of the form 
\begin{align}
 a^2(t)&=(\Lambda t^2)^n,
\end{align}
where the parameter $n$ is defined analogously to that appearing in equation~(\ref{ScaleFacFLRW}). The dual type IIA coupling constant (\ref{TyPeIIA coUplInG}) becomes 
\begin{align}
\overline{g}_{s}\equiv\frac{g_s}{\mathrm{H}(y)}&=\frac{1}{a(t)}=\frac{1}{(\Lambda t^2)^{n/2}}.
\end{align}
Correspondingly, the EFT condition (\ref{EFTconditION}) implies 
\begin{align}
    \frac{\partial \overline{g}_s}{\partial t }&=-\frac{n}{t}\overline{g}_{s}=-n{\Lambda^{1/2}\overline{g}_{s}^{1/n} }\overline{g}_{s}\propto \overline{g}_s^{(1+1/n)\geq 0}\nonumber\\
   &~~~~~~~~~~~~ \implies \frac{1}{n}\geq -1.
\end{align}
This is precisely, the null energy condition for a flat 3+1 dimensional FLRW cosmology as appearing in equation (\ref{NECforFLRWcos}). Thus, for the de Sitter configuration in type IIB theory \textit{the existence of well defined effective field theory description boils down to satisfying the null energy condition}. In the following subsections we study the null energy condition in heterotic $\mathrm{SO}(32)$ and $\mathrm{E}_8\times \mathrm{E}_8$ theory. 
\subsection{NEC in heterotic \texorpdfstring{$\mathrm{SO}(32)$}{SO(32)} string theory}
Recall from (\ref{HeTerotiCSO(32)MetrIc}),  the metric of heterotic $\mathrm{SO(32)}$ theory in string frame is given as
\begin{align}
ds^{2}_{\mathrm{Het}~\mathrm{SO}(32)}&=  \mathrm{F}_{1}(t)a^2(t)\left(-dt^2+\sum_{i=1}^3\left(dx^{i}\right)^2 \right) +\mathrm{H}^{4}(y)\mathrm{F}_{1}(t)\mathrm{F}_{2}(t)g_{mn}dy^{m}dy^{n}\nonumber\\
~&~~~~~~~~~~~+\delta_{\alpha\beta}\left(dy^{\alpha}+\mathcal{B}^{(1)\alpha}_{~~~m} dy^{m}\right)\left(dy^{\beta}+\mathcal{B}^{(1)\beta}_{~~~n} dy^{n}\right).
\end{align}
We therefore demand 
\begin{align}
\mathrm{F}_1(t)a^2(t)&=(\Lambda t^2)^n.
\end{align}
Substituting the dominant contribution of $\mathrm{F}_1(t)$ from equation (\ref{ScalingOfFi's}) and the scale factor from equation (\ref{TypeIIAcoUpliNg}) we get 
\begin{align}
    \left(\frac{g_s}{\mathrm{H}(y)}\right)^{\beta_{o}-2}&=(\Lambda t^2)^n\nonumber\\
    \implies {\overline{g}_s}&=\left(\Lambda t^2\right)^{\frac{n}{\beta_o -2}}.\label{SFrameCoup}
\end{align}
In Einstein frame the metric of heterotic $\mathrm{SO}(32)$ theory  is given as (\ref{EinsFrameSO32}) 
\begin{align}
    ds^2_{\mathrm{Het~SO(32),(E)}}&=\frac{\sqrt{\mathrm{F}_{1}(t)}a^2(t)}{\mathrm{H}(y)}\left(-dt^2+\sum_{i=1}^3\left(dx^{i}\right)^2 \right)+\mathrm{H}^{3}(y)\sqrt{\mathrm{F}_{1}(t)}\mathrm{F}_{2}(t)g_{mn}dy^{m}dy^{n} \nonumber\\
    ~&~~~~~~~+\frac{1}{\mathrm{H}(y)\sqrt{\mathrm{F}_{1}(t)}}\delta_{\alpha\beta}\left(dy^{\alpha}+\mathcal{B}^{(1)\alpha}_{~~~m} dy^{m}\right)\left(dy^{\beta}+\mathcal{B}^{(1)\beta}_{~~~n} dy^{n}\right).
\end{align}
As a result, we now have to demand 
\begin{align}
    \sqrt{\mathrm{F}_1(t) }a^2(t)&=(\Lambda t^2)^n\nonumber\\
    \implies \left(\frac{g_s}{\mathrm{H}(y)}\right)^{\frac{\beta_o }{2}-2} & =(\Lambda t^2)^n\nonumber\\
    \implies {\overline{g}_s}=&(\Lambda t^2)^{\frac{2n}{\beta_o -4}}.\label{EinFrameCoup}
\end{align}
Combining (\ref{SFrameCoup}) and (\ref{EinFrameCoup}) we can write
\begin{align}
    \overline{g}_s&=(\Lambda t^2)^{\frac{nv}{\beta_o -2v}},\label{typeIIANECHetSO32}
\end{align}
where $v=1,2$ for the string frame, the Einstein frame respectively. Computing its derivative with respect to $t$ we get 
\begin{align}
  \frac{\partial \overline{g}_s}{\partial t}&= \left( \frac{2 n v}{\beta_o -2v}\right) \frac{\overline{g}_s}{t}\nonumber\\
  &=\left( \frac{2 n v}{\beta_o -2v}\right)\Lambda^{1/2}\overline{g}_s^{\frac{2v-\beta_o}{2nv}} {\overline{g}_s}\propto \overline{g}_s^{\left(1+ \frac{2v-\beta_o}{2nv}\right)\geq 0}\nonumber\\
  \therefore ~~& 1+\frac{2v-\beta_o}{2nv}\geq 0 \implies \frac{1}{n}\geq -\frac{1}{1-\frac{ \beta_o }{2v}},
\end{align}
with $0< \beta_o < \frac{2}{3}$. Since, the bound should be independent of the frame at consideration as well as the  parameter $\beta_o$ of the theory, one can conclude that we must have $\frac{1}{n}\geq -1$, which in fact holds for either $v=1$ or $v=2$ and all values of $\beta_o$ lying within its range of validity. 

If we further consider the temporal variation of the parameter $\beta(t)$ as in equation (\ref{F1(t)withbeta(t)}) then the type IIA coupling (\ref{typeIIANECHetSO32}) gets modified as 
\begin{align}
    \overline{g}_s&=(\Lambda t^2)^{\frac{n v}{\beta(t)-2v}}. 
\end{align}
Taking its derivative with respect to $t$ gives 
\begin{align}
    \log \overline{g}_s&=\left(\frac{nv}{\beta(t)-2v}\right) \log (\Lambda t^2)\nonumber\\
    \implies \frac{1}{\overline{g}_s}\frac{\partial \overline{g}_s}{\partial t }&= -\frac{nv}{(\beta(t)-2v)^2}\dot{\beta}(t)\log(\Lambda t^2)+\left(\frac{nv}{\beta(t)-2v}\right) \frac{2 }{ t}\nonumber\\
    &=\left(\frac{\dot{\beta}(t) }{2v-\beta(t)}\right) \log \overline{g}_s - \frac{2nv\sqrt{\Lambda}}{(2v-\beta(t))}\overline{g}_s^{\frac{2v-\beta(t)}{2nv}}\nonumber\\
 \implies   \frac{\partial \overline{g}_s }{\partial t}&=\overline{g}_s^{1+\mathrm{B}(t)}-2nv \sqrt{\Lambda}~ \overline{g}_s^{1+\frac{1}{n}-\frac{\beta(t)}{2nv}+\mathrm{C}(t)}\label{tderofgsbar}
\end{align}
where 
\begin{align}
    \overline{g}_s^{\mathrm{B}(t)}&=\left(\frac{\dot{\beta}(t) }{2v-\beta(t)}\right) \log \overline{g}_s \implies \mathrm{B}(t)=\frac{\log\left[\left(\frac{\dot{\beta}(t) }{2v-\beta(t)}\right) \log \overline{g}_s\right]}{\log \overline{g}_s},\label{ExpofB(t)}\\ \mathrm{and}~~~\overline{g}_s^{\mathrm{C}(t)}&=\frac{1}{(2v-\beta(t))}~~\implies C(t)=-\frac{\log(2v-\beta(t))}{\log\overline{g}_s}.\label{ExpofC(t)}
\end{align}
Since $\overline{g}_s<<1$ and $2v-\beta(t)>1$ for $v=(1,2)$ and  $0<\beta(t)<\frac{2}{3}$ we find that
\begin{align}
\mathbf{sgn}~\mathrm{C}(t)=\mathbf{sgn}\left(\frac{\log(2v-\beta(t))}{|\log \overline{g}_s|} \right)>0.
\end{align}
For $\mathrm{B}(t)$ in equation (\ref{ExpofB(t)}) to be a well-defined function, $\beta(t)$ must be a decreasing function satisfying $\dot{\beta}(t)<0$, which in fact is consistent with the form of $\beta(t)$ given in equation (\ref{F1(t)withbeta(t)}). With this information, we can write $\mathrm{B}(t)$ as 
\begin{align}
    \mathrm{B}(t)&=-\frac{\log\left[\frac{|\dot{\beta}(t)|}{(2v-\beta(t))}|\log\overline{g}_s| \right] }{|\log\overline{g}_s|}.
\end{align}
However, to determine the sign of $\mathrm{B}(t)$ we need to know the precise value of $|\dot{\beta}(t)|.$ Let us consider,
\begin{align}
\frac{|\dot{\beta}(t)|}{(2v-\beta(t))}<\frac{1}{|\log \overline{g}_s|}.\label{EFTCondItionForSO32Het}
\end{align}
It is evident that for this case $\mathbf{sgn}~\mathrm{B}(t)>0.$ As a result, the first term on the right hand side of equation (\ref{tderofgsbar}) satisfies the effective field theory criterion (\ref{EFTconditION}). Further, the null energy condition is obtained from the second term on the right hand side of equation (\ref{tderofgsbar}), viz. $\frac{1}{n}\geq -1$ which is independent of the time dependent parameters of the theory. 

If we now consider $|\dot{\beta}(t)|>>1,$ then 
\begin{align}
\mathbf{sgn}~\mathrm{B}(t)=\mathbf{sgn}\left(\frac{\log(2v-\beta(t))-\log|\log \overline{g}_s|-\log|\dot{\beta}(t)|}{|\log\overline{g}_s |} \right)<0.
\end{align}
This follows from the fact that $1 < 2v - \beta(t) \leq 4$, and therefore the sum
\begin{equation}
\log\lvert \log \overline{g}_s \rvert + \log \lvert \dot{\beta}(t) \rvert
\end{equation}
dominates over $\log\big(2v - \beta(t)\big)$. In this regime, we cannot make a definitive statement about the validity of the effective field theory, since this would require detailed knowledge of the magnitude of $\lvert \mathrm{B}(t) \rvert$. Note, however, that in the late-time limit as $\overline{g}_s \to 0$, it is reasonable to expect that
$
\lvert \log \overline{g}_s \rvert > \log\!\big(2v - \beta(t)\big), 
\lvert \log \overline{g}_s \rvert > \log \lvert \log \overline{g}_s \rvert ~
\mathrm{along~with}~\log \lvert \log \overline{g}_s \rvert > \log\!\big(2v - \beta(t)\big).
$ This implies  $|\mathrm{B}(t)|$ is less than 1 provided  
\begin{align}
    |\log \overline{g}_s|&>\log|\dot{\beta}(t)|\nonumber\\
    \implies -\log \overline{g}_s&>\log|\dot{\beta}(t)|\nonumber\\
    \implies \log ~\overline{g}_s|\dot{\beta}&(t)|<0\nonumber\\
    \implies \overline{g}_s|\dot{\beta}(t)|&<1.\label{ConstEFTinHetSo32}
\end{align} 
As a result $1+\mathrm{B}(t)>0$ holds true, satisfying the effective field theory constraint (\ref{EFTconditION}) for the case of heterotic $SO(32)$ theory with time dependent parameter $\beta(t).$  
\subsection{NEC in heterotic \texorpdfstring{$\mathrm{E}_{8}\times \mathrm{E}_{8}$}{E8XE8}  string theory }
From (\ref{HetroTicE8XE8 metric}) the metric of $\mathrm{E}_8 \times \mathrm{E}_8$ heterotic theory is given as 
\begin{align} ds^{2}_{\mathrm{Het~E_8\times E_8}}&=a^2(t)\mathrm{F}^{1/2}_{1}(t)\mathrm{F}_{3}^{1/2}(t)\left(-dt^2+ \sum_{i}\left(dx^{i}\right)^2+\left(dx^3\right)^2\right)\nonumber\\
+&\left[\mathrm{F}_{1}^{-1/2}\mathrm{F}_{3}^{1/2} (d\theta_{1})^2+\mathrm{H}^{4}(y)\mathrm{F}^{1/2}_{1}\mathrm{F}_{3}^{1/2}\mathrm{F}_{2}g_{mn}dy^{m}dy^{n}\right]+\mathrm{F}_{1}^{-1/2}\mathrm{F}_{3}^{1/2}(d\tilde{x}_{11})^2.
    \end{align}
Following the similar procedure as in the previous subsection, now we need to consider
\begin{align}
\sqrt{\mathrm{F}_1(t)\mathrm{F}_3(t)} a^2(t)&= \left( \Lambda t^2\right)^n
\end{align}
Substituting $\mathrm{F}_1(t)$ and $\mathrm{F}_3(t)$ from equations (\ref{EXpofF1(t)}) and (\ref{EXpofF3(t)}) (without taking into account the time dependence of the parameters $\hat{\alpha}(t)$ and $\hat{\beta}(t)$) we obtain 
\begin{align}
    \left(\frac{g_s}{\mathrm{H}(y)}\right)^{\frac{\hat{\alpha}_o+\hat{\beta}_o}{3}-2}&=(\Lambda t^2)^n\nonumber\\
    \implies \overline{g}_s&= (\Lambda t^2)^{\frac{3n}{\hat{\alpha}_o+\hat{\beta}_o-6}}.\label{SframeinE8Th}
\end{align}
In Einstein frame, the metric of heterotic $\mathrm{E}_8\times \mathrm{E}_8$ theory takes the following form 
\begin{align}
ds^2_{\mathrm{Het}~\mathrm{E_8\times E_8 ,(E)}}&= \frac{1}{\sqrt{g_{\mathrm{Het}} }}ds^2_{\mathrm{Het}~\mathrm{E}_8\times\mathrm{E}_8}=\frac{1}{\mathrm{H}(y) \mathrm{F}_{1}^{1/8}(t)\mathrm{F}_3^{3/8}(t) }ds^2_{\mathrm{Het}~\mathrm{E}_8\times\mathrm{E}_8}\nonumber\\
&=\frac{a^2(t)}{\mathrm{H}(y)} \mathrm{F}_1^{3/8}(t)\mathrm{F}_3^{1/8}(t)\left(-dt^2+\sum_{i=1}^3(dx^i)^2 \right)\nonumber\\
&+\frac{1}{\mathrm{H}(y)}\left[\mathrm{F}_1^{-5/8}(t)\mathrm{F}_3^{1/8}(t)(d\theta_1)^2+\mathrm{H}^4(y)\mathrm{F}_1^{3/8}(t)\mathrm{F}_2(t)\mathrm{F}_3^{1/8}(t)g_{mn}dy^m dy^n \right]\nonumber\\
&+\frac{1}{\mathrm{H}(y)}\mathrm{F}_1^{-5/8}(t)\mathrm{F}_3^{1/8}(t)(d\tilde{x}_{11})^2,\label{EinframeinE8Th}
\end{align}
where we have used the coupling $g_{\mathrm{Het}}$ of heterotic $\mathrm{E}_8\times \mathrm{E}_8$ theory as given in equation (\ref{CouplIngE8E8HeT}). Thus, in the Einstein frame, we must have 
\begin{align}
    \mathrm{F}_1^{3/8}(t)\mathrm{F}_3^{1/8}(t)a^2(t)&=(\Lambda t^2)^n\nonumber\\
    \implies \left(\frac{g_s}{\mathrm{H}(y)} \right)^{\frac{3\hat{\alpha}_o+\hat{\beta}_o}{12}-2 }&=(\Lambda t^2)^n\nonumber\\
    \implies \overline{g}_s =(\Lambda t^2&)^{\frac{12n }{3\hat{\alpha}_o+\hat{\beta}_o-24}}
\end{align}
The above expression of $\overline{g}_s$ in Einstein frame can be compactly written along with the expression of $\overline{g}_s$ in the string frame (\ref{SframeinE8Th}) as 
\begin{align}
    \overline{g}_s&=(\Lambda t^2)^{\frac{n}{\frac{(5-v)\hat{\alpha}_o+(7-3v)\hat{\beta}_o}{12}-2}}
\end{align}
where $v=1,2$ for the string, the Einstein frame respectively. Computing its derivative with respect to $t$ gives 
\begin{align}
    \frac{\partial \overline{g}_s }{\partial t}&=\left(\frac{2n}{\frac{(5-v)\hat{\alpha}_o+(7-3v)\hat{\beta}_o}{12}-2}\right)\frac{\overline{g}_s}{t}\nonumber\\
    &\propto \overline{g}_s^{\left(\frac{2- \frac{(5-v)\hat{\alpha}_o+(7-3v)\hat{\beta}_o }{12}}{2n}+1\right)\geq 0}\nonumber\\
    \therefore~~~~ & \frac{2- \frac{(5-v)\hat{\alpha}_o+(7-3v)\hat{\beta}_o }{12}}{2n}+1\geq 0\implies \frac{1}{n}\geq -\frac{1}{1- \frac{(5-v)\hat{\alpha}_o+(7-3v)\hat{\beta}_o}{24} }.\label{EFTBound}
\end{align}
As explained in Section \ref{E8XE8ThrdistWarpf}, the range of $\hat{\alpha}_o ,\hat{\beta}_o$ is $0<\frac{\hat{\alpha}_{o}}{9}<\hat{\beta}_o<\hat{\alpha}_{o}<1 $. Since the bound (\ref{EFTBound}) must be independent of the frames as well as the parameters $\hat{\alpha}_o$ and $\hat{\beta}_o$, the effective field theory criterion can be written as 
\begin{align}
    \frac{1}{n} \ge -1,
\end{align}
which is the most stringent one.  
If we now take into account the time dependence of the parameters $\hat{\alpha}(t)$ and $\hat{\beta}(t)$ as in equations (\ref{EXpofF1(t)}) and (\ref{EXpofF3(t)}) then the string frame coupling constant (\ref{SframeinE8Th})   gets modified as  
\begin{align}
    \overline{g}_s&=({\Lambda}t^2)^{\frac{3n }{\hat{\alpha}(t) +\hat{\beta}(t)-6}}.
\end{align}
Taking its derivative with respect to $t$ would give us the similar answer as in \eqref{tderofgsbar} with $v=1$ and $\beta(t)$ replaced by $\frac{\hat{\alpha}(t)+\hat{\beta}(t)}{3}$. Below we write down the result
\begin{align}
    \frac{ \partial \overline{g}_s}{\partial t}&=\overline{g}_s^{1+ {\mathrm{B}}(t)}- 6n \sqrt{ \Lambda} \overline{g}_s^{1+ \frac{1}{n}- \frac{\hat{\alpha}(t)+\hat{\beta}(t)}{6n}+{\mathrm{C}}(t)}
\end{align}
where 
\begin{align}
{\mathrm{B}}(t)&=\frac{ \log\left[\left( \frac{ \dot{\hat{\alpha}}(t)+\dot{\hat{\beta}}(t) }{6-\hat{\alpha}(t)-\hat{\beta}(t)} \right) \log \overline{g}_s \right] }{\log\overline{g}_s},~~~\mathrm{C}(t)=-\frac{\log \left(6-\hat{\alpha}(t)-\hat{\beta}(t) \right) }{\log \overline{g}_s}.
\end{align}
Since $\overline{g}_s<<1$ and $\hat{\beta}(t)<\hat{\alpha}(t)<1$, the sign of $\mathrm{C}(t)$ is positive. Further, for $\mathrm{B}(t)$ to make sense it is necessary that we have $(\dot{\hat{\alpha}}(t)+\dot{\hat{\beta}}(t))\leq 0$, i.e. 
\begin{align}
    \mathrm{B}(t)&=-\frac{ \log\left[ \frac{\vert \dot{\hat{\alpha}}(t)+\dot{\hat{\beta}}(t) \vert }{\left( 6-\hat{\alpha}(t)-\hat{\beta}(t)\right) }  \vert \log \overline{g}_s \vert \right] }{\vert\log\overline{g}_s \vert}.
\end{align}
As in the case of heterotic $SO(32)$ theory, the sign of $\mathrm{B}(t)$ would require us to know the precise value of $| \dot{\hat{\alpha}}(t)+\dot{\hat{\beta}}(t) |.$ If the functional forms of $\hat{\alpha}(t)$ and $\hat{\beta}(t)$ are such that the following condition holds
\begin{align}
    \frac{ |\dot{\hat{\alpha}}(t)+\dot{\hat{\beta}}(t) | }{(6 -\hat{\alpha}(t)-\hat{\beta}(t) )}<\frac{1}{|\log\overline{g}_s|}
\end{align}
then $\mathbf{sgn}~\mathrm{B}(t)>0$, implying the existence of well defined effective field theory description.  

\vspace{5pt}
\noindent
In the next section we discuss the implication of the constraint $\overline{g}_s<1$. As we will see this leads to a \textit{temporal regime} in which we expect the effective field theory description to remain well defined. Further, we will investigate  the constraint (\ref{ConstEFTinHetSo32}) with specific functional choices of $\beta(t)$ in heterotic $SO(32)$ theory.   
\section{Trans-Planckian Censorship Conjecture bound}
\label{sec:TCCconsTraints}
\begin{figure}[h!]
    \centering
\includegraphics[width=0.70\linewidth]{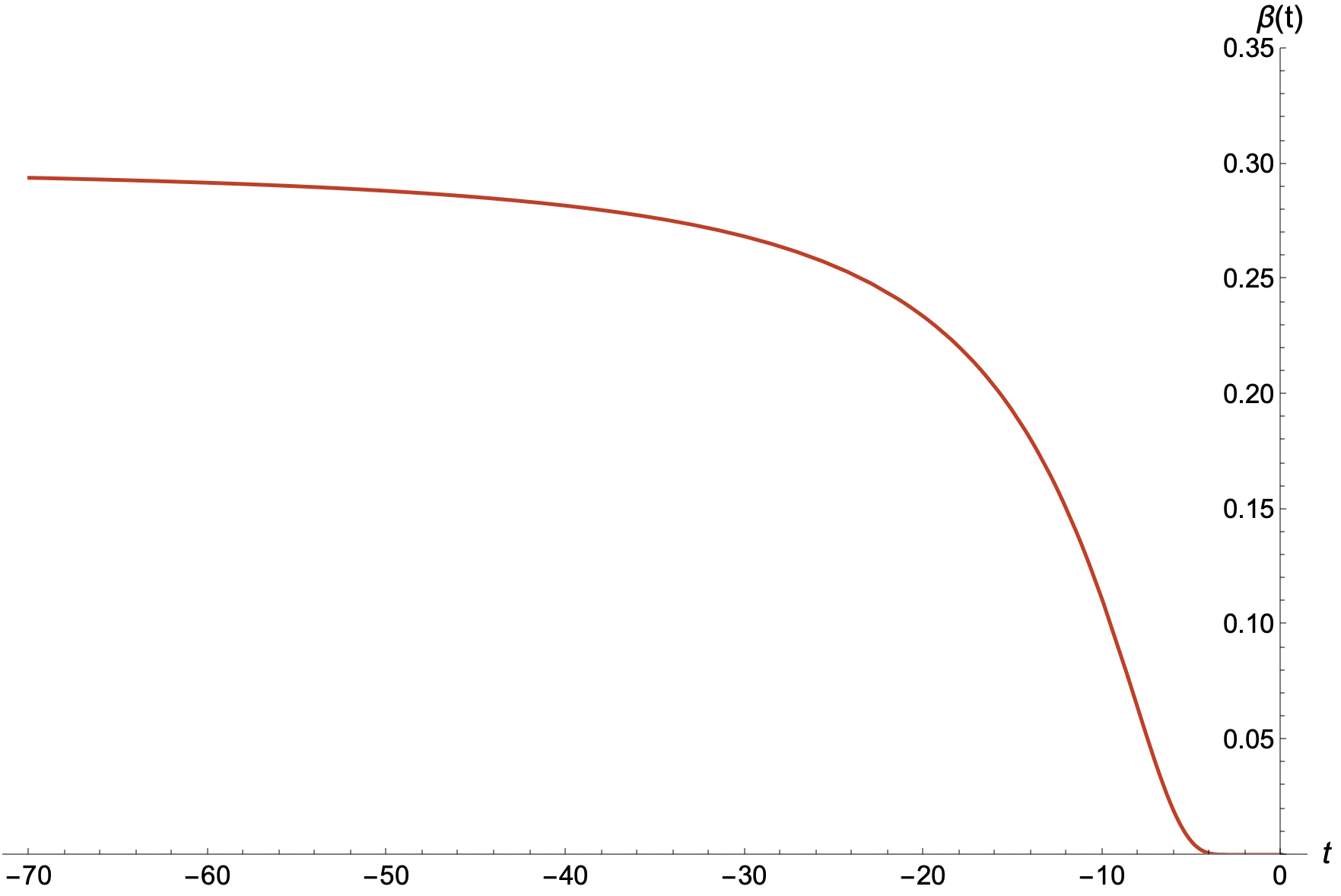}
    \caption{Plot of $\beta(t)$ as in equation \eqref{2ndbetatAnsatZ} with $\sigma=60,$ $\beta_o=0.3,~\mathrm{and}~|\epsilon |=\frac{1}{6}$. }
\label{2ndBetatPlot}
\end{figure}

According to the Trans-Planckian Censorship Conjecture, in any consistent quantum theory of gravity a sub-Planckian mode can never exit the Hubble horizon and freeze \cite{Bedroya_2020}. This gives rise to a bound on the duration of the accelerating phase. Surprisingly, the constraint $\overline{g}_s<1$, which is necessary for a well defined effective field theory description, gives us a temporal domain which is identical to the Trans-Planckian Censorship Conjecture (TCC) bound. In particular, consider the expression of $\overline{g}_s$ (\ref{IIAcouPlingdeSitterIIB}) in the case of de Sitter isometry in type IIB theory. Imposing the aforementioned constraint we get
\begin{align}
\overline{g}_s\equiv \frac{g_s}{\mathrm{H}(y)}=\sqrt{\Lambda }|t|<1\implies -\frac{1}{\sqrt{\Lambda}}<t<0.\label{TCCBounD}
\end{align}
Likewise, the expression of $\overline{g}_s$ (\ref{StringCCSFEF}) in the case of de Sitter isometry in heterotic $\mathrm{SO}(32)$ would give us the same temporal domain 
\begin{align}
    \overline{g}_s&=(\sqrt{\Lambda}|t|)^{\frac{ 2v }{2v-\beta(t)}}<1\implies -\frac{1}{\sqrt{\Lambda}}<t<0.
\end{align}
However, the constraint $\overline{g}_s|\dot{\beta}(t)|<1$ (see equation \ref{ConstEFTinHetSo32}), that allows the effective field theory description would now reduce the aforementioned temporal domain. To quantify this shorter temporal domain we consider the following functional form of $\beta(t)$ 
\begin{align}
\beta(t)&=\frac{\beta_o\exp \left[\frac{1}{\sqrt{\Lambda }}\left(\frac{1}{|\epsilon|}-\frac{1}{|t|} \right) \right] }{1+\exp \left[\frac{1}{\sqrt{\Lambda }}\left(\frac{1}{|\epsilon|}-\frac{1}{|t|} \right) \right]}\label{AnsatzofBetat1}
\end{align}
where $0<\beta_o< \frac{2}{3}$ and $\epsilon>0$ is as given in equation (\ref{ValueofEpsilonHetSO}). 
\begin{figure}
    \centering
\includegraphics[width=0.69\linewidth]{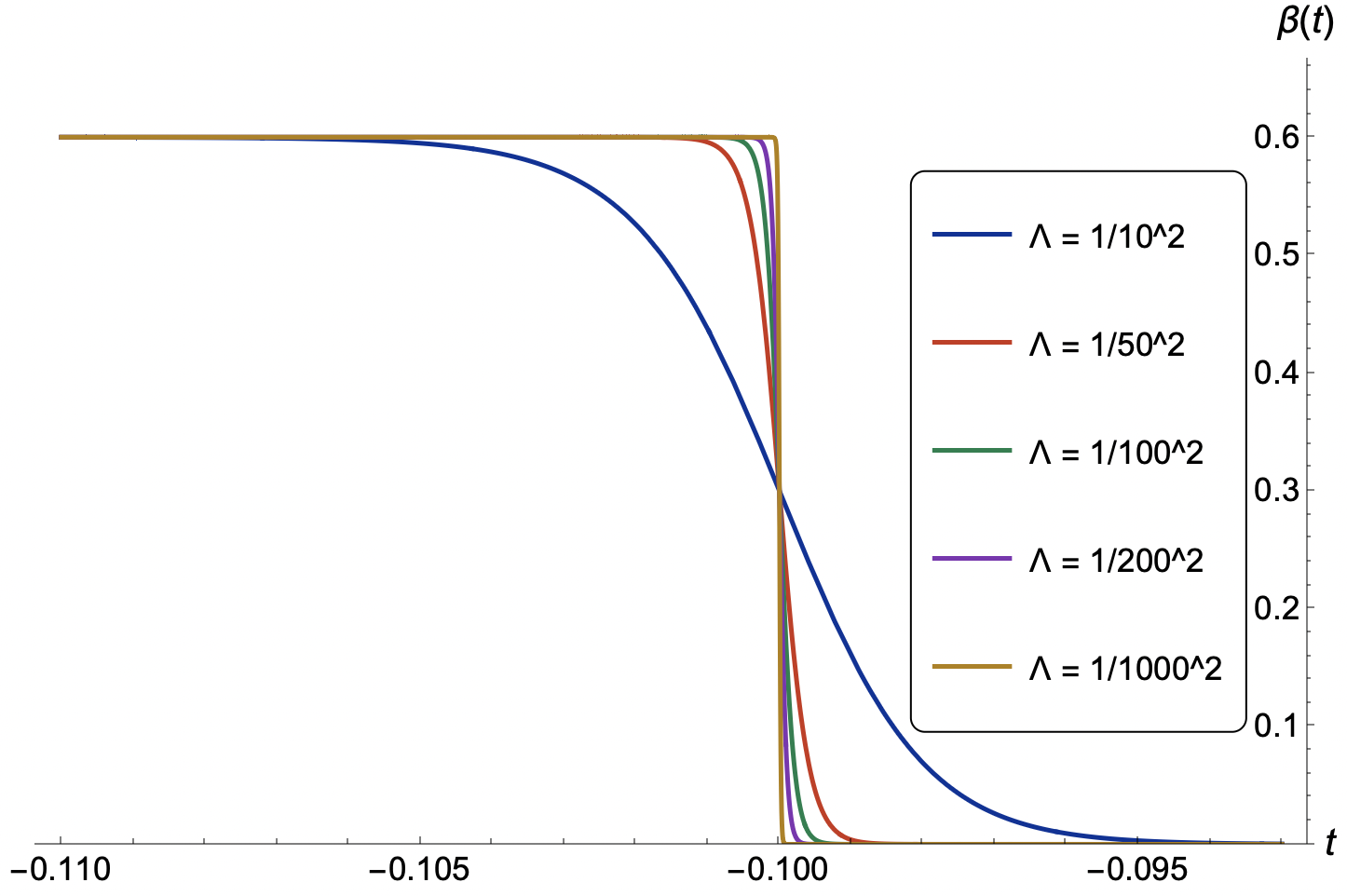}
    \caption{Plots of $\beta(t)$ (see equation (\ref{AnsatzofBetat1})) for different values of $\Lambda$ and $\beta_o=0.6,~|\epsilon |=0.1$}
\label{betat1}
\end{figure}
Plots of this function for different values of $\Lambda$ appears in Figure \ref{betat1}. Computing its derivative with respect to $t$ we get 
\begin{align}
    \dot{\beta}(t)&=-\frac{\beta_o}{\sqrt{\Lambda}|t|^2}\frac{\exp \left[\frac{1}{\sqrt{\Lambda }}\left(\frac{1}{|\epsilon|}-\frac{1}{|t|} \right) \right] }{\left[1+\exp \left[\frac{1}{\sqrt{\Lambda }}\left(\frac{1}{|\epsilon|}-\frac{1}{|t|} \right) \right]\right]^2}.\label{Ansatzdotbetat1}
\end{align}
\begin{figure}[h!]
    \centering
\includegraphics[width=0.74\linewidth]{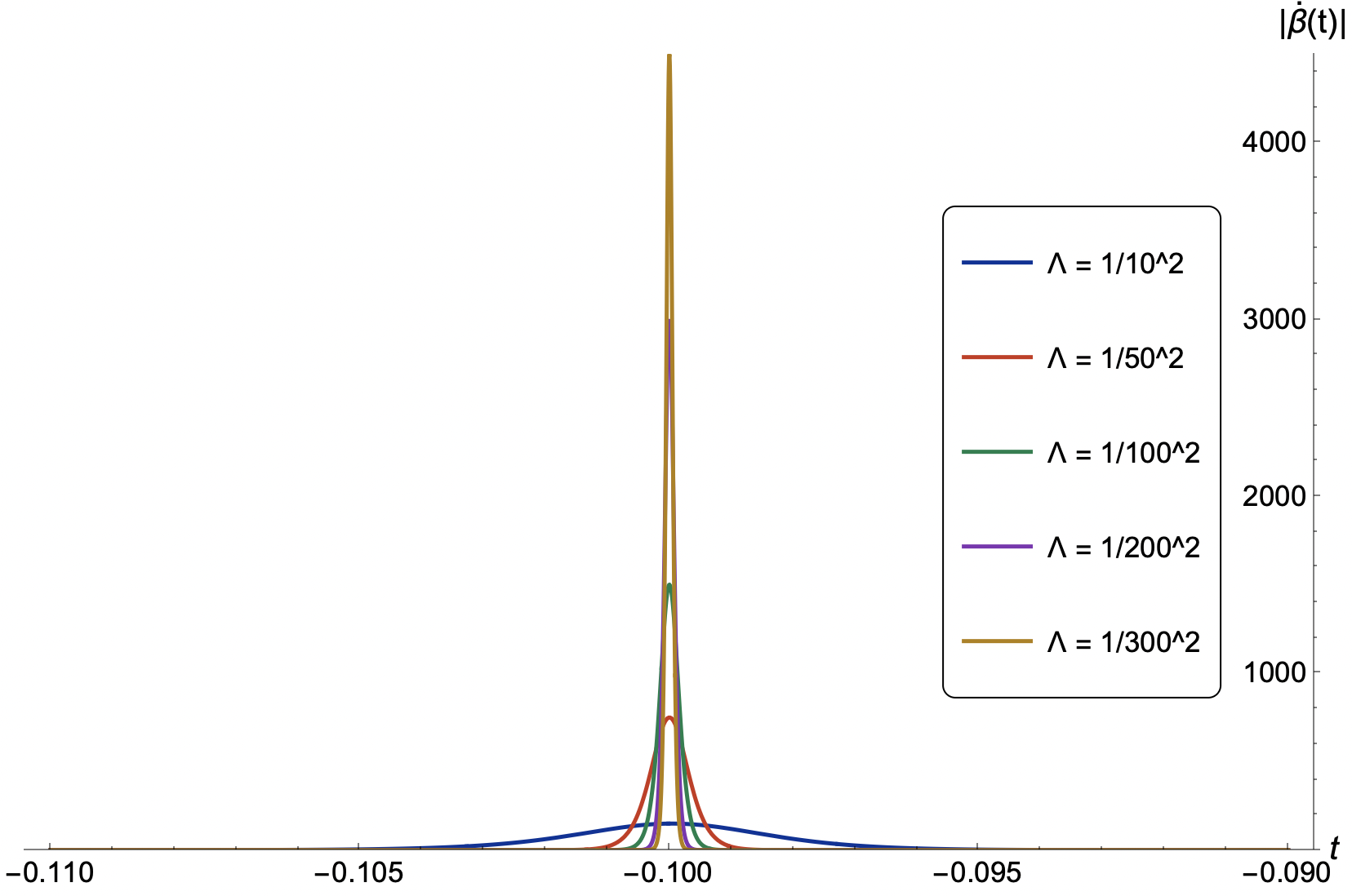}
    \caption{Plots of $|\dot{\beta}(t)|$ as in  equation (\ref{Ansatzdotbetat1}) for different values of $\Lambda$ and $\beta_o=0.6,~|\epsilon|=0.1$}
\label{dotbetat1}
\end{figure}
The absolute values of $\dot{\beta}(t)$, for different values of $\Lambda$, have been plotted in Figure \ref{dotbetat1}. We notice these are sharply peaked at $|t|=|\epsilon|$ with $\dot{\beta}(\epsilon)=-\frac{\beta_o}{4|\epsilon|^2\sqrt{\Lambda}}.$ 
\begin{figure}[h!]
    \centering
\includegraphics[width=0.74\linewidth]{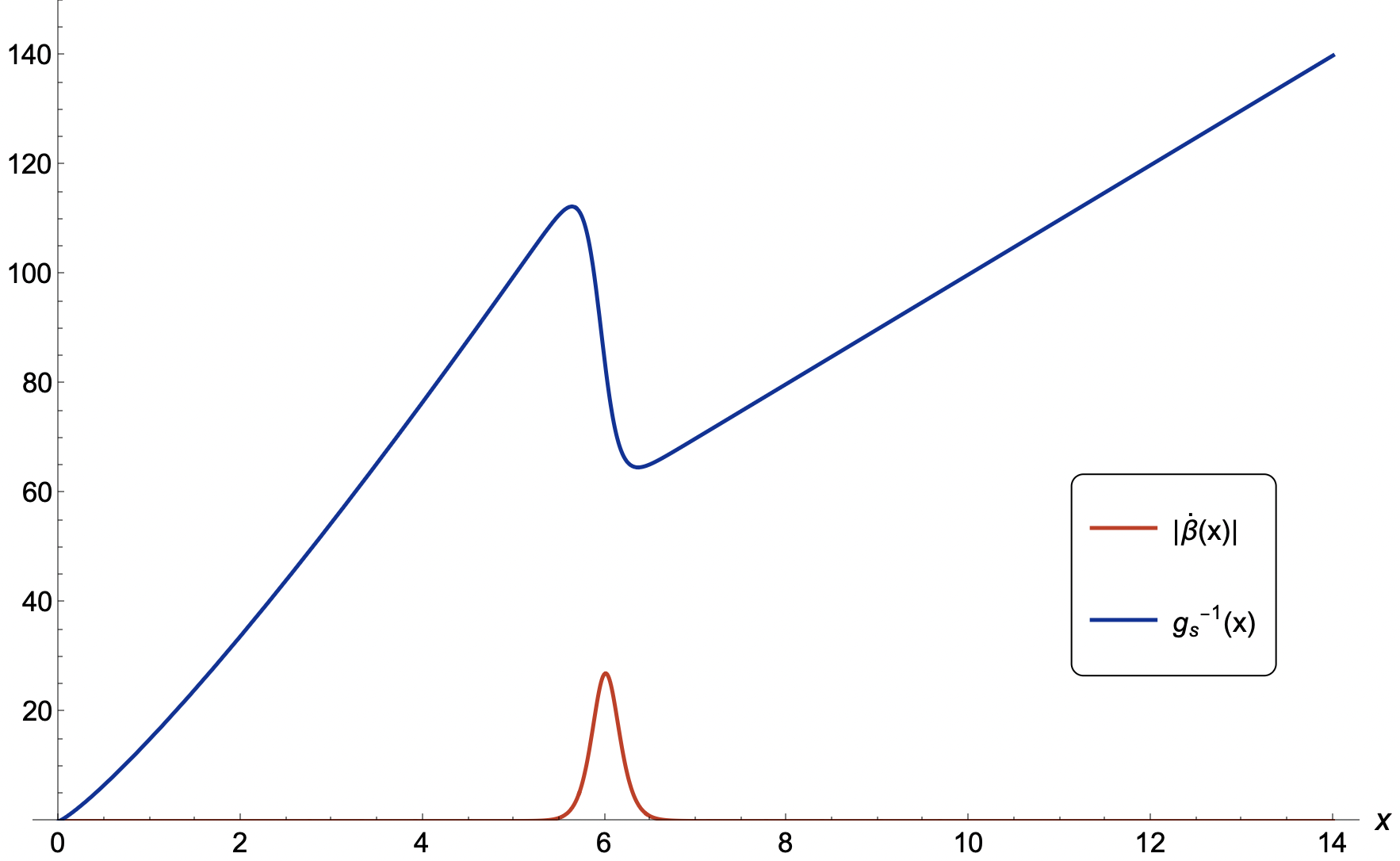}
    \caption{Plots of $|\dot{\beta}(x)|$ and $\overline{g}_{s}^{-1}(x)=(\sqrt{\Lambda} x^{-1})^{-\frac{2v}{(2v-\beta(x))}}$ with $\beta(x)$ as in equation (\ref{bEtaInVarx}) and $\Lambda=0.01,$ $\beta_o=0.3,~v=1,~|\epsilon |=\frac{1}{6}$. Here $x:=\frac{1}{|t|}\in (0,\infty)$. }
\label{betadOt&gsinV}
\end{figure}
\begin{figure}[h!]
    \centering
\includegraphics[width=0.74\linewidth]{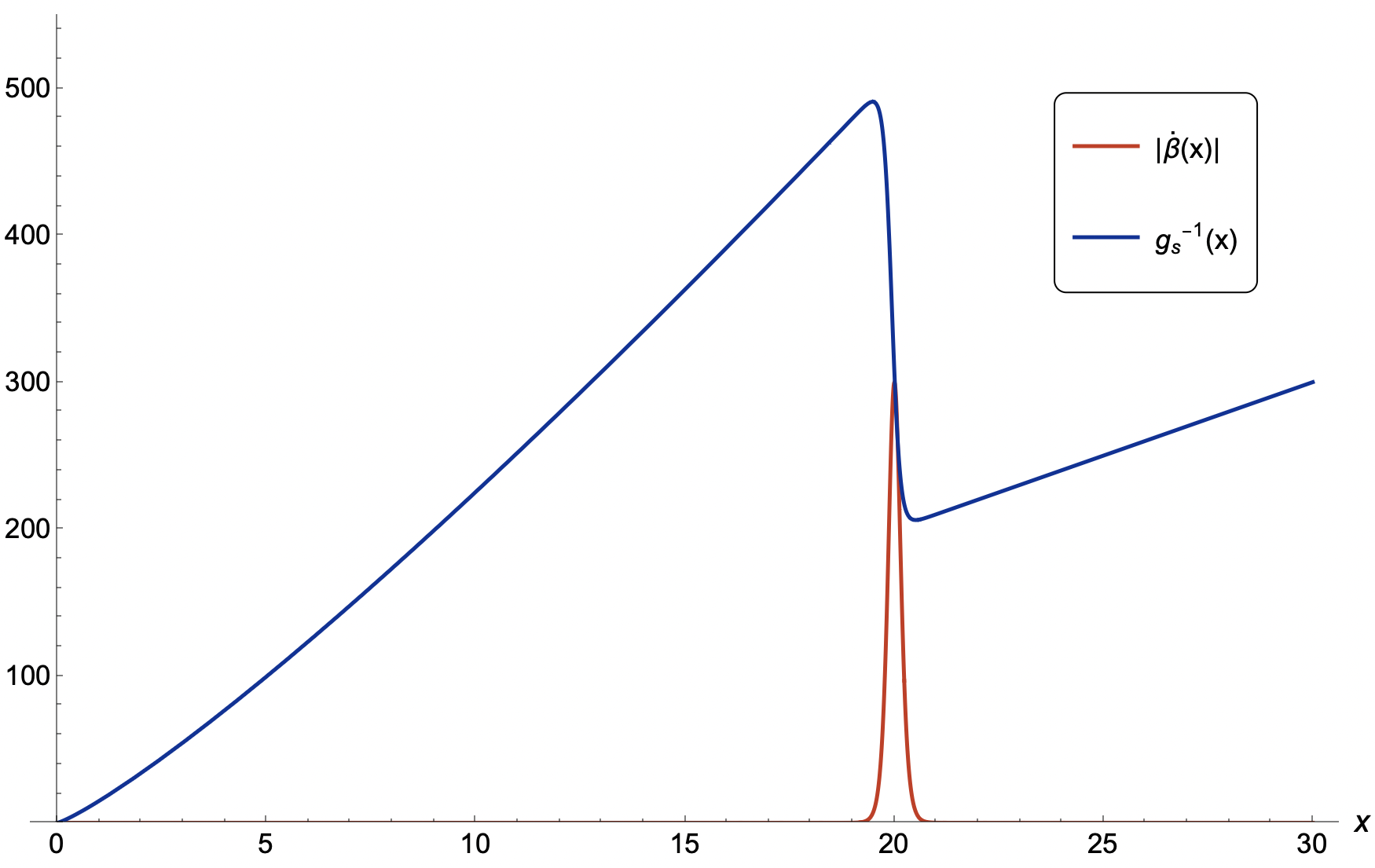}
    \caption{Plots of $|\dot{\beta}(x)|$ and $\overline{g}_{s}^{-1}(x)=(\sqrt{\Lambda} x^{-1})^{-\frac{2v}{(2v-\beta(x))}}$ with $\beta(x)$ as in equation (\ref{bEtaInVarx}) and $\Lambda=0.01,$ $\beta_o=0.3,~v=1,~|\epsilon|=\frac{1}{20}$. Here $x:=\frac{1}{|t|}\in (0,\infty)$. The two curves intersect at $x=20.01$ and $x=20.08$.}
\label{betadOt&gsinVmin20}
\end{figure}
With $\frac{1}{|t|}\equiv x\in(0,\infty)$ we can rewrite (\ref{AnsatzofBetat1}) as 
\begin{align}
    \beta(x)&=\frac{\beta_o \exp\left[ \frac{1}{\sqrt{\Lambda}}\left(\frac{1}{|\epsilon|}-x \right) \right] }{1+\exp\left[ \frac{1}{\sqrt{\Lambda}}\left(\frac{1}{|\epsilon|}-x\right) \right]}.\label{bEtaInVarx}
\end{align}
In Figure \ref{betadOt&gsinV} we have plotted $|\dot{\beta}(x)|$ and $\overline{g}_{s}^{-1}(x)$ in the string frame with $\Lambda=0.01,~\beta_o=0.3$ and $|\epsilon |=\frac{1}{6}$. We see that the two curves do not intersect at any value of $x$. For this case, it is easy to check numerically that $\overline{g}_s(x)|\dot{\beta}(x)|<1$ is satisfied for all $x$ lying in the range $\sqrt{\Lambda}<x<\infty$. If instead we choose the value of $|\epsilon|$ to be $\frac{1}{20}$, then the two curves $|\dot{\beta}(x)|$ and $\overline{g}_s^{-1}(x)$ intersect close to $x=\frac{1}{|\epsilon|}=20$ (see Figure \ref{betadOt&gsinVmin20}). Numerically we find that close to the intersection points the constraint $\overline{g}_s(x)|\dot{\beta}(x)|<1$ ceases to hold. In other words, we no longer have a well-defined effective field theory description in the entire domain $\sqrt{\Lambda }<x<\infty$. Instead, we can break this domain into two parts $\sqrt{ \Lambda}<x<\frac{1}{|\epsilon|}$ and $\frac{1}{|\epsilon|}<x<\infty$. Since the latter corresponds to the late-time regime and is essential for realizing de Sitter isometry, it is reasonable to say that the effective field theory is valid only in $-\epsilon<t<0$.  One might think that decreasing the value of $\Lambda$ can circumvent this issue. But we have checked that we can always find a suitable value of $\epsilon$ for which there is a breakdown of the effective field theory constraint (\ref{ConstEFTinHetSo32}) near $x=\frac{1}{|\epsilon|}$. This implies that the functional ansatz of $\beta(t)$ (\ref{Ansatzdotbetat1}) is not suitable and hence we look for a smoother function having the following form
\begin{align}
    \beta(t)&=\beta_o \exp\left( -\frac{\sigma^2 |\epsilon|^2}{|t|^2}\right),\label{2ndbetatAnsatZ}
\end{align}
where $\sigma>>1$ is a dimensionless constant. The plot of this function with $\sigma=60$ and $|\epsilon |=\frac{1}{6}$ appears in Figure \ref{2ndBetatPlot}. As $t\rightarrow 0^{-},$ $\beta(t)\rightarrow \beta_{o}\exp(-\infty)\approx 0$. At $|t|=t_{\star}=\sigma |\epsilon|,$ $\beta(t_{\star})=0.37\beta_o$. Thus, in the temporal domain $-\epsilon \leq t<0$, $\beta(t)= 0$ satisfying our requirement as in equation \eqref{F1(t)withbeta(t)}. Further, as $t\rightarrow -\frac{1}{\sqrt{\Lambda}},$ $\beta (t)\rightarrow \beta_o \exp\left(-\sigma^2|\epsilon|^2 \Lambda \right)\approx \beta_o$ provided $\sigma |\epsilon|\sqrt{\Lambda}<<1$.
The derivative of $\beta(t)$ \eqref{2ndbetatAnsatZ} with respect to $t$ gives 
\begin{align}
    \dot{\beta}(t)= -\frac{ 2 \beta_o \sigma^2 |\epsilon|^2}{|t|^3}\exp\left( -\frac{\sigma^2 |\epsilon|^2}{|t|^2}\right),\label{DotBeTAt2nd}
\end{align}
\begin{figure}[h!]
    \centering
\includegraphics[width=0.70\linewidth]{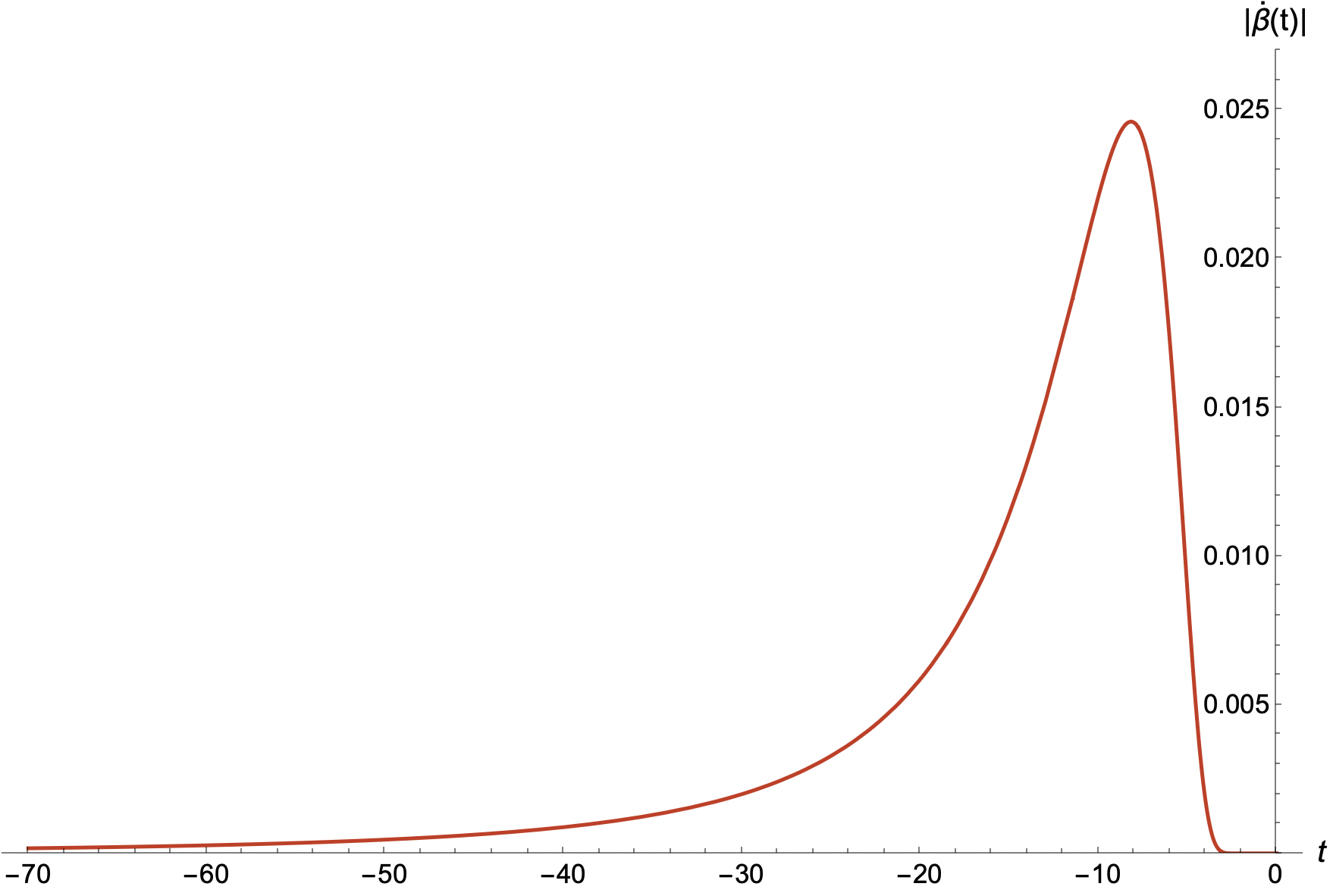}
    \caption{Plot of $|\dot\beta(t)|$ as in equation \eqref{DotBeTAt2nd} with $\sigma=60,$ $\beta_o=0.3,~\mathrm{and}~|\epsilon |=\frac{1}{6}$. }
\label{2ndBetadotPlot}
\end{figure}

which peaks at 
\begin{align}
    \ddot{\beta}(t)\big|_{|t|}&=\left(-\frac{ 6 \beta_o \sigma^2|\epsilon|^2 }{|t|^4}+ \frac{ 4\beta_o \sigma^4|\epsilon|^4 }{|t|^6} \right)\exp\left(- \frac{\sigma^2 |\epsilon|^2 }{|t|^2}\right)=0\nonumber\\
    \implies &|t|=\sqrt{ \frac{2\sigma^2|\epsilon|^2 }{3}}<<\sqrt{ \frac{2}{3 \Lambda } }.
\end{align}
The plot of $|\dot{\beta}(t)|$ as in Figure \ref{2ndBetadotPlot} shows that this is not as sharply peaked as the ones in Figure \ref{dotbetat1}.  
\begin{figure}[h!]
    \centering
\includegraphics[width=0.74\linewidth]{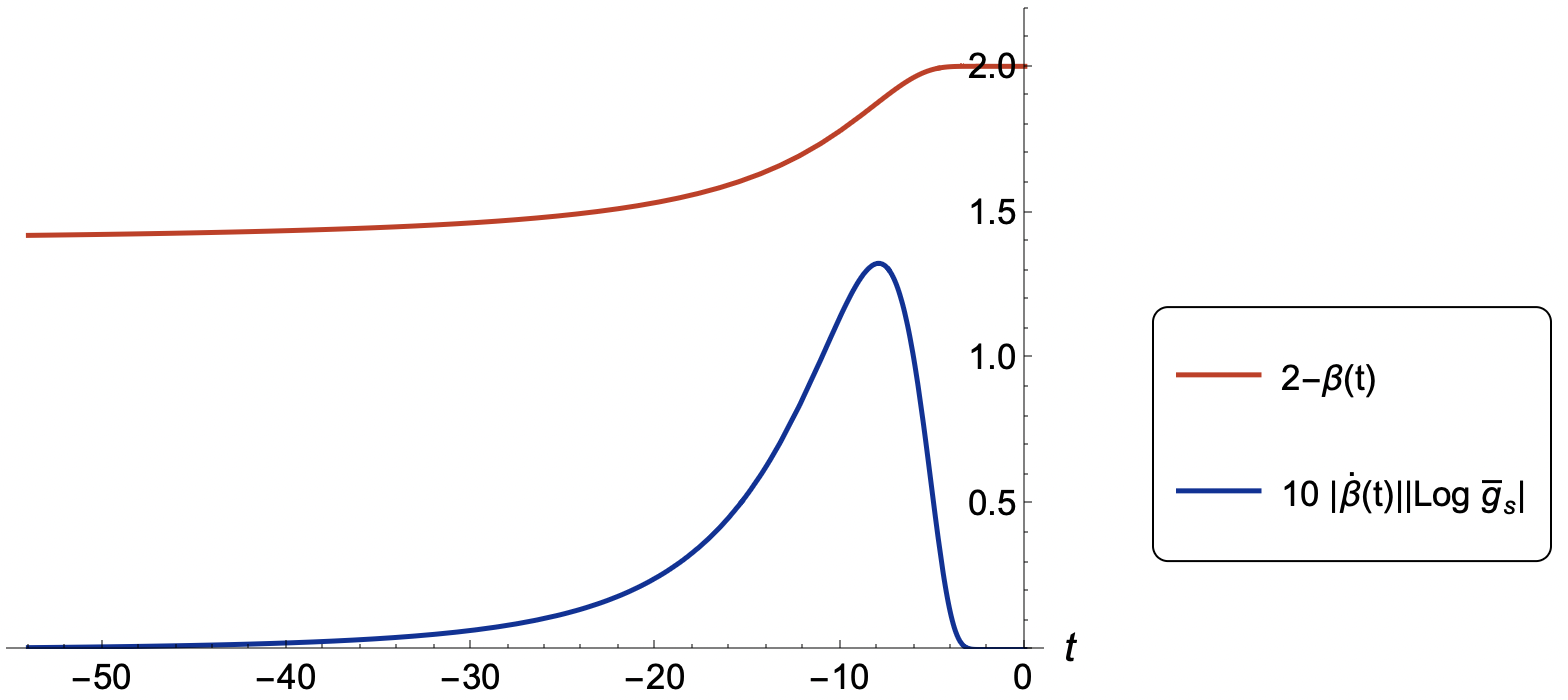}
\caption{Plots of $2-\beta(t)$ and $10 |\dot{\beta}(t)||\log\overline{g}_s|$ with $\beta(t),\dot{\beta}(t)~\mathrm{and}~\overline{g}_s$ as in equations \eqref{2ndbetatAnsatZ}, \eqref{DotBeTAt2nd} and \eqref{StringCCSFEF}. Various parameters have been taken to be  $\sigma=60,$ $\beta_o=0.6,~|\epsilon |=\frac{1}{6},~\Lambda=10^{-4}$. }
\label{Plotsof2minbetat&DotBetaLoggs}
\end{figure}
In Figure \ref{Plotsof2minbetat&DotBetaLoggs} we have plotted $2-\beta(t)$ and $10 |\dot{\beta}(t)||\log \overline{g}_s|$. It shows that for all values of $t$ the condition
\begin{align}
    2-\beta(t)> |\dot{\beta}(t)||\log \overline{g}_s|
\end{align}
holds true. From equation (\ref{EFTCondItionForSO32Het}), it implies that there is an well defined effective field theory description in the entire temporal domain $-\frac{1}{\sqrt{\Lambda}}<t<0$.  
One may worry whether these time-dependent solutions respect the constraints arising from axionic cosmology. We address this issue in the following section. 
\section{Axions in \texorpdfstring{$\mathrm{E}_{8}\times \mathrm{E}_{8}$}{E8XE8}  heterotic string theory}
\label{sec:AxionsinE8Thoery}
In string frame the metric of $\mathrm{E}_{8}\times \mathrm{E}_{8}$ heterotic string theory (\ref{HetroTicE8XE8 metric}) with de Sitter isometry can be written as
\begin{align}    ds^2_{10}&=\frac{1}{\Lambda t^2} ds^2_{\mathbb{R}^{3,1}}+\mathrm{H}^4(y)\mathrm{F}_{2}(t)\sqrt{\mathrm{F}_1(t)\mathrm{F}_3(t)}ds^2_{\mathcal{M}_4}+\sqrt{\frac{\mathrm{F}_3(t)}{\mathrm{F}_1(t)}}ds^2_{\mathbb{T}^2}\label{SframeMetE8xE8ThrwithdS}\\
&=\frac{1}{\Lambda t^2}ds^2_{\mathbb{R}^{3,1}}+\mathrm{H}^4(y)\left(\Lambda t^2\right)^{\frac{\hat{\alpha}(t)-\hat{\beta}(t)}{12-2\hat{\alpha}(t)-2\hat{\beta}(t)}}ds^2_{\mathcal{M}_4}+
    \left(\Lambda t^2\right)^{\frac{\hat{\beta}(t)-\hat{\alpha}(t)}{6-\hat{\alpha}(t)-\hat{\beta}(t)}}ds^2_{\mathbb{T}^2}
\end{align}
where to arrive at the second step we have substituted the expressions of $\mathrm{F}_1(t),~\mathrm{F}_3(t)$ and $\mathrm{F}_2(t)$ from (\ref{EXpofF1(t)}), (\ref{EXpofF3(t)}) and (\ref{ExpofF2inE8TheorY}), along with the value of $g_{s}/\mathrm{H}(y)$ from (\ref{TimedepgsinE8Thr}) and $\mathbb{T}^2\equiv \hat{\mathbf{S}}^1_{\theta_1}\times \mathbf{S}^1_{11}$.
The coupling of this theory is (\ref{CouplIngE8E8HeT})
\begin{align}
g_{\mathrm{het}}&=\mathrm{H}^2(y)\mathrm{F}_{1}^{{1}/{4}}(t)\mathrm{F}_{3}^{{3}/{4}}(t)=\mathrm{H}^2(y)\left( \frac{g_s}{\mathrm{H}(y)} \right)^{\frac{\hat{\alpha}(t)+3\hat{\beta}(t)}{6}}=\mathrm{H}^2(y)\left( \Lambda t^2\right)^{\frac{\hat{\alpha}(t)+3\hat{\beta}(t)}{12-2\hat{\alpha}(t)-2\hat{\beta}(t)}}.\label{HetCoUStringFrame}
\end{align}  
In Einstein frame, the metric of $\mathrm{E}_8\times \mathrm{E}_8$ heterotic string theory (\ref{EinframeinE8Th}) has the following form 
\begin{align}
ds^2_{\mathrm{Het}~\mathrm{E}_8\times \mathrm{E}_8,~(\mathrm{E})}&=g_{s}^{-2}\mathrm{H}(y)\mathrm{F}_1^{3/8}(t)\mathrm{F}_{3}^{1/8}(t) ds^2_{\mathbb{R}^{3,1}}+\mathrm{H}^3(y)\mathrm{F}_{1}^{3/8}(t)\mathrm{F}_2(t)\mathrm{F}_3^{1/8}(t)ds^2_{\mathcal{M}_4}+ \nonumber\\
&~~~~~~~~~+\mathrm{H}^{-1}(y)\mathrm{F}_1^{-5/8}(t)\mathrm{F}_3^{1/8}(t)ds^2_{\mathbb{T}^2}\label{MetrHetE8ThEinFrame}
\end{align}
For this theory to have de Sitter isometry in flat slicing the following condition must be satisfied 
\begin{align}
g_{s}^{-2}\mathrm{H}(y)\mathrm{F}_1^{3/8}(t)\mathrm{F}_3^{1/8}(t)&=\frac{1}{\Lambda t^2 \mathrm{H}(y)}\nonumber\\
\implies \left(\frac{g_{s}}{\mathrm{H}(y)}\right)^{-2+\frac{  2 \hat{\alpha}(t)}{3}\times \frac{3}{8}+ \frac{ 2\hat{\beta}(t) }{3}\times \frac{1}{8} }&=\frac{1}{\Lambda t^2}\nonumber\\
\implies \overline{g}_s=(\Lambda t^2)^{\frac{ 12}{24-3\hat{\alpha}(t)-\hat{\beta}(t)}}&.\label{ExpofgsinEinFrameE8Th}
\end{align}
Above result of $\overline{g}_s$ in Einstein frame and the string frame result (\ref{TimedepgsinE8Thr}) can be compactly written as
\begin{align}
\overline{g}_s&=( \Lambda t^2)^{\frac{3v^2}{ 6v^2 -(2v-1)\hat{\alpha}(t)-\hat{\beta}(t) }},\label{IIAinString&EinsFrame}
\end{align}
where $v=1,2$ for the string frame, the Einstein frame respectively. Hence, in the Einstein frame, the heterotic coupling (\ref{CouplIngE8E8HeT}) takes the form 
\begin{align} g_{\mathrm{Het}}&=\mathrm{H}^2(y)\mathrm{F}_1^{1/4}(t)\mathrm{F}_3^{3/4}(t)=\mathrm{H}^2(y)\left( {\overline{g}_s }\bigg|_{v=2}\right)^{\frac{\hat{\alpha}(t)+3\hat{\beta}(t)}{6}}=\mathrm{H}^2(y)\left(\Lambda t^2\right)^{\frac{ 2\hat{\alpha}(t)+6\hat{\beta}(t)}{24- 3\hat{\alpha}(t)-\hat{\beta}(t)}},\label{HetCoUEinsFrameE8Th}
\end{align}
and time independence of the volume of the six dimensional internal space implies 
\begin{eqnarray}
\left( \mathrm{F}_1^{3/8}(t)\mathrm{F}_2(t)\mathrm{F}_3^{1/8}(t)\right)^4 (\mathrm{F}_1^{-5/8}(t) \mathrm{F}_3^{1/8}(t))^2&=&1\nonumber\\
\implies \mathrm{F}_{2}(t)= \mathrm{F}_1^{-\frac{1}{16}}(t)\mathrm{F}_3^{-\frac{3}{16}}(t)&=& \left({ \overline{g}_s}\bigg|_{v=2}\right)^{\frac{ -\hat{\alpha}(t) -3\hat{\beta}(t)}{24} }. \label{ExpofF2(t)inEinstFrameE8Th}
\end{eqnarray}
Substituting (\ref{ExpofgsinEinFrameE8Th}) and (\ref{ExpofF2(t)inEinstFrameE8Th}) the metric (\ref{MetrHetE8ThEinFrame}) takes the form: 
\begin{align}
ds^2_{ \mathrm{Het}~\mathrm{E}_8\times \mathrm{E}_8,(\mathrm{E})}&=\frac{\mathrm{H}^{-1}(y)}{\Lambda t^2 } ds^2_{ \mathbb{R}^{3,1}}+ \mathrm{H}^3(y) \left(\Lambda t^2\right)^{ \frac{ 5 \hat{\alpha}(t)-\hat{\beta}(t)}{48- 6\hat{\alpha}(t)-2\hat{\beta}(t)}}ds^2_{\mathcal{M}_4 }+\mathrm{H}^{-1}(y)\left( \Lambda t^2 \right)^{\frac{\hat{\beta}(t)-5\hat{\alpha}(t)}{24- 3\hat{\alpha}(t)-\hat{\beta}(t)}}ds^2_{\mathbb{T}^2}.\label{EinsFrameMetDSisOE8Th} 
\end{align}
It shows that in the Einstein frame as $t\rightarrow 0^-$, $\hat{\beta}(t)$ has to be equal to $5 \hat{\alpha}(t)$ for there to be no late time singularities of the internal manifolds $\mathcal{M}_4$ and $\mathbb{T}^2$. So, we can define 
\begin{align}
\hat{\alpha}(t)&=\begin{cases}   \hat{\alpha}_o,~~~~-\frac{1}{\sqrt{\Lambda }}<t<-\epsilon_1\\
\hat{\gamma}_{o},~~~~~-\epsilon_1< t<0
\end{cases},~~~~~\hat{\beta}(t)=\begin{cases}
\hat{\beta}_o,~~~~~~~~~~~~~~-\frac{1}{\sqrt{\Lambda }}<t<-\epsilon_1\\
(4v-3)\hat{\gamma}_{o},~~~~~-\epsilon_1< t<0 
\end{cases}.\label{FuncAnsTzhatAlphahatBeta}
\end{align}
In the rest of this section, we are going to analyse the constraints coming from axionic cosmology in order to better understand the behavior of $\hat{\alpha}(t)$ and $\hat{\beta}(t)$ proposed above.  

\vspace{5pt}
\noindent
In string frame, the action of the heterotic string theory is as follows 
\begin{align}
    S&=\frac{1}{2\kappa_{10}^2}\int d^{10}x \sqrt{-\mathrm{det}G}~e^{-2\phi}\left[\mathrm{R}+4\partial_{\mu}\phi \partial^{\mu}\phi-\frac{1}{2}\vert \mathbf{H}_{3}\vert^2-\frac{\alpha^\prime}{120}\mathrm{Tr}(\vert \mathbf{F}\vert^2) \right]
\end{align}
where the trace is taken with respect to the 496 dimensional adjoint representation of $\mathrm{E}_{8}\times \mathrm{E}_{8}$ theory and $|{\rm A}_p|^2=\frac{1}{p!}{\rm A}_{M_1\cdots M_p}{\rm A}_{N_1\cdots N_p}G^{M_1N_1}\cdots G^{M_pN_p}.$ Taking into account the string frame metric (\ref{SframeMetE8xE8ThrwithdS}), the zero mode action of the graviton turns out to be 
\begin{align}
    S_{G}^{(0)}&=\frac{1}{2\kappa_{10}^2 }\int d^{10}x \sqrt{-\mathrm{det}G  }~e^{-2\phi} ~\mathrm{R}\nonumber\\
    &\rightarrow \frac{1}{2\kappa_{10}^2}\int d^4x d^6y\sqrt{(\mathrm{H}^4(y))^4}~ \sqrt{\mathrm{det}g_{\mathcal{M}_4}\mathrm{det}g_{\mathbb{T}^2}}\frac{1}{\mathrm{H}^4(y)\left( \Lambda t^2\right)^{\frac{\hat{\alpha}(t)+3\hat{\beta}(t)}{6-\hat{\alpha}(t)-\hat{\beta}(t)}}}\mathrm{R}^{(4)}\nonumber\\
    &=\frac{1}{2\kappa_{10}^2}\int d^4x  \frac{1}{\left( \Lambda t^2\right)^{\frac{\hat{\alpha}(t)+3\hat{\beta}(t)}{6-\hat{\alpha}(t)-\hat{\beta}(t)}}}\mathrm{R}^{(4)}\int d^6y~ \mathrm{H}^4(y)\sqrt{\mathrm{det}g_{\mathcal{M}_4}\mathrm{det}g_{\mathbb{T}^2}}\nonumber\\
    &=\frac{\mathrm{M}_p^2}{2}\int d^4x  \frac{1}{\left( \Lambda t^2\right)^{\frac{\hat{\alpha}(t)+3\hat{\beta}(t)}{6-\hat{\alpha}(t)-\hat{\beta}(t)}}}\mathrm{R}^{(4)}
\end{align}
where 
\begin{align}
\mathrm{M}_{p}^2&=\frac{1}{\kappa_{10}^2}\int d^6y ~\mathrm{H}^4(y)\sqrt{\mathrm{det}g_{\mathcal{M}_4}\mathrm{det}g_{\mathbb{T}^2}}\nonumber\\
    &=4\pi \mathrm{M}_{s}^{8}\int d^6y ~\mathrm{H}^4(y)\sqrt{\mathrm{det}g_{\mathcal{M}_4}\mathrm{det}g_{\mathbb{T}^2}}.\label{Mp2ExpResSION}
\end{align}
Likewise, the action due to the zero mode of the NS-NS field turns out to be 
\begin{align}
    S_{\rm NS}^{(0)}&=   -\frac{1}{4\kappa_{10}^2}\int d^{10}x \sqrt{-\mathrm{det}G} ~e^{-2\phi}~ \vert \mathbf{H}_3\vert^2 
    \nonumber\\
    &\rightarrow-\frac{1}{4\kappa_{10}^2}\int d^4 x d^6y\sqrt{(\mathrm{H}^4(y))^4}~ \sqrt{\mathrm{det}g_{\mathcal{M}_4}\mathrm{det}g_{\mathbb{T}^2}}\frac{1}{\mathrm{H}^4(y)\left( \Lambda t^2\right)^{\frac{\hat{\alpha}(t)+3\hat{\beta}(t)}{6-\hat{\alpha}(t)-\hat{\beta}(t)}}}\vert\mathbf{H}_{3}\vert^2\nonumber\\
    &=-\frac{1}{4\kappa_{10}^2}\int d^4x \left( \Lambda t^2\right)^{\frac{\hat{\alpha}(t)+3\hat{\beta}(t)}{\hat{\alpha}(t)+\hat{\beta}(t)-6}}\vert \mathbf{H}_3\vert^2\int d^6y~{\rm H}^{4}(y)\sqrt{\mathrm{det}g_{\mathcal{M}_4}\mathrm{det}g_{\mathbb{T}^2}}\nonumber\\
    &=-\frac{\mathrm{M}_{p}^2}{4}\int d^4x ~\beta(t)~\vert \mathbf{H}_3\vert^2
\end{align}
where in the second step we have used the coupling of heterotic theory in string frame as given in equation (\ref{HetCoUStringFrame}) and
\begin{align}
    \beta(t)&=\left(\Lambda t^2 \right)^{\frac{\hat{\alpha}(t)+3\hat{\beta}(t)}{\hat{\alpha}(t)+\hat{\beta}(t)-6}}.\label{AxiOnbetatStringFrame}
\end{align}
In Einstein frame, the action of the heterotic string theory is given as 
\begin{align}
S_{\mathrm{E}}&=\frac{1}{2 \kappa_{10}^2} \int d^{10}x \sqrt{ -\mathrm{det}g}\left[\mathrm{R}+ 4 \partial_{\mu}\phi \partial^{\mu}\phi -\frac{1}{2}e^{-\phi} \vert \mathbf{H}_3\vert^2- \frac{\alpha^\prime}{120}e^{-\phi /2}\mathrm{Tr}(\vert \mathbf{F}\vert^2) \right].
\end{align}
Now, we take into account the Einstein frame metric (\ref{EinsFrameMetDSisOE8Th}) to write down the zero mode action due to the graviton as 
\begin{align}
S_{g,~\mathrm{E}}^{(0)}&=\frac{1}{2 \kappa_{10}^2} \int d^{10}x \sqrt{-\mathrm{det}g} ~\mathrm{R}\nonumber\\
    &\rightarrow \frac{1}{2\kappa_{10}^2}\int d^4xd^6y \sqrt{ {(\mathrm{H}^{-1}(y) )^4} (\mathrm{H}^3(y))^4{(\mathrm{H}^{-1}(y))^2}}\sqrt{ \mathrm{det}g_{\mathcal{M}_4 } \mathrm{det}g_{\mathbb{T}^2}}~\mathrm{H}(y) \mathrm{R}^{(4)}\nonumber\\
    &=\frac{1}{2 \kappa_{10}^2}\int d^4x~\mathrm{R}^{(4)} \int d^6y~\mathrm{H}^4(y)\sqrt{ \mathrm{det}g_{\mathcal{M}_4 } \mathrm{det}g_{\mathbb{T}^2}}\nonumber\\
    &=\frac{ \mathrm{M}_p^2 }{2}\int d^4x~ \mathrm{R}^{(4)}
\end{align}
where $\mathrm{M}_p^2$ is as given in equation (\ref{Mp2ExpResSION}). In a similar way, the zero mode  action due to the NS-NS field gives 
\begin{align}
    S^{(0)}_{\mathrm{NS}, ~\mathrm{E}}&=-\frac{1}{4\kappa_{10}^2}\int d^{10}x \sqrt{ -\mathrm{det}g} ~e^{-\phi} \vert \mathbf{H}_3\vert^2\nonumber\\
    &\rightarrow -\frac{1}{4 \kappa_{10}^2} \int d^4x d^6 y \sqrt{(\mathrm{H}^{-1}(y))^4(\mathrm{H}^3(y))^4(\mathrm{H}^{-1}(y))^2}\sqrt{ \mathrm{det} g_{\mathcal{M}_4 }\mathrm{det} g_{\mathbb{T}^2}} \nonumber\\
    &~~~~~~~~~~~~~~~~\times \frac{1}{ \mathrm{H}^2(y) (\Lambda t^2 )^{ \frac{2 \hat{\alpha}(t)+6\hat{\beta}(t)}{24-3 \hat{\alpha}(t)-\hat{\beta}(t) }} } \times (\mathrm{H}^{-1}(y))^{-3} \vert \mathbf{H}_3\vert^2 \nonumber\\
    &= -\frac{1}{  4\kappa_{10}^2}\int d^4x ~(\Lambda t^2 )^{ \frac{2 \hat{\alpha}(t)+6\hat{\beta}(t)}{3 \hat{\alpha}(t)+\hat{\beta}(t)-24 }} \vert \mathbf{H}_3\vert^2 \int d^6y~ \mathrm{H}^{4}(y)~ \sqrt{ \mathrm{det} g_{\mathcal{M}_4 }\mathrm{det} g_{\mathbb{T}^2}}\nonumber\\
    &=-\frac{ \mathrm{M}_p^2 }{4}\int d^4 x ~\beta(t) ~\vert \mathbf{H}_3\vert^2 ,  
\end{align}
to arrive at the second step we have substituted the coupling of heterotic theory in Einstein frame (\ref{HetCoUEinsFrameE8Th}) and 
\begin{align}
    \beta(t)=(\Lambda t^2)^{ \frac{2 \hat{\alpha}(t)+ 6\hat{\beta}(t) }{3 \hat{\alpha}(t)+\hat{\beta}(t)-24}}.\label{AxiOnbetatEinFrame}
\end{align}
The results in the string frame (\ref{AxiOnbetatStringFrame}) and the Einstein frame (\ref{AxiOnbetatEinFrame}) can be compactly  written as 
\begin{align}
    \beta(t)&=( \Lambda t^2 )^{ \frac{v \hat{\alpha}(t)+ 3 v \hat{\beta}(t) }{(2v-1)\hat{\alpha}(t)+\hat{\beta}(t) -6v^2 } }
\end{align}
where $v=1,2$ corresponds to the string frame and the Einstein frame respectively. We can rewrite the dimensionally reduced heterotic action due to the NS-NS field as
\begin{align}
    S_{4}&=-\frac{ \mathrm{M}_p^2}{4}\int \beta(t) \vert \mathbf{H}_3\vert^2 + \int a \left[d\mathbf{H}_3 -\frac{\alpha^\prime}{4}\left(\mathrm{tr} \mathrm{R}^{(4)}\wedge \mathrm{R}^{(4)}- \frac{1}{30}\mathrm{Tr}\mathbf{F}\wedge \mathbf{F}\right) \right],
\end{align}
here $a$ is a Lagrange multiplier which in fact is related to the axionic field. We can integrate out $\mathbf{H}_3$ to express the above action in terms of $a:$
\begin{align}
    S_{4}&=- \frac{ \mathrm{M}_p^2}{4 \times 3!}\int d^4 x ~\beta(t)~g^{\mu \mu^\prime }g^{\nu \nu^\prime }g^{\rho \rho^\prime } \mathbf{H}_{3,\mu\nu\rho}\mathbf{H}_{3,\mu^\prime \nu^\prime\rho^\prime} + \frac{1}{3!} \int d^4 x ~a ~\partial_{\sigma}\mathbf{H}_{3,\mu\nu\rho}\epsilon^{\sigma \mu \nu \rho}\nonumber\\
    &~~~~~~~~-\frac{\alpha^\prime}{4}\int a \left( \mathrm{tr}\mathrm{R}^{(4)}\wedge \mathrm{R}^{(4)}-\frac{1}{30}\mathrm{Tr}\mathbf{F}\wedge \mathbf{F}\right)\nonumber\\
    &=- \frac{ \mathrm{M}_p^2}{4 \times 3!}\int d^4 x ~\beta(t)~g^{\mu \mu^\prime }g^{\nu \nu^\prime }g^{\rho \rho^\prime } \mathbf{H}_{3,\mu\nu\rho}\mathbf{H}_{3,\mu^\prime \nu^\prime\rho^\prime} - \frac{1}{3!} \int d^4 x ~(\partial_{\sigma} a) ~\mathbf{H}_{3,\mu\nu\rho}\epsilon^{\sigma \mu \nu \rho}\nonumber\\
    &~~~~~~~~-\frac{\alpha^\prime}{4}\int a \left( \mathrm{tr}\mathrm{R}^{(4)}\wedge \mathrm{R}^{(4)}-\frac{1}{30}\mathrm{Tr}\mathbf{F}\wedge \mathbf{F}\right)\label{4diMActIoNHetThEorY}
\end{align}
Setting the variation of the above action with respect to $\mathbf{H}_3$ to zero we get the equation of motion
\begin{align}
- \frac{\mathrm{M}_p^2 }{2}\beta(t)g^{\mu \mu^\prime }g^{\nu \nu^\prime }g^{\rho \rho^\prime} \mathbf{H}_{3,\mu^\prime \nu^\prime \rho^\prime}- (\partial_{\sigma}a)\epsilon^{\sigma \mu \nu \rho}&=0\nonumber\\
\implies -\frac{\mathrm{M}_p^2}{2}\beta(t)\mathbf{H}_{3}^{\mu\nu\rho}-\left(\partial_{\sigma} a\right)\epsilon^{\sigma\mu\nu\rho}&=0\nonumber\\
    \implies -\frac{\mathrm{M}_p^2}{2}\beta(t)\epsilon_{\lambda \mu\nu\rho}\mathbf{H}_{3}^{\mu\nu\rho}-\left(\partial_{\sigma}a \right)\epsilon_{\lambda \mu\nu\rho}\epsilon^{\sigma\mu\nu\rho}&=0\nonumber\\
    \implies  -\frac{\mathrm{M}_{p}^2}{2}\beta(t)\epsilon_{\lambda \mu\nu\rho}\mathbf{H}_{3}^{\mu\nu\rho}-\left(\partial_\sigma a\right) (-3! \delta^{\sigma}_{\lambda})&=0\nonumber\\
    \implies \frac{1}{3!}\epsilon_{\lambda \mu\nu\rho}\mathbf{H}_{3}^{\mu\nu\rho}=\frac{2}{\beta(t)\mathrm{M}_p^2}\partial_{\lambda}&a\nonumber\\
    \implies \star_{4}\mathbf{H}_{3}=\frac{2}{\beta(t)\mathrm{M}_p^2}da
\end{align}
This also implies that in four dimensions the axion is Hodge dual of the anti-symmetric Kalb-Ramond field. Substituting it back into the action (\ref{4diMActIoNHetThEorY}) we get 
\begin{align}
   {S}_{4}&=-\frac{\mathrm{M}_{p}^2}{4\times 3!} \int d^4 x  ~\beta(t)g^{\mu\mu^\prime} g^{\nu\nu^\prime}g^{\rho\rho^\prime}\times \left(- \frac{2}{\beta(t)\mathrm{M}_p^2}\right)^2\left(\partial^{\sigma}a\right)\left(\partial^{\lambda}a\right)\epsilon_{\sigma\mu\nu\rho}\epsilon_{\lambda \mu^\prime \nu^\prime \rho^\prime}\nonumber\\
    -&\frac{1}{3!} \int d^4 x ~(\partial_{\sigma}a)\left(-\frac{2}{\beta(t)\mathrm{M}_p^2} \right) \left(\partial^{\lambda}a\right)\epsilon_{\lambda \mu\nu\rho} \epsilon^{\sigma\mu\nu\rho}-\frac{\alpha^\prime}{4}\int a\left(\mathrm{tr}\mathrm{R}^{(4)}\wedge \mathrm{R}^{(4)}-\frac{1}{30}\mathrm{Tr}\mathbf{F}\wedge \mathbf{F}\right)\nonumber\\
    &=-\frac{1}{3! \mathrm{M}_p^2}\int d^4x  \frac{1}{\beta(t)}(\partial_{\sigma}a) (\partial^\lambda a)\epsilon_{\sigma\mu\nu\rho}\epsilon^{\lambda \mu\nu\rho}+\frac{2}{3!\mathrm{M}_p^2}\int d^4x \frac{1}{\beta(t)}(\partial_{\sigma}a)(\partial^\lambda a) \epsilon_{\lambda \mu\nu\rho}\epsilon^{\sigma\mu\nu\rho}\nonumber\\
    &~~~~~~-\frac{\alpha^\prime}{4}\int a\left(\mathrm{tr}\mathrm{R}^{(4)}\wedge \mathrm{R}^{(4)}-\frac{1}{30}\mathrm{Tr}\mathbf{F}\wedge \mathbf{F}\right)\nonumber\\
    &=\frac{1}{3!\mathrm{M}_p^2}\int d^4x \frac{1}{\beta(t)}(\partial_{\sigma}a) (\partial^{\lambda}a )\left(-3!\delta^{\lambda}_{\sigma} \right)-\frac{\alpha^\prime}{4}\int a\left(\mathrm{tr}\mathrm{R}^{(4)}\wedge \mathrm{R}^{(4)}-\frac{1}{30}\mathrm{Tr}\mathbf{F}\wedge \mathbf{F}\right)\nonumber\\
    &=\frac{2}{\mathrm{M}_p^2}\int \frac{d^4x}{\beta(t)}\left(-\frac{1}{2}\partial_{\mu}a \partial^\mu a \right)-\int {\alpha^\prime a \over 4}\left(\mathrm{tr}\mathrm{R}^{(4)}\wedge \mathrm{R}^{(4)}-\frac{1}{30}\mathrm{Tr}\mathbf{F}\wedge \mathbf{F}\right).\label{milenjon}
\end{align}
We can rescale $a$ as 
\begin{align}
    \tilde{a}&=\sqrt{ \frac{ 2}{ \mathrm{M}_p^2~ \beta(t)}}~a,
\end{align}
to rewrite the action in the following way 
\begin{align}
    S_{4}&=\int d^4 x \left( -\frac{1}{2}\partial_{\mu}\tilde{a}~\partial^{\mu}\tilde{a} \right)-\int \frac{\tilde{a} }{\sqrt{2/(\mathrm{M}_p^2 \beta(t) ) } }{\alpha^\prime  \over 4}\left(\mathrm{tr}\mathrm{R}^{(4)}\wedge \mathrm{R}^{(4)}-\frac{1}{30}\mathrm{Tr}\mathbf{F}\wedge \mathbf{F}\right)\nonumber\\
    &=\int d^4 x \left( -\frac{1}{2}\partial_{\mu}\tilde{a}~\partial^{\mu}\tilde{a} \right)-\int \frac{\tilde{a} }{\sqrt{2/(\mathrm{M}_p^2 \beta(t) ) } }{(\mathrm{M}_s^{-2}/(4\pi^2) )  \over 4}\left(\mathrm{tr}\mathrm{R}^{(4)}\wedge \mathrm{R}^{(4)}-\frac{1}{30}\mathrm{Tr}\mathbf{F}\wedge \mathbf{F}\right)\nonumber\\
    &=\int d^4 x \left( -\frac{1}{2}\partial_{\mu}\tilde{a}~\partial^{\mu}\tilde{a} \right)-\int \frac{\tilde{a} }{f_{a} }\frac{1}{16\pi^2} \left(\mathrm{tr}\mathrm{R}^{(4)}\wedge \mathrm{R}^{(4)}-\frac{1}{30}\mathrm{Tr}\mathbf{F}\wedge \mathbf{F}\right)
\end{align}
where $f_a$, referred to as the axion coupling constant, has the following form 
\begin{align}
    f_{a}&=\sqrt{ \frac{ 2}{\beta(t)} }\frac{ \mathrm{M}_s^2 }{ \mathrm{M}_p }\nonumber\\
    &=\frac{1}{ \sqrt{2\pi } } (\sqrt{\Lambda}|t|)^{\frac{ v\hat{\alpha}(t)+ 3 v \hat{\beta}(t)}{ 6v^2 -(2v-1)\hat{\alpha}(t)-\hat{\beta}(t)}} \left( \mathrm{M}_s^{4}\int d^6y \mathrm{H}^4(y)\sqrt{ \mathrm{det}g_{\mathcal{M}_4}(y)\mathrm{det}g_{\mathbb{T}^2}(y)} \right)^{-1/2}.\label{AxioNicCoupConst}
\end{align}
As $t\rightarrow 0^-$ in the late time limit, $f_{a}(t)\rightarrow 0$. This is as opposed to the range that $f_{a}(t)$ has to satisfy viz. $10^9 ~\mathrm{GeV}<f_{a}(t)<10^{12}~\mathrm{GeV}$. In order to set an upper bound of $f_{a}(t)$ we make $\hat{\gamma}_{o}$ in equation (\ref{FuncAnsTzhatAlphahatBeta}) time dependent such that 
\begin{align}
    \left(\sqrt{\Lambda} |\epsilon_1|\right)^{\frac{v \hat{\alpha}(-\epsilon_1)+3 v\hat{\beta}(-\epsilon_1)}{6v^2 -(2v-1)\hat{\alpha}(-\epsilon_1)-\hat{\beta}(-\epsilon_1)}}&=\left(\sqrt{\Lambda}|t|\right)^{\frac{v \hat{\gamma}_o(t) +3v(4v-3)\hat{\gamma}_o(t)}{6v^2-(2v-1)\hat{\gamma}_o(t)-(4v-3)\hat{\gamma}_o(t)}},~(\mathrm{for}~-\epsilon_1<t<0 ).
\end{align}
Taking logarithm on both sides we get 
\begin{align}
c_{o}:=\frac{(v\hat\alpha(-\epsilon_1) + 3v\hat\beta(-\epsilon_1))} {6v^2 - (2v-1)\hat\alpha(-\epsilon_1) -\hat\beta(-\epsilon_1)}  \big\vert {\rm log}(\sqrt{\Lambda} \vert\epsilon_1\vert)\big\vert &={\frac{ (v \hat{\gamma}_o(t) +3v(4v-3)\hat{\gamma}_o(t))}{6v^2-(2v-1)\hat{\gamma}_o(t)-(4v-3)\hat{\gamma}_o(t)}} \nonumber\\
&~~~\times \big\vert {\rm log}(\sqrt{\Lambda} \vert t\vert)\big\vert\nonumber\\
\implies c_{o}&=\frac{ (12 v^2 -8 v )\hat{\gamma}_o(t)}{6v^2 -(6v-4)\hat{\gamma}_{o}(t)}\big\vert {\rm log}(\sqrt{\Lambda} \vert t\vert)\big\vert \nonumber\\
\implies c_{o}&=\frac{ (3 v^2 - 2 v )\hat{\gamma}_o(t)}{3v^2 -(3v-2)\hat{\gamma}_{o}(t)}\big\vert {\rm log}({\Lambda}  t^2)\big\vert \nonumber\\
\implies\hat{\gamma}_o(t)&=\left( \frac{3v^2}{3v-2}\right)\left(\frac{c_o}{c_o+v\big\vert {\rm log}({\Lambda}  t^2)\big\vert} \right).\label{ExpRsco}
\end{align}
Incorporating the aforementioned modification we express (\ref{FuncAnsTzhatAlphahatBeta}) as
\begin{align}
    \hat{\alpha}(t)&=\begin{cases}   \hat{\alpha}_o,~~~~-\frac{1}{\sqrt{\Lambda }}<t<-\epsilon_1\\
\hat{\gamma}_{o}(t),~~~~~-\epsilon_1< t<0
\end{cases},~~~~~\hat{\beta}(t)=\begin{cases}
\hat{\beta}_o,~~~~~~~~~~~~~~-\frac{1}{\sqrt{\Lambda }}<t<-\epsilon_1\\
(4v-3)\hat{\gamma}_{o}(t),~~~~~-\epsilon_1< t<0 
\end{cases}.\label{AxionicCCTimeDepFactors}
\end{align}
Let us consider the following trial functions
\begin{align}\label{TIhostsmey}
&\hat\alpha(t) = \alpha_o\Theta(-t - \vert\epsilon_1\vert) 
{{\rm exp}\left[{ \frac{1}{ \sqrt{\Lambda}}}\left({1\over \sigma_1} - {a_1\over \vert t \vert}\right)\right] \over 
1 +{\rm exp}\left[{1\over \sqrt{\Lambda}}\left({1\over \sigma_1} - {a_1\over \vert t \vert}\right)\right]} + 
\Theta(t + \vert\epsilon_1\vert) {\alpha_v c_o \over  c_o + v \vert {\rm log}(\Lambda t^2)\vert}, \nonumber\\
& \hat\beta(t) = \beta_o\Theta(-t - \vert\epsilon_1\vert) 
{{\rm exp}\left[{1\over \sqrt{\Lambda}}\left({1\over \sigma_2} - {a_2\over \vert t \vert}\right)\right] \over 
1 +{\rm exp}\left[{1\over \sqrt{\Lambda}}\left({1\over \sigma_2} - {a_2\over \vert t \vert}\right)\right]} + 
\Theta(t + \vert\epsilon_1\vert) {\alpha_v (4v-3) c_o \over  c_o + v \vert {\rm log}(\Lambda t^2)\vert},\end{align}
where 
\begin{align}
\hat\alpha_o =  {\alpha_o~{\rm exp}\left[{1\over \sqrt{\Lambda}}\left({1\over \sigma_1} - {a_1\sqrt{\Lambda}}\right)\right] \over 
1 +{\rm exp}\left[{1\over \sqrt{\Lambda}}\left({1\over \sigma_1} - {a_1\sqrt{\Lambda}}\right)\right]}&, ~~
\hat\beta_o =  {\beta_o~{\rm exp}\left[{1\over \sqrt{\Lambda}}\left({1\over \sigma_2} - {a_2\sqrt{\Lambda}}\right)\right] \over 
1 +{\rm exp}\left[{1\over \sqrt{\Lambda}}\left({1\over \sigma_2} - {a_2\sqrt{\Lambda}}\right)\right]},\\
\alpha_{v}=\frac{3v^2}{3v-2}.&
\end{align}  
Imposing the continuity of $\hat{\alpha}(t)$ at $t=-\epsilon_1$ we find 
\begin{align}
{{\alpha}_{o} ~{\rm exp}\left[{1\over \sqrt{\Lambda}}\left({1\over \sigma_1} - {a_1\over \vert \epsilon_{1} \vert}\right)\right] \over 
1 +{\rm exp}\left[{1\over \sqrt{\Lambda}}\left({1\over \sigma_1} - {a_1\over \vert \epsilon_{1}\vert}\right)\right]}&=\frac{\alpha_{v}c_{o}}{c_{o}+v|\log (\Lambda \epsilon_1^2)|}\nonumber\\
\implies \exp \left[-\frac{1}{\sqrt{\Lambda}}\left( \frac{1}{\sigma_{1}}-\frac{a_{1}}{|\epsilon_1|}\right) \right]+1&=\frac{{\alpha}_{o}}{\alpha_{v}}\left(1+\frac{v}{c_{o}}|\log (\Lambda \epsilon_{1}^2)|\right)\label{ConTForhatAlpha}\\
\implies \frac{1}{\sigma_1}=\frac{a_{1}}{|\epsilon_1|}-\sqrt{\Lambda}&~\log\left[\frac{{\alpha}_{o}}{\alpha_{v}}\left(1+\frac{v}{c_{o}}|\log (\Lambda \epsilon_{1}^2)|\right)-1\right]\label{ExpOfSigma1}
\end{align}
We also want to impose the continuity of the derivative of $\hat{\alpha}(t)$ near $t=-\epsilon_{1}$:
\begin{align}
\frac{d}{dt} \left[{\alpha_o {\rm exp}\left[{1\over \sqrt{\Lambda}}\left({1\over \sigma_1} - {a_1\over \vert t \vert}\right)\right] \over 
1 +{\rm exp}\left[{1\over \sqrt{\Lambda}}\left({1\over \sigma_1} - {a_1\over \vert t \vert}\right)\right]}\right]&
=-{ \alpha_o{\rm exp}\left[{2\over \sqrt{\Lambda}}\left({1\over \sigma_1} - {a_1\over \vert t \vert}\right)\right] \over 
\left(1 +{\rm exp}\left[{1\over \sqrt{\Lambda}}\left({1\over \sigma_1} - {a_1\over \vert t \vert}\right)\right]\right)^2}\times \left(\frac{1}{\sqrt{\Lambda}}\frac{a_1}{|t|^2}\times(-1)\right) \nonumber\\
&+{\alpha_o {\rm exp}\left[{1\over \sqrt{\Lambda}}\left({1\over \sigma_1} - {a_1\over \vert t \vert}\right)\right] \over 
\left(1 +{\rm exp}\left[{1\over \sqrt{\Lambda}}\left({1\over \sigma_1} - {a_1\over \vert t \vert}\right)\right]\right)}\times \left(\frac{1}{\sqrt{\Lambda}}\frac{a_1}{|t|^2}\times(-1)\right)\nonumber\\
&=-\frac{\alpha_o a_{1}}{\sqrt{\Lambda}|t|^2}\left( \frac{\exp\left[\frac{1}{\sqrt{\Lambda}} \left(\frac{1}{\sigma_{1}}-\frac{a_{1}}{|t|} \right)\right]}{\left( 1+\exp\left[\frac{1}{\sqrt{\Lambda}} \left(\frac{1}{\sigma_{1}}-\frac{a_{1}}{|t|} \right)\right]\right)^2}\right)
\end{align}
\begin{align}
\frac{d}{dt}\left[\frac{\alpha_v c_o}{c_{o}+v|\log (\Lambda t^2)|}\right]&=-\frac{\alpha_v c_o}{\left(c_{o}+v|\log (\Lambda t^2)|\right)^2}\left( -v\frac{1}{\Lambda t^2}(2\Lambda t)\right)\nonumber\\
&=\frac{2\alpha_{v}c_{o} v}{t\left(c_{o}+v|\log (\Lambda t^2)|\right)^2}=-\frac{2\alpha_{v}c_{o} v}{|t|\left(c_{o}+v|\log (\Lambda t^2)|\right)^2}
\end{align}
Therefore, at $t=-\epsilon_{1}$ we have 
\begin{align}
\frac{{\alpha}_{o} a_{1}}{\sqrt{\Lambda}|\epsilon_1|^2}\left( \frac{\exp\left[\frac{1}{\sqrt{\Lambda}} \left(\frac{1}{\sigma_{1}}-\frac{a_{1}}{|\epsilon_1|} \right)\right]}{\left( 1+\exp\left[\frac{1}{\sqrt{\Lambda}} \left(\frac{1}{\sigma_{1}}-\frac{a_{1}}{|\epsilon_1|} \right)\right]\right)^2}\right)&= \frac{2\alpha_{v}c_{o} v}{|\epsilon_1|\left(c_{o}+v|\log (\Lambda \epsilon_1^2)|\right)^2}. \label{t=-epsilon1Exprsn}
\end{align}
From the continuity relation (\ref{ConTForhatAlpha}) we substitute 
\begin{align}
   \exp\left[\frac{1}{\sqrt{\Lambda}} \left(\frac{1}{\sigma_{1}}-\frac{a_{1}}{|\epsilon_1|} \right)\right]&=\frac{1} {\frac{{\alpha}_{o}}{\alpha_{v}}\left(1+\frac{v}{c_{o}}|\log (\Lambda \epsilon_{1}^2)|\right)-1},
\end{align}
in equation (\ref{t=-epsilon1Exprsn}) to get 
\begin{align}
    \frac{
   {\alpha}_{o} a_{1}}{\sqrt{\Lambda }|\epsilon_{1}|^2}\left(\frac{\frac{1} {\frac{{\alpha}_{o}}{\alpha_{v}}\left(1+\frac{v}{c_{o}}|\log (\Lambda \epsilon_{1}^2)|\right)-1}}{\left(1+\frac{1} {\frac{{\alpha}_{o}}{\alpha_{v}}\left(1+\frac{v}{c_{o}}|\log (\Lambda \epsilon_{1}^2)|\right)-1} \right)^2}\right)&=\frac{2\alpha_{v}c_{o} v}{|\epsilon_1|\left(c_{o}+v|\log (\Lambda \epsilon_1^2)|\right)^2}\nonumber\\
    \implies 
    \frac{
   {\alpha}_{o} a_{1}}{\sqrt{\Lambda }|\epsilon_{1}|} \frac{ \left( \frac{{\alpha}_{o}}{\alpha_{v}} \left(1+\frac{v}{c_o} |\log(\Lambda \epsilon_{1}^2)| \right)-1 \right)}{\left( \frac{{\alpha}_{o}}{\alpha_{v}} \left(1+\frac{v}{c_o} |\log(\Lambda \epsilon_{1}^2)| \right) \right)^2}&= \frac{2\alpha_{v} v}{c_{o}\left(1+\frac{v}{c_{o}}|\log (\Lambda \epsilon_1^2)|\right)^2}\nonumber\\
    \implies \frac{
   {\alpha}_{o} a_{1}}{\sqrt{\Lambda}|\epsilon_1|} \frac{\alpha_v^2}{{\alpha}_{o}^2} \left(\frac{{\alpha}_{o}\left(c_o+v|\log(\Lambda \epsilon_{1}^2)| \right)-\alpha_{v}c_{o}}{\alpha_{v}c_{o}}\right)&=\frac{2\alpha_{v} v}{c_{o}}\nonumber\\
    \implies a_{1}=\frac{2 {\alpha}_{o} v |\epsilon_{1}| \sqrt{\Lambda} }{{\alpha}_{o} (c_{o}+v|\log (\Lambda \epsilon_{1}^2) |)-\alpha_{v}c_{o}}&.
\end{align}
Likewise, imposing the continuity of $\hat{\beta}(t)$ at $t=-\epsilon_{1}$ gives us the expression of the parameter $\sigma_2:$
\begin{align}
    \beta_o \frac{ \exp \left[\frac{1}{\sqrt{\Lambda }} \left(\frac{1}{\sigma_2}-\frac{a_2}{|t|} \right)  \right] }{1+\exp \left[ \frac{1}{\sqrt{\Lambda }}\left(\frac{1}{\sigma_2}-\frac{a_2}{|t|}\right) \right] }\bigg|_{t=-\epsilon_1}&= \frac{\alpha_v (4v-3) c_o}{c_o+ v|\log ( \Lambda t^2)|}\bigg|_{t=-\epsilon_1}\nonumber\\
    \implies \exp \left[ -\frac{1}{\sqrt{\Lambda }} \left( \frac{1}{\sigma_2}-\frac{a_2}{|\epsilon_1|} \right)\right]+1&=\frac{\beta_o}{\alpha_v (4v-3)}\left(1+\frac{v}{c_o}|\log (\Lambda \epsilon_1^2 )|\right) \nonumber\\
    \implies \frac{1}{\sigma_2}=\frac{a_2}{|\epsilon_1|} -\sqrt{\Lambda}\log &\left[ \frac{\beta_o }{\alpha_v (4v-3)} \left(1+ \frac{v}{ c_o} |\log (\Lambda \epsilon_1^2)|\right)-1\right].\label{ExpOfSigma2}
\end{align}
Continuity of the derivative of $\hat{\beta}(t)$ at $t=-\epsilon_1$ gives
\begin{align}
   \frac{d}{dt} \left[{\beta_o {\rm exp}\left[{1\over \sqrt{\Lambda}}\left({1\over \sigma_2} - {a_2\over \vert t \vert}\right)\right] \over 
1 +{\rm exp}\left[{1\over \sqrt{\Lambda}}\left({1\over \sigma_2} - {a_2\over \vert t \vert}\right)\right]}\right]\bigg|_{t=-\epsilon_1}&=\frac{ d}{ dt}\left[\frac{ \alpha_v (4v-3)c_o }{ c_o+ v |\log (\Lambda t^2 )| }\right]\bigg|_{t=-\epsilon_1}\nonumber\\
\implies    \frac{{\beta}_{o}a_{2}}{\sqrt{\Lambda}|\epsilon_1|^2}\left( \frac{\exp\left[\frac{1}{\sqrt{\Lambda}} \left(\frac{1}{\sigma_{2}}-\frac{a_{2}}{|\epsilon_1|} \right)\right]}{\left( 1+\exp\left[\frac{1}{\sqrt{\Lambda}} \left(\frac{1}{\sigma_{2}}-\frac{a_{2}}{|\epsilon_1|} \right)\right]\right)^2}\right)&= \frac{2\alpha_{v}(4v-3)c_{o} v}{|\epsilon_1|\left(c_{o}+v|\log (\Lambda \epsilon_1^2)|\right)^2}\nonumber\\
\implies \frac{\beta_o a_{2}}{\sqrt{\Lambda } |\epsilon_1|} \left(\frac{ \frac{1 }{\frac{\beta_o}{\alpha_v (4v-3)}\left(1+\frac{v}{c_o}|\log (\Lambda \epsilon_1^2 )|\right)-1 }   }{ \left(1+ \frac{1 }{\frac{\beta_o}{\alpha_v (4v-3)}\left(1+\frac{v}{c_o}|\log (\Lambda \epsilon_1^2 )|\right)-1 } \right)^2 }  \right)  &=\frac{2\alpha_{v}(4v-3)c_{o} v}{\left(c_{o}+v|\log (\Lambda \epsilon_1^2)|\right)^2}\nonumber\\
\implies \frac{\beta_o a_{2}}{\sqrt{\Lambda } |\epsilon_1|} \frac{\left( {\frac{\beta_o}{\alpha_v (4v-3)}\left(1+\frac{v}{c_o}|\log (\Lambda \epsilon_1^2 )|\right)-1 }  \right) }{ \left(\frac{\beta_o}{\alpha_v (4v-3)}\left(1+\frac{v}{c_o}|\log (\Lambda \epsilon_1^2 )|\right) \right)^2 }   &=\frac{2\alpha_{v}(4v-3) v}{c_{o}\left(1+\frac{v}{c_o} |\log (\Lambda \epsilon_1^2)|\right)^2}\nonumber\\
\implies a_2= \frac{ 2\beta_o v |\epsilon_1|\sqrt{\Lambda} }{\beta_o \left(c_o+v |\log (\Lambda |\epsilon_1|^2 ) | \right)- \alpha_v c_o (4v-3)}&.
\end{align}
The expressions of the parameters $\sigma_1$  and $\sigma_2$ as in equations  (\ref{ExpOfSigma1}) and (\ref{ExpOfSigma2}) constrains $\epsilon_1$, since we must have 
\begin{align}
    \frac{\alpha_o}{\alpha_v }\left( 1+ \frac{v}{c_o}|\log (\Lambda \epsilon_1^2 )| \right)>1 &\implies |\log (\Lambda \epsilon_1^2)|> \frac{c_o}{v}\left( \frac{ \alpha_v }{\alpha_o }-1 \right),\label{ConstSigma1}\\
    \mathrm{or}, ~~\frac{\beta_o}{\alpha_v (4v-3) } \left(1+\frac{v}{c_o} |\log (\Lambda \epsilon_1^2) | \right)>1&\implies |\log (\Lambda \epsilon_1^2)|>\frac{c_o}{v}\left( \frac{\alpha_v (4v-3) }{\beta_o}-1 \right).\label{ConstSigma2}
\end{align}
In string frame, the parameters $\alpha_o,~\beta_o,~\sigma_1,~\sigma_2,~a_1,~a_2$ should be such that the condition $1>\hat{\alpha}_{o}>\hat{\beta}_{o}>\frac{\hat{\alpha}_o}{9}>0$ (as described in Subsection \ref{E8XE8ThrdistWarpf})
holds true. If we consider $\alpha_o>\beta_o$, so that $\hat{\alpha}_o>\hat{\beta}_o$ then equation (\ref{ConstSigma2}) imposes the stringent condition on $\epsilon_1$.      

\section{Conclusion and Outlook}
\label{sec:Summary&Con}
In this article, we have investigated the existence of four-dimensional de Sitter space as an excited state over a supersymmetric Minkowski background. In Section~\ref{dSspacePoincareCo}, we reviewed the basic properties of de Sitter space, describing it from the embedding-space point of view and parametrizing it using Poincar\'e coordinates or flat-slicing coordinates. This led to the four-dimensional de Sitter metric given in equation~(\ref{dSFlAtSlic}).

In Section~\ref{dSinVarSTheOrieS}, we elaborated on three distinct duality sequences, illustrated in Figures~\ref{Type IIB tree diagram}, \ref{Het SO(32) tree diagram}, and \ref{Het E8XE8 tree diagram}, which realize de Sitter isometry in type IIB, heterotic $SO(32)$, and heterotic $\mathrm{E}_8 \times \mathrm{E}_8$ string theories, respectively. Starting from a generic M-theory configuration, each duality sequence proceeds dynamically and yields de Sitter isometry in the corresponding dual string theory in the late-time limit of Poincar\'e coordinates. We provided explicit expressions for the metric configurations, coupling constants, and fluxes at each intermediate step of these duality chains.

An important distinction arises in the realization of de Sitter space in heterotic $\mathrm{E}_8 \times \mathrm{E}_8$ string theory, where the generic M-theory metric configuration~(\ref{MthThreewarpFac}) involves three time-dependent warp factors $\mathrm{F}_1(t)$, $\mathrm{F}_2(t)$, and $\mathrm{F}_3(t)$. This is in contrast to the heterotic $SO(32)$~(\ref{MthHetSO32}) and type IIB~(\ref{MthEorYmeTric}) cases, each of which contains only two time-dependent warp factors, $\mathrm{F}_1(t)$ and $\mathrm{F}_2(t)$.

In Section~\ref{sec:PIinMth}, we reviewed the computation of the expectation value of the metric operator in M-theory using path-integral techniques, following the analysis of~\cite{Brahma:2022wdl}. The computation was carried out using scalar fields with an all-order derivative interaction term~(\ref{SinTeraCtiOn}). Employing functional integral methods, we evaluated the tree-level contribution, the first-order quantum correction, and the $\mathrm{N}$th-order correction, expressing the result as an asymptotic series~(\ref{ExpValMOp}). Upon performing the Borel transform and subsequent Borel resummation, we obtained the finite result quoted in equation~(\ref{BORELRSummed}).

We further demonstrated that the four-dimensional de Sitter solutions constructed in type IIB, heterotic $SO(32)$, and heterotic $\mathrm{E}_8 \times \mathrm{E}_8$ string theories satisfy the criteria of a well-defined effective field theory. In Section~\ref{sec:NEC}, we showed that satisfying the effective field theory constraint~(\ref{EFTconditION}) in these theories is equivalent to satisfying the Null Energy Condition~(\ref{NECforFLRWcos}) for a $(3+1)$-dimensional FLRW spacetime.

Section~\ref{sec:TCCconsTraints} contains discussion of the temporal regime over which the effective field theory description remains valid, commonly referred to as the Trans-Planckian Censorship Conjecture (TCC) bound~(\ref{TCCBounD}). By considering explicit functional forms for the dominant time dependence of the warp factor $\beta(t)$ in heterotic $SO(32)$ theory, we illustrated how different ans\"atze affect the duration of validity of the effective field theory. We found that a smoother functional ansatz, as given in equation~(\ref{2ndbetatAnsatZ}), preserves the same domain of validity as dictated by the TCC bound.

Finally, in Section~\ref{sec:AxionsinE8Thoery}, we analyzed constraints arising from axionic cosmology. We have shown that the bound on the axionic coupling constant $f_a(t)$ in equation~(\ref{AxioNicCoupConst}) modifies the dominant scaling behavior of the time-dependent warp factors $\hat{\alpha}(t)$ and $\hat{\beta}(t)$ from those given in equation~(\ref{FuncAnsTzhatAlphahatBeta}) to those in equation~(\ref{AxionicCCTimeDepFactors}) within heterotic $\mathrm{E}_8 \times \mathrm{E}_8$ string theory.
 To approximate the behavior of these warp factors, we introduced a trial function~(\ref{TIhostsmey}) and fixed the associated parameters by imposing continuity of the functions and their first time derivatives at $t=-\epsilon_1$. This procedure further led to the constraint~(\ref{ConstSigma2}) on $\epsilon_1$ in terms of the parameters of the ans\"atze.

The isometry group of the underlying time-independent supersymmetric Minkowski background is the super-Poincare group acting along the external spacetime directions. It would be interesting to investigate the isometry group associated with the de Sitter excited state constructed over this supersymmetric Minkowski background. Since our state breaks supersymmetry spontaneously, is transient in nature and asymptotically approaches Minkowski spacetime, its isometry group cannot be the exact $SO(1,4)$ group of a de Sitter space. Moreover, in \cite{Brahma:2024zuk, Dasgupta:2024aif}, attempts have been made to relate the Glauber--Sudarshan state to the wavefunctional of the Universe satisfying the Wheeler--DeWitt (WDW) equation. This turned out to be an {\it emergent} equation, although an explicit solution to this equation was not obtained therein. It would therefore be interesting to explore how the technique used in solving the WDW equation in \cite{chakraborty2023holographyinformationsitterspace},
that uses information theoretic approaches to de Sitter space, may be applied to our case for solving the corresponding {\it emergent} WDW equation.

\section{A brief overview of coherent states}
\label{GS state}
The importance of coherent states for the study of photon statistics of radiation fields is widely known \cite{PhysRev.131.2766}. The property that makes these quantum states of considerable interest is their well defined classical limit. In fact, they  are typically known as the \textit{minimum uncertainty states} -  the expectation value of any observable computed with respect to such states follow the classical trajectory with minimum uncertainty.  Even though the coherent states are not orthogonal to each other, they form a complete set, to be precise, an overcomplete set. We characterise them as eigenstate $|z \rangle$  of annihilation operator $\hat{a}$, in the context of harmonic oscillator problem of quantum mechanics:
\begin{align}
    \hat{a}|z \rangle &=z |z\rangle, 
\end{align}
where $z$ is the corresponding eigenvalue. As $\hat{a}$ is a not a hermitian operator, $z=|z| e^{i \varphi }\in \mathbb{C}$ is generically a complex number. Let us express the state $|z\rangle$ as a linear combination of the basis states $\{|n\rangle, ~n=0,1,...\}$, viz. the eigenstates of the number operator of harmonic oscillator
\begin{align}
    |z \rangle &=\sum_{n=0}^{\infty}c_{n}|n\rangle=c_{0}|0\rangle +c_{1}|1\rangle +c_{2}|2\rangle +c_3 |3\rangle+\cdots\nonumber\\
  \implies  \hat{a}|z \rangle &=c_{1}|0\rangle +\sqrt{2}c_{2}|1\rangle +\sqrt{3}c_{3}|2\rangle+\cdots=z(c_{0}|0\rangle +c_{1}|1\rangle +c_{2}|2\rangle +\cdots)
\end{align}
where we have used $\hat{a}|n\rangle=\sqrt{n}|n-1\rangle$. Comparing both sides we see that the coefficients satisfy the recurrence relation 
\begin{align}
c_{n+1}=\frac{z}{\sqrt{n+1} } c_n.
\end{align}
Hence the coherent state is as follows
\begin{align}
     |z \rangle &=c_{0}\sum_{n=0}^{\infty}\frac{z^n}{\sqrt{n!}}|n\rangle
\end{align}
We can fix the overall constant $c_{0}$ by demanding the state $|z\rangle$ to be normalised
\begin{align}
    \langle z | z \rangle &=|c_{0}|^2 \sum_{m,n}\frac{(z^\star)^m z^n}{\sqrt{ m! n!}}\langle m|n\rangle\nonumber\\
    \implies |c_{0}|&=\exp\left( -\frac{|z|^2}{2}\right)
\end{align}
Upon expressing the $n$-th excited state in terms of the vacuum state, the coherent state
\begin{align}
    |z \rangle &=\exp \left(-\frac{|z|^2}{2} \right) \sum_{n=0}^{\infty} \frac{z^n}{\sqrt{n!}}\frac{(\hat{a}^{\dagger})^n}{\sqrt{n!}}|0\rangle\equiv\hat{\mathrm{D}}(z)|0\rangle ,\label{CoHeREntStatE}
\end{align}
 can be interpreted as a state obtained by displacing the vacuum configuration $|0\rangle$ where the displacement operator 
\begin{align}
    \hat{\mathrm{D}}(z)&=\exp\left( -\frac{|z|^2}{2}\right)\exp\left( z \hat{a}^{\dagger}\right) \exp \left( -z^{\star}\hat{a}\right)=\exp\left( z \hat{a}^{\dagger}-z^{\star}\hat{a}\right),\label{DispOp}
\end{align}
is a unitary operator satisfying
\begin{align}
    \hat{\mathrm{D}}(z)\hat{\mathrm{D}}^{\dagger}(z)=\hat{\mathrm{D}}^{\dagger}(z)\hat{\mathrm{D}}(z)=1.
\end{align}
In $3+1$ dimensional quantum field theory where we have one harmonic oscillator for each momentum mode, we can  express the coherent state (\ref{CoHeREntStatE}) in the following way \cite{zhang1999coherent} 
\begin{align}
    |\{z_{k}\}\rangle &=\exp\left(\frac{1}{\mathrm{V}}\sum_{n} \left(z_{k_n} \hat{a}^{\dagger}_{k_n}-z^{\star}_{k_n}\hat{a}_{k_n}\right)\right)|0\rangle
\end{align}
where we have considered the four dimensional space-time to be a square lattice of volume $\mathrm{V}=L^{4}$ such that $k_{n}^{\mu}=\frac{2\pi n^{\mu}}{L}$, $n^\mu$ are integers.    
As mentioned in the introduction, the state that we construct differs from the usual coherent states. We spell out the differences in the following. 
\begin{itemize}
    \item[(a)]  Rather than a free vacuum $|0\rangle$, the state of our consideration is obtained by displacing an \textit{interacting vacuum} $|\Omega\rangle$. Suppose we consider a double well potential of the form $V(x)=(x^2-a^2)^2$. 
\begin{figure}[h!]
    \centering
\includegraphics[width=0.80\textwidth]{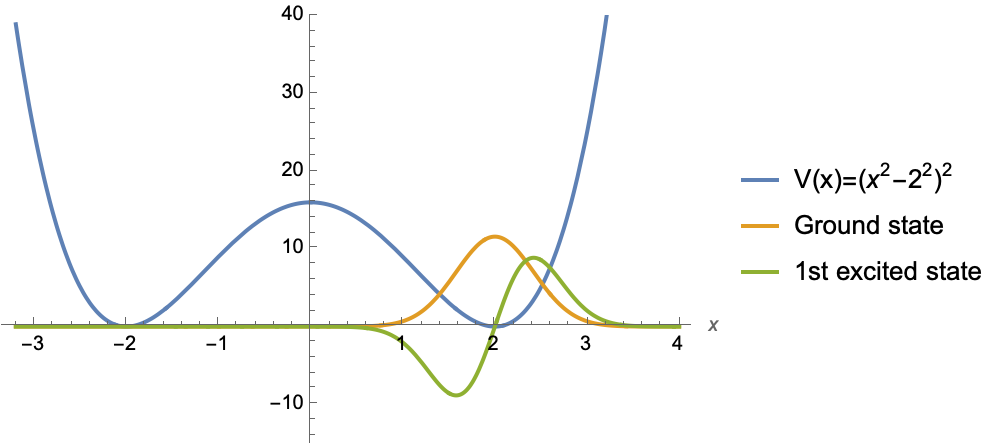}
    \caption{Double well potential $V(x)=(x^2-2^2)^2$ with ground state and 1st excited state around the minima at $x=+2$.}
    \label{DWellPot}
\end{figure}
It contains two minima at $x=\pm a$. The ground state corresponding to any of these two minima we refer to as the free vacuum. In figure \ref{DWellPot}, the free vacuum corresponding to the minima at $x=+a$ is shown by the orange curve. 
\begin{figure}[h!]
    \centering
\includegraphics[width=0.80\textwidth]{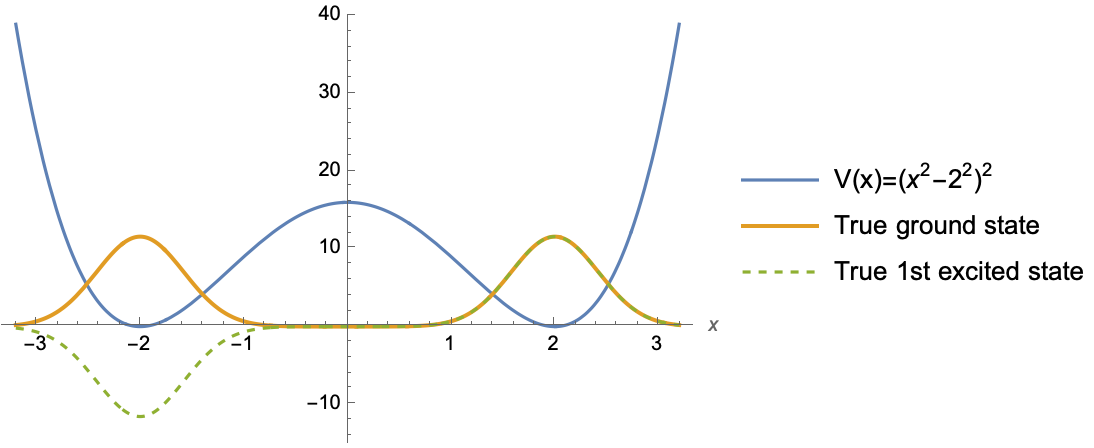}
    \caption{Double well potential $V(x)=(x^2-2^2)^2$ with the true ground state and 1st excited state.}
    \label{DWPotTGS}
\end{figure}
The true ground state of the double well potential is depicted by the orange curve in figure \ref{DWPotTGS}. We refer to this particular state as the interacting vacuum.  The interacting vacuum can be expressed using the total Hamiltonian $H=H_{0}+H_{\mathrm{int}}$ of a system as 
\begin{align}
    |\Omega \rangle &=\lim_{{T}\rightarrow \infty (1-i\epsilon)}\left(e^{-iE_{0}T}\langle \Omega |0 \rangle \right)^{-1} e^{-iHT}|0\rangle 
\end{align}
where $E_0= \langle \Omega | H | \Omega \rangle$ is the true ground state energy.

\item[(b)] For such an interacting vacuum configuration the form of the corresponding displacement operator must differ from (\ref{DispOp}). We introduce effective creation and annihilation operators $\hat{a}^\dagger_{\mathrm{eff},k}$ and $\hat{a}_{\mathrm{eff},k}$ respectively. The interacting vacuum is annihilated by $\hat{a}_{\mathrm{eff},k}$ \cite{Brahma:2020tak} 
\begin{align}
\hat{a}_{\mathrm{eff},k}|\Omega\rangle&=0
\end{align} 
The best one can do is express the displacement operator corresponding to the interacting vacuum of a quantum field theory in terms of effective creation and annihilation operators as 
\begin{align}
    \hat{\mathbb{D}}(\sigma(k))&=\exp\left(\frac{1}{\mathrm{V}}\sum_{i} \left(\sigma(k_i) \hat{a}_{\mathrm{eff},k_i}^{\dagger} +\sigma^{\star}(k_i)\hat{a}_{\mathrm{eff},k_i}\right)\right)\label{DisPlOpIntVAcuuM}
\end{align} 
where $\hat{\mathbb{D}}(\sigma(k))$ is a non-unitary operator. In terms of the fourier modes of the fields, the operator (\ref{DisPlOpIntVAcuuM}) can be expressed as
\begin{align}
    \hat{\mathbb{D}}(\sigma(k))&=\exp\left(\frac{1}{\mathrm{V}}\sum_{i} \left(\sigma(k_{i}) \tilde{\varphi}^\star (k_i)+\sigma^{\star}(k_{i})\tilde{\varphi}(k_{i})  \right) \right)\nonumber\\
    &=\exp\left(\frac{2}{\mathrm{V}}\sum_{i}\left(\mathbf{Re}~\sigma(k_i)~\mathbf{Re}~\tilde{\varphi}(k_i)+\mathbf{Im}~\sigma(k_i)~\mathbf{Im}~\tilde{\varphi}(k_i)~\right) \right).
\end{align}
At this point one might be worried that because of the non unitarity of the displacement operator, the state $|\sigma \rangle$ is unnormalised. However, note that 
\begin{align}
    \langle \sigma |\sigma \rangle&=\frac{\langle \Omega | \hat{ \mathbb{D}}^{\dagger} (\sigma)\hat{\mathbb{D}}(\sigma)|\Omega \rangle}{\langle \Omega |\Omega \rangle} \langle \Omega |\Omega \rangle \nonumber\\
    &=\langle \hat{\mathbb{D}}^{\dagger}(\sigma)\hat{\mathbb{D}}(\sigma)\rangle_{\Omega}  ~\langle \Omega |\Omega \rangle\nonumber\\
    &=\mathcal{N} \langle \Omega |\Omega \rangle .
\end{align}
Thus we can in principle define the normalised state as 
\begin{align}
    |\sigma^{\prime}\rangle &= \frac{1}{\sqrt{\mathcal{N}}}|\sigma \rangle .
\end{align}
But we will stick to the former definition.
\end{itemize}
To make distinction from coherent states, we refer to these states as \textit{Glauber-Sudarshan states}. Since the set-up of our consideration does not contain any free vacuum we are bound to consider an interacting vacuum and correspondingly a Glauber Sudarshan state. 

\section{Functional integral approach for \texorpdfstring{$\varphi^4$}{phi4}  theory}
\label{FIntApproachPhi4}
\vskip.1in
In this appendix, we elaborate the two point correlator of $\varphi^{4}$ theory in 3+1 dimensional quantum field theory for a Klein Gordon field using the method of functional integral. We follow closely the method given in Chap. 9 of Peskin \& Schroeder \cite{Peskin:1995ev}:
\begin{align}
    \langle \varphi(x_1)\varphi(x_2)\rangle &= \frac{\int [\mathcal{D}\varphi(x)]e^{i\mathbf{S}_{\mathrm{tot}}}\varphi(x_1)\varphi(x_{2})}{\int [\mathcal{D}\varphi(x)]e^{i\mathbf{S}_{\mathrm{tot}}}}\label{2ptCorrelator}
\end{align}
where 
\begin{align}
\mathbf{S}_{\mathrm{tot}}&=\mathbf{S}_{0}+\mathbf{S}_{\mathrm{int}}=\int d^{4}x \left(\frac{1}{2}\partial_{\mu}\varphi(x) \partial^{\mu} \varphi (x)-\frac{1}{2}m^2\varphi^2(x)\right)+\frac{\lambda}{4!} \int d^{4}x \varphi^{4}(x).\label{TotalAction}
\end{align}
The first step is to map the space-time into a square lattice having discrete points and thus the field $\varphi(x)$ defined on a continuum of points has to be mapped to $\varphi(x_{i})$ defined at the points of the aforementioned lattice. Consequently, the integral $\mathcal{D}\varphi$ over the space of all field configurations can be replaced in terms of a finite but large number of integrals over the lattice points 
\begin{align}
    \int [\mathcal{D}\varphi(x)]\rightarrow \prod_{i}\int d\varphi(x_{i})\label{Dvarphi}
\end{align}
The discrete Fourier transform of the field $\varphi(x_i)$ is given by
\begin{align}
    \varphi(x_{i})&=\frac{1}{\mathrm{V}}\sum_{n} e^{-ik_{n}\cdot x_i}\widetilde{\varphi}(k_{n}),
\end{align}
where the four dimensional space-time volume $\mathrm{V}=L^{4}$. The periodic boundary condition $\varphi(x_{i})=\varphi(x_{i}+L)$ imposed on the fields implies $k_{n}^{\mu}=2\pi n^{\mu}/L$, $n^\mu$ are the integers.  Since we are considering a real Klein Gordon field 
\begin{align}
    \varphi^{\star}(x_{i})&=\frac{1}{\mathrm{V}}\sum_{n}e^{ik_{n}\cdot x_{i}}\widetilde{\varphi}^{\star}(k_{n})=\varphi(x_{i}),
\end{align}
the Fourier coefficients has to obey the constraint $\widetilde{\varphi}^{\star}(k_{n})=\widetilde{\varphi}(-k_{n})$. Therefore, we can consider $\mathbf{Re}\widetilde{\varphi}(k_n)$ and $\mathbf{Im}\widetilde{\varphi}(k_{n})$ with $k_{n}^{0}>0$ as independent variables. In terms of these fourier coefficients we can rewrite the integral in (\ref{Dvarphi}) as 
\begin{align}
    \int[\mathcal{D}\varphi(x)]\rightarrow  \prod_{k^{0}
_{n}>0}\int d~ \mathbf{Re}\widetilde{\varphi}(k_{n}) d ~\mathbf{Im}\widetilde{\varphi}(k_{n})
\end{align}
Likewise, the free part of the action (\ref{TotalAction}) can be expressed as
\begin{align}
  \mathbf{S}_{0}&=  \int d^{4}x \frac{1}{2}\left(\partial_{\mu}\varphi(x)\partial^{\mu}\varphi(x)-m^2 \varphi^2(x)\right)\nonumber\\
  &= -\frac{1}{\mathrm{V}^2}\sum_{m,n} \int d^{4}x \frac{1}{2}~e^{-i(k_m+k_n)\cdot x} (k_{m\mu}k_{n}^{\mu} +m^2)\widetilde{\varphi}(k_{m})\widetilde{\varphi}(k_{n})\nonumber\\
    &=-\frac{1}{\mathrm{V}}\sum_{n}\frac{1}{2}(m^2-k_{n}^2) |\widetilde{\varphi}(k_{n})|^2=-\frac{1}{\mathrm{V}}\sum_{k_{n}^{0}>0}(m^2-k_{n}^{2})|\widetilde{\varphi}(k_{n})|^2,
\end{align}
and the interaction part can be expressed as 
\begin{align}
    \mathbf{S}_{\mathrm{int}}=\frac{\lambda}{4!}\int d^{4}x \varphi^{4}(x)&=\frac{\lambda}{4!}\frac{1}{\mathrm{V}^4}\sum_{m,n,p,q}\int d^4 x e^{-i(k_{m}+k_{n}+k_{p}+k_{q})\cdot x} \widetilde{\varphi}(k_{m})\widetilde{\varphi}(k_{n})\widetilde{\varphi}(k_{p})\widetilde{\varphi}(k_{q})\nonumber\\
    &=\frac{\lambda}{4!}\frac{1}{\mathrm{V}^3}\sum_{m,n,p}\widetilde{\varphi}(k_{m})\widetilde{\varphi}(k_{n})\widetilde{\varphi}(k_{p})\widetilde{\varphi}^{\star}(k_{m}+k_{n}+k_{p}).
\end{align}
With these ingredients we now analyse the numerator of (\ref{2ptCorrelator}) 
\begin{align}
&\mathrm{Num}[\langle \varphi(x_1)\varphi(x_2)\rangle]=\int [\mathcal{D}\varphi(x)] e^{i\mathbf{S}_{\mathrm{tot}}}\varphi(x_1)\varphi(x_2)\nonumber\\
&=\left(\prod_{k_{n}^{0}>0}\int d~\mathbf{Re}\widetilde{\varphi}(k_{n}) d~\mathbf{Im}\widetilde{\varphi}(k_{n})  \right) \left(\frac{1}{\mathrm{V}^2}\sum_{q,r}e^{-i(k_{q}\cdot x_1+k_{r}\cdot x_2)}\widetilde{\varphi}(k_{q})\widetilde{\varphi}(k_{r}) \right)\nonumber\\
&\times \exp \bigg[-\frac{i}{\mathrm{V}}\sum_{k_{n}^0>0}(m^2-k_{n}^2) |\widetilde{\varphi}(k_n)|^2+\frac{\lambda}{4!}\frac{i}{\mathrm{V}^3}\sum_{m,n,p}\widetilde{\varphi}(k_{m})\widetilde{\varphi}(k_{n})\widetilde{\varphi}(k_{p})\widetilde{\varphi}^{\star}(k_{m}+k_{n}+k_{p}) \bigg]\nonumber\\
    &=\left(\prod_{k_{n}^{0}>0}\int d~\mathbf{Re}\widetilde{\varphi}(k_{n}) d~\mathbf{Im}\widetilde{\varphi}(k_{n})  \right)\exp \bigg[-\frac{i}{\mathrm{V}}\sum_{k_{n}^0>0}(m^2-k_{n}^2) \left[(\mathbf{Re}\widetilde{\varphi}(k_{n}))^2+(\mathbf{Im}\widetilde{\varphi}(k_{n}))^2\right]\bigg]\nonumber\\
    &~~~~~\times \left(\frac{1}{\mathrm{V}^2}\sum_{q,r}e^{-i(k_{q}\cdot x_1+k_{r}\cdot x_2)} (\mathbf{Re}\widetilde{\varphi}(k_q)+i\mathbf{Im}\widetilde{\varphi}(k_q))(\mathbf{Re}\widetilde{\varphi}(k_r)+i\mathbf{Im}\widetilde{\varphi}(k_r))\right)\nonumber\\
    &~~~~~\times \bigg(1+\frac{\lambda}{4!}\frac{i}{\mathrm{V}^3}\sum_{m,n,p}\left(\mathbf{Re}\widetilde{\varphi}(k_{m})+i\mathbf{Im}\widetilde{\varphi}(k_{m})\right)\left(\mathbf{Re}\widetilde{\varphi}(k_{n})+i\mathbf{Im}\widetilde{\varphi}(k_{n})\right)\nonumber\\
&~~~~~\left(\mathbf{Re}\widetilde{\varphi}(k_{p})+i\mathbf{Im}\widetilde{\varphi}(k_{p})\right)\left(\mathbf{Re}\widetilde{\varphi}(k_{m}+k_{n}+k_{p})-i\mathbf{Im}\widetilde{\varphi}(k_{m}+k_{n}+k_{p})\right)+\mathcal{O}(\lambda^2)\bigg)
\end{align}
The tree level contribution to the above integral is given by 
\begin{align}
    &\mathrm{Num}[\langle \varphi(x_1)\varphi(x_{2})\rangle]\bigg|_{\mathcal{O}(\lambda^0)}\nonumber\\
    &=\left(\prod_{k_{n}^{0}>0}\int d~\mathbf{Re}\widetilde{\varphi}(k_{n}) d~\mathbf{Im}\widetilde{\varphi}(k_{n})  \exp \bigg[-\frac{i}{\mathrm{V}}(m^2-k_{n}^2) \left[(\mathbf{Re}\widetilde{\varphi}(k_{n}))^2+(\mathbf{Im}\widetilde{\varphi}(k_{n}))^2\right]\bigg]\right)\nonumber\\
    &~~~~~\times \left(\frac{1}{\mathrm{V}^2}\sum_{q,r}e^{-i(k_{q}\cdot x_1+k_{r}\cdot x_2)} (\mathbf{Re}\widetilde{\varphi}(k_q)+i\mathbf{Im}\widetilde{\varphi}(k_q))(\mathbf{Re}\widetilde{\varphi}(k_r)+i\mathbf{Im}\widetilde{\varphi}(k_r))\right)
\end{align}
From the structure of the integrand it is obvious that for the values of $k_{q}\neq k_{r}$ the integral evaluates to zero. The result may become non vanishing for $k_{q}=\pm k_{r}$. Below we analyse these two cases separately.
\vskip.2in
\noindent{\textbf{Case 1:} {$\boldsymbol{k_{q} =+ k_{r}}$}
\begin{align}
    &\mathrm{Num}[\langle \varphi(x_1)\varphi(x_{2})\rangle]\bigg|^{\mathrm{Case ~1}}_{\mathcal{O}(\lambda^0)}\nonumber\\
    &=\left(\prod_{k_{n}^{0}>0}\int d~\mathbf{Re}\widetilde{\varphi}(k_{n}) d~\mathbf{Im}\widetilde{\varphi}(k_{n})  \exp \bigg[-\frac{i}{\mathrm{V}}(m^2-k_{n}^2) \left[(\mathbf{Re}\widetilde{\varphi}(k_{n}))^2+(\mathbf{Im}\widetilde{\varphi}(k_{n}))^2\right]\bigg]\right)\nonumber\\
    &~~\times \left(\frac{1}{\mathrm{V}^2}\sum_{q}e^{-ik_{q}\cdot (x_1+x_2)} ((\mathbf{Re}\widetilde{\varphi}(k_q))^2-(\mathbf{Im}\widetilde{\varphi}(k_q))^2+2i\mathbf{Re}\widetilde{\varphi}(k_q)\mathbf{Im}\widetilde{\varphi}(k_q))\right)
\end{align}
The third term $\mathbf{Re}\widetilde{\varphi}(k_q)\mathbf{Im}\widetilde{\varphi}(k_{q})$ makes the integrand odd and hence doesn't contribute to the integral. The remaining two terms viz. $(\mathbf{Re}\widetilde{\varphi}(k_{q}))^2$ and $(\mathbf{Im}\widetilde{\varphi}(k_{q}))^2$ contribute equally to the integral and hence the net result vanishes. 
\vskip.2in
\noindent{\textbf{Case 2:} {$\boldsymbol{k_{q} =- k_{r}}$}
\begin{align}
    &\mathrm{Num}[\langle \varphi(x_1)\varphi(x_{2})\rangle]\bigg|^{\mathrm{Case ~2}}_{\mathcal{O}(\lambda^0)}\nonumber\\
    &=\left(\prod_{k_{n}^{0}>0}\int d~\mathbf{Re}\widetilde{\varphi}(k_{n}) d~\mathbf{Im}\widetilde{\varphi}(k_{n})  \exp \bigg[-\frac{i}{\mathrm{V}}(m^2-k_{n}^2) \left[(\mathbf{Re}\widetilde{\varphi}(k_{n}))^2+(\mathbf{Im}\widetilde{\varphi}(k_{n}))^2\right]\bigg]\right)\nonumber\\
    &~~~~~~~~~~~~\times \left(\frac{1}{\mathrm{V}^2}\sum_{q}e^{-ik_{q}\cdot (x_1-x_2)} ((\mathbf{Re}\widetilde{\varphi}(k_q))^2+(\mathbf{Im}\widetilde{\varphi}(k_q))^2)\right)\nonumber\\
    &=\frac{1}{\mathrm{V}^2}\sum_{q}e^{-ik_{q}\cdot (x_1-x_2)}\left(\prod_{k_{n}^0>0} \frac{-i\pi \mathrm{V}}{m^2-k_{n}^2}\right) \frac{-i \mathrm{V}}{m^2-k_{q}^2}
\end{align}
The factor similar to that in the parentheses of last line also appears from the denominator of the path integral (\ref{2ptCorrelator}) and hence gets canceled out once we take into account the denominator as well. Now, if we take the continuum limit (of the remaining terms), $\mathrm{V}\rightarrow \infty$ we get the tree level contribution to be
\begin{align}
    \mathrm{Num}[\langle \varphi(x_1)\varphi(x_2)\rangle]\bigg|_{\mathcal{O}(\lambda^0)}&=\left(\prod_{k_{n}^0>0} \frac{-i\pi \mathrm{V}}{m^2-k_{n}^2}\right) \int \frac{d^4 k}{(2\pi)^4}\frac{i e^{-ik\cdot(x_1-x_2)}}{k^2-m^2}. 
\end{align}
which is nothing but the Feynman propagator upto an overall factor. In a similar vein one can analyse the quantum corrections. We follow this method extensively in Section \ref{sec:PIinMth} to evaluate one point function with respect to an excited state.

\section{Duality rules}
\label{DualitYRulES}
\textbf{T-duality:}
Let us consider a closed string theory where a string is moving in 9-dimensional Minkowski space-time ($\mathbb{R}^{8,1}$) times a circle ($S^1$) of radius $R$. There is a $\mathbb{Z}_{2}$  $\mathrm{T}-$duality transformation that inverts the radius of the circle $R\rightarrow\tilde{R}=\frac{\alpha^\prime}{R}$. This mapping of $R\rightarrow \tilde{R}$ leaves the mass formula of the string invariant provided the Kaluza-Klein excitation number is exchanged with the string winding number. 
The above transformation is usually referred to as Buscher's transformation. There is a generalization of it which transforms solution of type $\mathrm{IIA}$ string on a background with one isometry to a type $\mathrm{IIB}$ string solution on a background with one isometry. For the Neveu-Schwarz/Neveu-Schwarz sector this transformation is essentially that of Buscher's but it interchanges the R-R fields of type $\mathrm{IIA}$ with those of the R-R fields of type $\mathrm{IIB}$ string theory. 

We denote the isometry direction to be $y$, the remaining directions as $x^{\mu}$ and all the directions as $x^{\alpha}=(x^{\mu},y)$. All fields have to be independent of $y$ coordinate. Below we list how the fields of type $\mathrm{IIA}$ superstring theory get mapped to that of the fields of type $\mathrm{IIB}$ superstring theory under a $\mathrm{T}-$duality performed along $y$ direction  \cite{bergshoeff1995duality}:
    \begin{align}
 \tilde{{g}}_{\mu \nu}&=\hat{g}_{\mu \nu}-\left(\hat{g}_{y\mu}\hat{g}_{y\nu}-{B}^{}_{y\mu}{B}^{}_{y\nu}\right)/\hat{g}_{yy}\\
\tilde{g}_{y\mu}&={B}^{}_{y\mu}/\hat{g}_{yy}\\
\tilde{g}_{yy}&=1/\hat{g}_{yy}\\
{D}_{y\mu \nu \rho}&=\frac{3}{8}\bigg[{C}_{\mu \nu \rho}-{A}_{[\mu}{B}^{}_{\nu \rho ]}+\hat{g}_{y[\mu}{B}^{}_{\nu \rho]}{A}_{y}/\hat{g}_{yy}-\frac{3}{2}\hat{g}_{y[\mu}{C}_{\nu \rho]y}/\hat{g}_{yy}\bigg] \\
{\mathcal{B}}^{(2)}_{\mu \nu}&=\frac{3}{2}{C}_{\mu \nu y}-2 {A}_{[\mu}{B}^{}_{\nu]y}+2\hat{g}_{y[\mu}{B}^{}_{\nu]y}{A}_{y}/\hat{g}_{yy}\\
{\mathcal{B}}^{(2)}_{y\mu}&=-{A}_{\mu}+{A}_{y}\hat{g}_{y\mu}/\hat{g}_{yy}\\
{\mathcal{B}}^{(1)}_{\mu \nu}&={B}^{}_{\mu \nu}+2 \hat{g}_{y[\mu}{B}^{}_{\nu]y}/\hat{g}_{yy}\\
{\mathcal{B}}^{(1)}_{y\mu}&=\hat{g}_{y\mu}/\hat{g}_{yy}\\
{l}&={A}_{y}\\
{\varphi}&={\phi}-\frac{1}{2}\log (\hat{g}_{yy})
    \end{align}
Likewise, a  $\mathrm{T}-$duality performed along $y$ direction maps the fields of type $\mathrm{IIB}$ superstring theory onto the fields of type $\mathrm{IIA}$ superstring theory in the following way: 
    \begin{align}
        \hat{{g}}_{\mu \nu}&=\tilde{g}_{\mu \nu}-\left(\tilde{g}_{y\mu}\tilde{g}_{y\nu}-{\mathcal{B}}^{(1)}_{y\mu}{\mathcal{B}}^{(1)}_{y\nu}\right)/\tilde{g}_{yy}\\
    \hat{g}_{y\mu}&={\mathcal{B}}^{(1)}_{y\mu}/\tilde{g}_{yy}\\
\hat{g}_{yy}&=1/\tilde{g}_{yy}\\
{C}_{y\mu \nu}&=\frac{2}{3}\left[{\mathcal{B}}^{(2)}_{\mu\nu}+2 {\mathcal{B}}^{(2)}_{y[\mu}\tilde{g}_{\nu]y}/\tilde{g}_{yy}\right]\\
{C}_{\mu \nu \rho}&=\frac{8}{3}{D}_{y\mu \nu \rho}+\epsilon^{ij}{\mathcal{B}}^{(i)}_{y[\mu}{\mathcal{B}}^{(j)}_{\nu \rho]}+\epsilon^{ij}{\mathcal{B}}^{(i)}_{y[\mu}{\mathcal{B}}^{(j)}_{|y|\nu}\tilde{g}_{\rho]y}/\tilde{g}_{yy}\\
{B}^{}_{\mu \nu}&={\mathcal{B}}^{(1)}_{\mu \nu}+2 {\mathcal{B}}^{(1)}_{y[\mu}\tilde{g}_{\nu]y}/\tilde{g}_{yy}\\
{B}^{}_{y\mu}&=\tilde{g}_{y\mu}/\tilde{g}_{yy}\\
{A}_{\mu}&=-{\mathcal{B}}^{(2)}_{y\mu}+{l}{\mathcal{B}}^{(1)}_{y\mu}\\
{A}_{y}&={l}\\
{\phi}&={\varphi}-\frac{1}{2}\log (\tilde{g}_{yy})
    \end{align} 
In the above rules, fields of type $\mathrm{IIA}$ and $\mathrm{IIB}$ theory are denoted as
\begin{align}
    \mathrm{NS-NS~sector ~of ~type~ IIA :}~&(\hat{g}_{\alpha\beta},~B^{}_{\alpha\beta},~\phi)\nonumber\\
    \mathrm{R-R ~sector~ of ~type~ IIA :}~&(A_{\alpha},~C_{\alpha\beta\gamma})\nonumber\\
        \mathrm{NS-NS~sector ~of ~type~ IIB :}~&(\tilde{g}_{\alpha\beta},~\mathcal{B}^{(1)}_{\alpha\beta},~\varphi)\nonumber\\
    \mathrm{R-R ~sector~ of ~type~ IIB :}~&(l,~\mathcal{B}^{(2)}_{\alpha\beta},~D_{\alpha\beta\gamma\delta}).\nonumber
\end{align}

\noindent
    \textbf{S duality:} S-duality is a transformation that relates a strongly coupled theory to a weakly coupled one. Unlike T-duality, it is a non-perturbative symmetry. For example, the type $\mathrm{I}$ and $SO(32)$ superstring theories are $\mathrm{S}$ dual to each other as they are equivalent when the following conditions hold
    \begin{align}\Phi_{\mathrm{Het}}=-&\Phi_{\mathrm{I}},~~~g_{\mu\nu}^{\mathrm{Het}}=e^{-\Phi_{\mathrm{I}}}g^{\mathrm{I}}_{\mu\nu},~B_{\alpha\beta}^{\mathrm{Het}}=C_{\alpha\beta}^{\mathrm{I}},~~A_{\alpha}^{\mathrm{Het}}=A_{\alpha}^{\mathrm{I}}.
    \end{align}
It implies that the coupling constant of type $\mathrm{I}$ superstring theory is the reciprocal of the coupling constant of heterotic $SO(32)$ theory $g^{\mathrm{I}}=1/g^{\mathrm{Het}},$ in other words when type $\mathrm{I}$ theory is strongly coupled, the $\mathrm{S}$ dual $SO(32)$ heterotic theory is weakly coupled and vice versa and the Weyl rescaled metric of type $\mathrm{I}$ theory is the metric of the dual heterotic $SO(32)$ theory 
\begin{align}
ds^{2}_{\text{Het}}&=e^{-\phi_{\mathrm{I}}}ds^{2}_{\text{I}}=\frac{1}{g^{\mathrm{I}}}ds^2_{\mathrm{I}}.
    \end{align}
  \noindent 
  \textbf{Dimensional reduction:} Low-energy limit of the type $\mathrm{IIA}$ superstring theory is the type $\mathrm{IIA}$ supergravity theory and it can be obtained from $\mathcal{N}=1,~d=11$ supergravity theory by dimensional reduction.  Bosonic fields of the 11 dimensional theory consist of the metric $\mathbf{g}_{\mathrm{MN}}$ and a three-form potential $\mathbf{C}_{\mathrm{MNP}}$. The coordinates are denoted as $x^{\mathrm{M}}=(x^\alpha, z),~\alpha=0,1,...9.$ In order to dimensionally reduce the 11th direction (viz. the $z$ coordinate) one has to compactify that direction on a circle of radius $R_{11}$ and assume all fields to be independent of $z$ and consequently the fields of ten-dimensional type $\mathrm{IIA}$ supergravity can be expressed in terms of the eleven-dimensional ones as
    \begin{align}
    \hat{g}_{\alpha\beta}&=(\mathbf{g}_{zz})^{1/2}~\left(\mathbf{g}_{\alpha\beta}-\mathbf{g}_{\alpha z } \mathbf{g}_{\beta z }/\mathbf{g}_{zz}\right),\\
{A}_{\alpha}&=\mathbf{g}_{\alpha z}/\mathbf{g}_{zz}\\
    {\phi}&=\frac{3}{4}\log \left(\mathbf{g}_{zz}\right)\label{DilatonIIA}\\
{C}_{\alpha\beta\gamma}&=\mathbf{C}_{\alpha\beta\gamma}\\
{B}^{}_{\alpha\beta}&=\frac{3}{2}\mathbf{C}_{\alpha\beta z}.
    \end{align}
That is, dimensional reduction of the metric gives rise to a ten -dimensional metric $\hat{g}_{\alpha \beta}$, a vector field $A_{\alpha}$, and a scalar field (dilaton) $\phi$ while the dimensional reduction of the three-form potential gives rise to a ten-dimensional three form $C_{\alpha\beta\gamma}$, and a two-from $B_{\alpha\beta}$. From (\ref{DilatonIIA}) it turns out that 
\begin{align}
R_{11}=\sqrt{\mathbf{g}_{zz}}l_{p}=e^{2\phi/3}l_{p}=l_{s}g_{s},\label{R_{11}}    
\end{align}
 where $g_{s}=e^{\phi}$ is the type $\mathrm{IIA}$ coupling constant. Thus the perturbative regime ($g_{s}\rightarrow 0$) of type $\mathrm{IIA}$ theory corresponds to the $R_{11}\rightarrow 0$ limit. 

\noindent
        \textbf{Dimensional Uplift:} The strong coupling limit of type $\mathrm{IIA}$ theory, that is the limit $g_{s}\rightarrow \infty$, leads to decompactification of the 11th circle of radius $R_{11}$ (\ref{R_{11}}). Hence in this limit the ten dimensional type $\mathrm{IIA}$ supergravity theory gets uplifted to the eleven dimension
        al supergravity theory. The bosonic fields of the eleven dimensional theory can be expressed in terms of the ten dimensional ones as 
    \begin{align}
\mathbf{g}_{\alpha\beta}&= e^{-\frac{2}{3}{\phi}} \hat{g}_{\alpha\beta} + e^{\frac{4}{3}{\phi}}{A}_{\alpha}{A}_{\beta}\\
\mathbf{g}_{\alpha z}&=e^{\frac{4}{3}{\phi}}{A}_{\alpha}\\
\mathbf{g}_{zz}&=e^{\frac{4}{3}{\phi}}\\
\mathbf{C}_{\alpha\beta\gamma}&=C_{\alpha\beta\gamma}\\
\mathbf{C}_{\alpha\beta z}&=\frac{2}{3}{B}_{\alpha\beta}.
\end{align}

\section{Volume of the internal manifold}
\label{sec:AppendixD}
Consider the six-dimensional internal space metric (\ref{HeTerotiCSO(32)MetrIc}) written in Kaluza--Klein (KK) form
\begin{align}
   ds^2_{6}&= \delta_{\alpha\beta}\left(dy^{\alpha}+\mathcal{B}^{(1)\alpha}_{~~~m} dy^{m}\right)\left(dy^{\beta}+\mathcal{B}^{(1)\beta}_{~~~n} dy^{n}\right)+\mathrm{H}^{4}(y)\mathrm{F}_{1}(t)\mathrm{F}_{2}(t)g_{mn}dy^{m}dy^{n}
\end{align}
where $\alpha,\beta=4,5$ and $m,n=6,\ldots,9$.
The vectors $\mathcal{B}^{(1)\alpha}{}_m$ are the KK gauge fields describing the twisting of
the two-dimensional fiber ($\tilde{\mathbf{S}}^1_{4}\times \tilde{\mathbf{S}}^1_5$) over the four-dimensional base $\mathcal{M}_4$. Let us define
$$
A(y,t) \equiv \mathrm{H}^4(y)\,\mathrm{F}_1(t)\, \mathrm{F}_2(t),~~~\mathrm{where}~y\equiv y^m\in \mathcal{M}_4.
$$
The six dimensional metric can be written in block form as
$$
g_{MN}
=
\begin{pmatrix}
\delta_{\alpha \beta} &~~ \mathcal{B}^{(1)\gamma}{}_n \delta_{\alpha \gamma}\\
\mathcal{B}^{(1)~\gamma}_{~~m} \delta_{\gamma \beta} & ~~A g_{mn} + \mathcal{B}^{(1)}_{~~m\gamma }\delta^{\gamma \tilde{\gamma} } \mathcal{B}^{(1)}_{~~\tilde{\gamma} n} 
\end{pmatrix},~~\mathrm{where}~M,N=4,5,6,\cdots ,9.
$$
Using the standard block determinant identity
\begin{align}
\det
\begin{pmatrix}
\mathrm{P} & \mathrm{Q} \\
\mathrm{Q}^{\mathrm{T}} &~~ \mathrm{S}
\end{pmatrix}= 
\det(\mathrm{P} )\,\det( \mathrm{S} - \mathrm{Q}^{\mathrm{T}} \mathrm{P}^{-1} \mathrm{Q}),
\end{align}
for invertible $\mathrm{P}$ and identifying 
\begin{align}
&\mathrm{P}_{2\times 2}=\delta_{\alpha \beta},~~\mathrm{Q}_{2\times 4}=\mathcal{B}^{(1)}_{~~\alpha n},~~\mathrm{S}_{4\times 4}=A g_{mn} + \mathcal{B}^{(1)}_{~~m\gamma }\delta^{\gamma \tilde{\gamma} } \mathcal{B}^{(1)}_{~~\tilde{\gamma} n}
\end{align}
we find 
\begin{align}
    \mathrm{det}(g_{MN})&=\mathrm{det}(\delta_{\alpha \beta }) \mathrm{det}(A g_{mn} + \mathcal{B}^{(1)}_{~~m\gamma }\delta^{\gamma \tilde{\gamma} } \mathcal{B}^{(1)}_{~~\tilde{\gamma} n}-\mathcal{B}^{(1)}_{~~m\alpha }\delta^{\alpha {\beta} } \mathcal{B}^{(1)}_{~~{\beta} n} )\nonumber\\
    &=\mathrm{det}(\delta_{\alpha \beta } ) \mathrm{det}(A g_{mn})=A^4 \mathrm{det}(g_{mn}).
\end{align}
It shows that the contribution due to the KK vector cancels out. Therefore, the internal six-dimensional volume is
$$
\mathrm{Vol}_6(t)
= \int d^6y\,\sqrt{g_6(y,t)}
= \int d^6y\;\,\big(\mathrm{H}^4(y) \mathrm{F}_1(t)\mathrm{F}_2(t)\big)^2\,\sqrt{g_4(y)}.
$$
\subsubsection*{Differential-form interpretation}
With the introduction of the shifted one-forms
$$
\eta^\alpha = dy^\alpha + \mathcal{B}^{(1)\alpha}_{~~~m} dy^m,
$$
the metric takes the form
$$
ds_6^2 =\delta_{\alpha\beta}\,\eta^\alpha \eta^\beta+ A(y,t) g_{mn}(y) dy^m dy^n 
$$
The six-dimensional volume form is proportional to 
\begin{align}
    \mathrm{Vol}_6 &\sim  \eta^4 \wedge \eta^5 \wedge dy^6\wedge dy^7 \wedge dy^8 \wedge dy^9 \nonumber\\
    &= (dy^4 + \mathcal{B}^{(1)\alpha}_{~~~m} dy^m )\wedge (dy^5 + \mathcal{B}^{(1)5}_{~~~n} dy^n)\wedge dy^6\wedge dy^7 \wedge dy^8\wedge dy^9\nonumber\\
    &=dy^4 \wedge dy^5 \wedge dy^6\wedge dy^7 \wedge dy^8 \wedge dy^9,
\end{align}
since the wedge product is completely anti-symmetric, the terms involving the KK vectors do not contribute to the volume form.

\subsubsection*{Geometric interpretation}

The KK vectors $B^\alpha{}_m$ act as connection one-forms on the fiber bundle.
They describe how the internal two-torus is twisted over the four-dimensional
base manifold. A connection affects parallel transport and curvature, but not
local volumes. Consequently, the internal volume depends only on the base metric
$A g_{mn}$ and the fiber metric $\delta_{\alpha\beta}$, and not on the KK
connection.

    \bibliographystyle{unsrt}
\bibliography{references.bib}

\end{document}